\documentclass[lettersize,journal]{IEEEtran}
\usepackage[figuresright]{rotating}
\usepackage{graphicx}
\usepackage{enumerate}
\usepackage{caption}
\usepackage{algorithm}
\usepackage{algorithmic}
\usepackage{color}
\usepackage{subfigure}
\usepackage{graphicx}
\usepackage{stfloats}
\usepackage{float}
\usepackage{amsthm}
\usepackage{amsmath}
\usepackage{amsfonts}
\usepackage{booktabs}
\usepackage{multirow}
\usepackage{amssymb}
\usepackage{diagbox}
\usepackage{color}
\usepackage{pifont}
\usepackage{balance}
\usepackage{picins}
\usepackage{bbding}
\usepackage[utf8]{inputenc}

\newtheorem{defi}{Definition}
\newtheorem{thm}{Theorem}

\begin{document}

\title{Geo-MOEA: A Multi-Objective Evolutionary Algorithm with Geo-obfuscation for Mobile Crowdsourcing Workers}

\author{S.~Zhang,
	T.~Zhang,
    Z.~Chen,
	N. Xiong
\IEEEcompsocitemizethanks{
\IEEEcompsocthanksitem S. Zhang and T. Zhang are with School of Computer Science and Technology, Anhui University, Hefei 230601, China. 
E-mail: szhang@ahu.edu.cn (S. Zhang), e20301311@stu.ahu.edu.cn (T. Zhang)
\IEEEcompsocthanksitem Z. Chen is with Software Engineering Institute, East China Normal University, Shanghai 200062, China.
E-mail: zhlchen@sei.ecnu.edu.cn
\IEEEcompsocthanksitem N. Xiong is with the National Engineering Research Center for ELearning, Central China Normal University, Wuhan 430079, China. 
E-mail: nnxiong10@yahoo.com}}

\markboth{Journal of \LaTeX\ Class Files,~Vol.~14, No.~8, October~2022}%
{Shell \MakeLowercase{\textit{et al.}}: A Sample Article Using IEEEtran.cls for IEEE Journals}

\maketitle

\begin{abstract}
The rapid development of mobile Internet and sharing economy brings the prosperity of Spatial Crowdsourcing (SC). SC applications assign various tasks according to reported location information of task's requesters and outsourced workers (such as DiDi, MeiTuan and Uber). However, SC-servers are often untrustworthy and the exposure of users' locations raises privacy concerns. In this paper, we design a framework called Geo-MOEA (Multi-Objective Evolutionary Algorithm with Geo-obfuscation) to protect location privacy of workers involved on SC platform in mobile networks environment. We propose an adaptive regionalized obfuscation approach with inference error bounds based on geo-indistinguishability (a strong notion of differential privacy), which is suitable for the context of large-scale location data and task allocations. This enables each worker to report a pseudo-location that is adaptively generated with a personalized inference error threshold. Moreover, as a popular computational intelligence method, MOEA is introduced to optimize the trade-off between SC service availability and privacy protection while ensuring theoretically the most general condition on protection location sets for larger search space. Finally, the experimental results on two public datasets show that our Geo-MOEA approach achieves up to $20\%$ reduction in service quality loss while guaranteeing differential and geo-distortion location privacy.
\end{abstract}


\section{Introduction}
In recent years, due to the vigorous development of mobile networks (4G/5G) and dramatic proliferation of mobile devices, \emph{Spatial Crowdsourcing} (SC) (a.k.a. mobile crowdsourcing) \cite{KS12,TZZ20} has emerged
as a prevailing paradigm of crowdsourcing and has stimulated various
applications such as taxi calling (e.g., Didi Chuxing and Uber), food delivery (e.g., MeiTuan) and social network platforms  (e.g., Twitter and Microblogs) in the real world \cite{KS12}

Compared with conventional crowdsourcing, SC based on location services has led to an exponential growth in data collection and sharing of intelligent terminals. For the scenario of large-scale SC, a larger number of tasks are requested and allocated at the SC-server side among a wider range of potential workers for providing faster and efficient services.
However, the SC-server may be untrustworthy, and disclosing individual locations has serious
privacy implications \cite{TGF17,GZF18,HWC20}. In recent years, many techniques appear to protect workers' location information \cite{CEP17} while bringing about some more issues, like computational complexity and consuming time, to remedy.

Most of the current mechanisms utilize centralized differential privacy, which requests a third-party trusted entity to collect workers' actual locations. To et al. \cite{TGF17} introduced a trusted Cellular Service Provider (CSP) to partition the domain of worker locations. Indeed, CSP has also the risk of being invaded, which destroys individual sensitive information. To avoid relying on trusted third-party entities, some authors pay attentions to obfuscation mechanisms with \emph{Local Differential Privacy} (LDP) guarantee \cite{TSX18,Sho15,ZDC22}.

However, the existing solutions with LDP still have several limitations, which are summarized as follows.

(1) \textbf{Ignoring adversary's inference attack.} Some obfuscation mechanisms generate pseudo-locations with LDP guarantee via geo-indistinguishability \cite{TSX18}, while not resisting against the Bayesian inference attack by prior knowledge \cite{STT12}.

(2) \textbf{Single-objective optimization.} Generally, current privacy mechanisms adopt
single-objective optimization \cite{Sho15}.  In fact, we have to take into account the trade-off between two conflicting aspects: privacy level and service quality.

 (3) \textbf{Small-scale location domain.}  Recent LDP protection mechanisms \cite{Sho15,WYH19} mainly adopt linear programming to improve the allocation efficiency that result in much higher computational complexity, while a previous scheme DPIVE \cite{ZDC22} with Protection Location Sets (PLSs) can only be applied separately in small-scale disjoint domains.

\par To address the aforesaid issues, we design an adaptive obfuscation approach called `Geo-MOEA' (Multi-Objective Evolutionary Algorithm with Geo-obfuscation) based on geo-indistinguishability,
which achieves the distortion privacy guarantee on each worker location. Under our privacy framework, workers report their pseudo-locations with LDP directly to the server. 
The mechanism involving expected inference errors can effectively resist Bayesian attacks via prior knowledge. Moreover, the obfuscation scheme employs the genetic operator \cite{Holland92} to optimize the two conflicting metrics, expected inference error and service quality loss, for producing multiple optimized recommendations.

The main contributions are as follows:
\begin{enumerate}
\item 
    We consider a large-scale SC scenario where the worker wants to protect individual true location without third-party trusted entities. For this, we propose ``Geo-MOEA'', an obfuscation approach preserving differential and geo-distortion location privacy
    in all regions.
\item To the best of our knowledge, this is the first work to introduce a solution approach combining location privacy and MOEA to simultaneously optimize two conflicting metrics: service quality loss and average expected inference error. 
\item We confirm theoretically the most general condition on qualified protection location sets for larger search space, by which service quality loss can keep mainly decreasing trend for ascending privacy budget.
\item A series of experimental results show that our proposal is more effective than the existing obfuscation mechanisms. Providing multiple recommendation mechanisms for practical applications can reduce workers travel distance while protecting workers' location privacy in a global and scalable sense.
\end{enumerate}

The remainder of this paper is structured as follows. In Section \ref{sec:rela_work}, we conduct a survey of related work. Section \ref{sec:backg} introduces system model, goal and adversary model. Section \ref{sec:framework} describes our designed privacy approach Geo-MOEA. Experimental  evaluations are presented in Section \ref{sec:evaluation}. Finally, we conclude this
paper in Section \ref{sec:conclusion}.

\section{Related Work}\label{sec:rela_work}

We review related works in the direction of classical and differential privacy protection methods as well as multi-objective optimization.

\textbf{Classical privacy protection methods}. With the rapid development of mobile technologies, Spatial Crowdsourcing (SC) has attracted extensive attention from both the academia and the industry while the arising location privacy problem has also attracted extensive attention \cite{KS12,TZZ20}. 
For this, traditional privacy protection methods mainly include spatial anonymity and cryptography. Specifically, spatial anonymity technology often uses a spatial anonymity area to replace the user's accurate location information. As a common kind of spatial anonymity models in SC, 
this model requires that each of the released locations be indistinguishable from at least $k-1$ other locations \cite{YYC22}. It has the fatal weakness that $k$-anonymity is vulnerable to attacks of background knowledge and reference 
\cite{LS20}. Homomorphic encryption \cite{ZCQ20} is a good model to ensure the confidentiality of task's location policy \cite{YLL19} while it has disadvantages of large key size and low calculation efficiency. Hence, it is not practical for large-scale scenarios \cite{SW17}.

\textbf{Differential privacy protection}. The Differential Privacy (DP) method \cite{Dwo06} is a strong privacy concept that has been widely used for privacy protection recently. In particular, To et al. \cite{TGF17} develop centralized DP based Private Space Decomposition (PSD) to protect location privacy. Gong et al. \cite{GZF18} protect workers' location and reputation privacy by using reputation-based DP method. Both schemes rely on a trusted third-party entity to protect location privacy and some attacks may lead to the disclosure of true locations.

\textbf{Geo-indistinguishability and distortion privacy}. Local Differential Privacy (LDP) can help users to report their pseudo-locations. Based on geo-indistinguishability \cite{ABC13}, some LDP frameworks are proposed for SC scenario \cite{TSX18, WYH19, WZY20}. Expected inference error \cite{STL11} is a concept of distortion privacy complementary to geo-indistinguishability, and can provide strict privacy protection against Bayesian attacks. Shokri \cite{Sho15} combines both concepts to construct the Joint mechanism and optimizes utility using linear programming. Yu et al. \cite{YLP17} formally study their relationship and combine them by adding personalized error lower bounds. Later, DPIVE mechanism is presented and solves the previous privacy theory problem of intersections among protection location sets \cite{ZDC22}. Recently, Liu et al. \cite{LZS21} propose an obfuscation mechanism with the Gamma distribution and use a game-theoretic approach to maximize two-users' utilities while preserving desired location privacy radius. However, these works focus mainly on the optimization of utility and have no concern with global trade-off between the conflicting utility and distortion privacy.

\textbf{Multi-objective optimization}. With the rapid development of scientific and engineering areas, a variety of optimization problems containing multiple conflicting objectives have appeared \cite{BKS01}. This means that there does not exist a single solution optimizing all of the objectives. It is natural to find the Pareto optimal set that contains multiple solutions trading off between all of the objectives, where the improvement of one objective cannot be achieved without the deterioration of some other objectives. A number of swarm intelligence algorithms have been proposed to search for well-converged solutions by using conventional reproduction operators \cite{DPA02,TSZ21}, such as the genetic algorithm \cite{Holland92} and Particle Swarm Optimization (PSO) \cite{EK95,CK02}. 
To achieve better privacy protection and higher service quality, Zhang et al. \cite{ZYL18} provide a PSO anonymization method to accelerate the process of finding similar attributes. However, in many occasions the convergence of the canonical PSO is premature and it suffers from local minima \cite{DKC05}.
Zhang et al. \cite{ZXC19} present a Multi-Objective Evolutionary Algorithm (MOEA) framework to protect private information based on the hybrid elite selection strategy.
In order to obtain optimal task allocation with the differential-and-distortion geo-obfuscation, Wang et al. \cite{WYH19} use linear scaling to execute optimization of two concordant objectives while ignoring local inference error bounds.
In this paper, for the large-scale SC data scenario,
the MOEA approach is adopted to design Geo-MOEA,
which realizes the trade-off between conflicting distortion privacy and service quality for the first time.

\section{System Model and Definitions}\label{sec:backg}

This section first presents our system model, goal and properties
to be achieved, as well as an adversary model, then we provide the problem statement and the concepts of multi-objective optimization.

\subsection{Spatial Crowdsourcing Model}

\emph{Spatial Crowdsourcing} (SC) \cite{KS12} is a new type of platform for efficient and scalable data collection of online crowdsourcing in the era of mobile Internet and sharing economy, which requires the worker to travel to a specific location to complete the task. Usually, there are two categories of task assignment modes based on how workers are matched to tasks, Worker Selected Tasks (WST) and Server Assigned Tasks (SAT). WST  doesn't require workers to share their locations with the SC-server and is more friendly to users' privacy. The SAT mode can better optimize the overall task efficiency while requiring the SC-server to know the workers' locations, which poses a privacy threat.

We attempt to combine the advantages of both models. The SAT model is used to assign tasks for efficient running performance, while still protecting users' location privacy. Requesters submit tasks that include locations, online workers send their pseudo-locations to the server that assigns tasks to nearby workers. Table \ref{tlb:notation} summarizes the notations used in our work.

\begin{table} \small
\caption{Summary of Notations}
\label{tlb:notation}
        \centering
		\begin{tabular}{ll}
			\toprule
			Symbol  & \quad\quad\quad\quad\quad\quad\quad Definition \\
			\midrule
                        $\epsilon_0,\ \epsilon_k$      & Total privacy budget and privacy level on $\Phi_{k}$
\\
						$f(x'|x)$       & Probability of reporting location $x'$ for the actual $x$
\\
                        $d(x,y)$        & Travel distance between the locations $x$ and $y$
                        \\
                        $\epsilon_g,\ \theta$ & Geo-indistinguishability parameter and its deviation
                        \\
                        $\Phi$    & Protection Location Set (PLS)
                        \\
                        $D(\Phi)$ & Diameter of $\Phi$ (the largest distance between points)
                        \\
                        $\Delta q$   & Sensitivity of the scoring function $q$
                        \\
                        $\pi$         & Prior distribution
                        \\
                        $ExpEr(x')$    & Conditional expected inference error for reported $x'$
                        \\
                        $E_m$         & Minimum (local) inference error
                        \\
                        $\hat{x}$    & The location estimated by optimal inference attack
                        \\
                        ExpErr        & Unconditional expected inference error
                        \\
                        QLoss          & Service quality loss
                        \\
                        HV           & Hypervolume indicator
                        \\
			\bottomrule
		\end{tabular}
	\end{table}

\subsection{Goal - Achieving Differential Privacy}
\emph{Differential Privacy} (DP) \cite{Dwo06} provides provable privacy protection for users. Regardless of the adversary's prior knowledge, it ensures that any adversary can not confirm the presence of a particular individual in the processed data set. As for location privacy notion, geo-indistinguishability \cite{ABC13} has been widely used to achieve differential privacy. In order to achieve DP on each Protection Location Set (PLS), we adopt the loose definition on the PLS introduced by \cite{ZDC21}.

\begin{defi}[$(\epsilon_g,\theta)$-Geo-indistinguishability within PLS \cite{ZDC21}]\label{def:geoI}
Suppose the probability distribution $f(\cdot|\cdot)$ for a mechanism $\mathcal{A}$ satisfies, for any $x, y$ in PLS $\Phi$ with the reporting range $\mathcal{Y}\subset\mathcal{X}$,
\begin{equation}
\frac{f(x^\prime|x)}{f(x^\prime|y)}\leq  e^{\epsilon_g \left(d(x,y)+\theta\right)},\ \ \ \  x^\prime\in \mathcal{Y},
\end{equation}
then $\mathcal{A}$ is $(\epsilon_g,\theta)$-geo-indistinguishable on $\Phi$. If $\theta=0$, we say that $\mathcal{A}$ gives $\epsilon_g$-geo-indistinguishability on $\Phi$ without deviation.
\end{defi}

This means that any two locations geographically close to each other have similar probability distributions, so that they are theoretically indistinguishable to each other for the adversary. Particularly, the geo-indistinguishability parameter $\epsilon_g$ is related to the privacy budget and the circular region usually centered at the user's location. If all locations in a region have similar release distributions $f$, then the true location can be hidden in this region and the whole locations in this region are called Protection Location Set (PLS). Afterwards,  we define $\epsilon=\epsilon_g\cdot (D+\theta)$, where $D$ denotes the diameter of the PLS region. By Definition \ref{def:geoI}, the mechanism $f$ satisfies $\epsilon$-DP on PLS as follows.

\begin{defi}[$\epsilon$-DP on PLS \cite{YLP17}]\label{def:eps_DPPLS}
A randomized location obfuscation mechanism
$f(\cdot|\cdot)$ satisfies $\epsilon$-differential privacy on protection location
set $\Phi$, if for any locations $x, y \in\Phi$, and any output $x'\in\mathcal{Y}\subset\mathcal{X}$, we have
\begin{equation}\label{defi:eps-DP}
\frac{f(x^\prime|x)}{f(x^\prime|y)}
\leq  e^{\epsilon}.
\end{equation}
\end{defi}


As above, the local DP is achieved via geo-indistinguishability operating on a single dataset $\mathcal{X}$,
we use the notion $\epsilon$-DP instead of $\epsilon$-LDP as in the previous literatures \cite{Sho15,WYH19}. This paper continues to consider geo-indistinguishability and local DP on each PLS and some more general cases. Any two locations from the same PLS are regarded to be neighboring.
Further, as a quite
general approach preserving DP, the exponential mechanism usually involves a scoring function $q:\ \Phi\times\mathcal{Y}\rightarrow \mathbb{R}$ which assigns a real-valued score to each point-point pair, ideally such that each point $x'$ from the reporting range $\mathcal{Y}$ having better utility receives a higher real score for a given $x\in\Phi$.

\begin{defi}[Sensitivity on PLS \cite{DMN06}]\label{def:Sensitivity}
Let $x_1,\ x_2\in\mathcal{X}$ be any pair of neighboring locations (in PLS $\Phi$). The sensitivity of the scoring function $q$ on $\Phi$ is given by, its maximal change, with $x'\in\mathcal{Y}$,
\begin{equation}
\Delta q = \sup_{x_1,\, x_2,\, x'} \left| {q(x_1,x') - q(x_2,x')} \right|.
\end{equation}
\end{defi}

\begin{defi}[Exponential Mechanism on PLS \cite{MT07,DR14}]\label{def:Exponential}
Given a scoring function $q$ on $\Phi\times\mathcal{Y}$,
 the exponential mechanism
$\mathcal{M}(x,q)$ outputs $x' \in \mathcal{Y}$  with probability proportional to $\exp \left(\frac{\epsilon q(x,x')}{2\Delta q}\right)$.
\end{defi}

\subsection{Bayesian Adversary Model}
As all the Location-Based Service (LBS) providers require the access permission to users' location data, the location privacy is  potentially disclosed to untrusted entities. Knowing user's locations, an adversary can perform a broad spectrum of attacks. Thus, ensuring location privacy is foremost for LBS applications. A common method is location perturbation, which generates a pseudo-location based on the true location and the user sends it to the server.
Following \cite{HBM17, YLP17}, we suppose that the discretized location set $\mathcal{X}$ represents the user's possible locations.
An obfuscation mechanism takes the user's real location $x$ from $A$ as input and randomly chooses a
pseudo-location $x^\prime$ from $O$ with the probability $f(x^\prime|x)$:

\begin{equation}
f(x^\prime|x)=\text{Pr}(O=x^\prime|A=x), \ \ \ \ \ x,\ x^\prime\in \mathcal{X}.
\end{equation}

In general, the objective of obfuscation mechanisms is mainly to design suitable probability distribution $f(\cdot|\cdot)$.
Similar to \cite{YLP17}, we assume that the adversary has prior knowledge about user's location, which is regarded as background knowledge to perform inference attacks \cite{STT12}. The adversary collects background knowledge by building a prior probability distribution  $\pi$  on  $\mathcal{X}$. The prior probability $\pi$ can be obtained via population density, historical locations and so on. The adversary is also informed of the obfuscation mechanism $f$. 
In the current scenario, the adversary infers the user's real location $x$ under the Bayesian adversary model. After the user reports her/his pseudo-location $x'\in\mathcal{X}$, the adversary computes the probability that each apriori location $x\in\mathcal{X}$ is the true location in the condition of generating $x'$, i.e., the posterior probability distribution $\text{Pr}(x|x')$, by

\begin{equation}\label{formu:post-dist}
\text{Pr}(x|x')=\frac{\text{Pr}(x,x')}{\text{Pr}(x')} =\frac{\pi(x)f(x^\prime|x)}{\sum_{x\in \mathcal{X}}\pi(x)f(x^\prime|x)}.
\end{equation}

Afterwards, a Bayesian adversary can launch optimal inference attack \cite{STT12} to get the estimated location $\hat{x}$ which has the minimal expected distortion, i.e.,

\begin{equation}\label{eq:attack x1}
\hat{x}=\mathop{\arg\min}\limits_{y\in \mathcal{X}}\sum_{x\in \mathcal{X}}\text{Pr}(x|x^\prime)d(y,x).
\end{equation}


Now we come to the conflicting metrics. The distortion privacy level is measured by unconditional expected inference error \cite{STL11},


\vspace{-2mm}
\begin{equation}
ExpErr=\sum_{x'\in \mathcal{X}}\mathop{\min}\limits_{\hat{x}\in \mathcal{X}}\sum_{x\in \mathcal{X}}\pi (x)f(x'|x)d(\hat{x},x).
\end{equation}
The service quality loss is  defined by the unconditional expected distance between true and perturbed locations \cite{Sho15},
\begin{equation}
QLoss=\sum_{x\in \mathcal{X}}\sum_{x'\in \mathcal{X}}\pi (x)f(x'|x)d(x',x).
\end{equation}

\subsection{Problem Statement}
With the application of mobile devices, location-based SC services are spreading to many cities. The rapid growth of service data brings two serious challenges.

Firstly, 
task allocation is often oriented to large-scale user location privacy data (such as DiDi, MeiTuan and Uber).  
The existing geo-distortion mechanisms (without a trusted third party)
 are only suitable to generate pseudo-locations with probability distributed in a small-scale domain \cite{ZDC22,Sho15}. Adding the domain scale would increase greatly the quality losses while boosting the computational complexity.

It is desired to develop a LDP framework to avoid the constraint of small-region boundaries during location reporting and task allocations.
In detail, a new domain partition method
for generating families of PLSs is necessary for achieving LDP based on geo-indistinguishability and implementing distortion privacy against Bayesian attacks.

Secondly, most of existing privacy mechanisms ignore the balance between privacy protection and practical quality. Currently, there is no standard optimization of conflicting metrics for SC in a global sense when workers have personalized requirements of distortion  privacy \cite{ZDC22,WYH19}.

This motivates us to combine location privacy with Multi-Objective Evolutionary Algorithm (MOEA) involving Geo-obfuscation. The problem formulation is as follows:

Given some discrete settings of two privacy knobs, differential privacy budget $\epsilon_0$ and inference error bound $E_m$, each idle worker in the large-scale domain of discrete locations is required to select one privacy setting and submit it to the server/platform for recommendation of multiple mechanisms. After this, the worker choose an obfuscation mechanism to generate a pseudo-location for possible allocation of SC task.
We focus on how to generate the multiple solutions at the SC-sever side, which should achieve (Pareto-)optimal trade-off between service quality loss and average expected inference error globally as well the above differential and geo-distortion privacy requirements.

\subsection{Multi-Objective Optimization}

\begin{figure}[tb]
\centering
\includegraphics[scale=0.32]{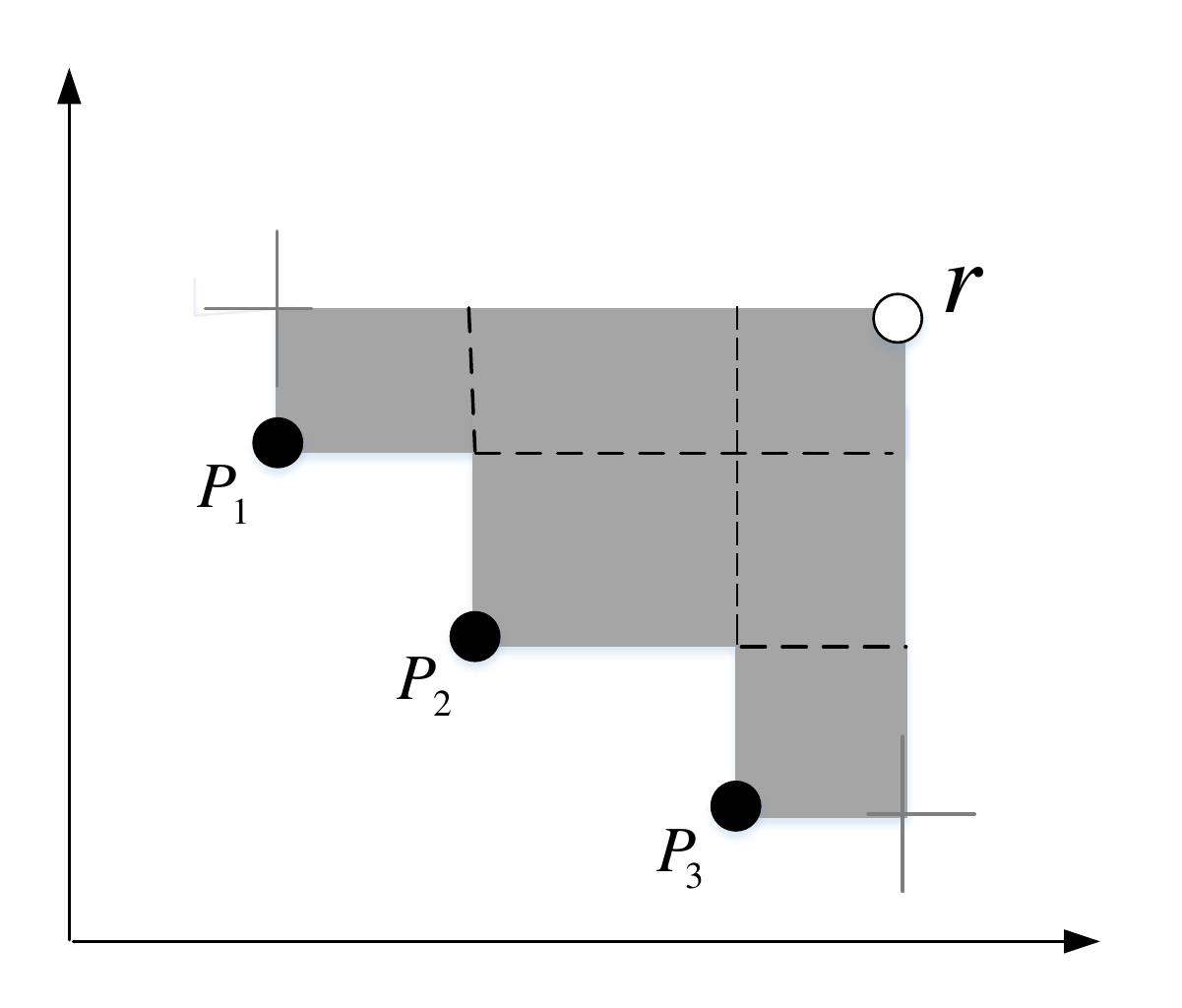}
\caption{The hypervolume indicator in the two-objective case.}
\label{fig:HV_model}
\end{figure}

In order to address the trade-off issue between privacy protection and practical quality, we introduce some basic concepts and methods related to Multi-Objective Optimization (MOO) \cite{BKS01}, which has attracted much attention in the computational intelligence community, see \cite{TSZ21} for a survey.

The definition of MOO problems is 
given by
\begin{equation}\label{eq:moo}
\min {\bf f(x)} =(f_1({\bf x}),f_2({\bf x}),\cdots,f_m({\bf x})),
\end{equation}
where ${\bf x} = (x_1,x_2,\cdots,x_d)$ denotes the $d$-dimensional decision vector of a solution from decision space $\Omega$, and $\bf{f(x)}$ is a objective vector involving $m$ conflicting functions.

\begin{defi}[Pareto Dominance \cite{DPA02}]\label{def:pareto}
For any two solutions ${\bf x}$ and $\bf{y}$, if $f_i({\bf x})\leq f_i({\bf y})$ for all $i = 1, 2,\cdots,m$ and $f_i({\bf x}) \textless f_i({\bf y})$ for at least one $i$, then $\bf{x}$ Pareto dominates $\bf{y}$. 
\end{defi}


Given a Pareto dominance set $S = \{P_{1},P_{2},...,P_{n}\}$ consisting of $n$ non-dominated two-objective vectors, the unary hypervolume indicator HV is a measure of the region which is dominated by $S$ and bounded from above by a reference point $r \in \mathbb{R}^2$ such that $r \ge (\max_P p_1, \max_P p_2)$, where $P = (p_1,p_2) \in S \subset \mathbb{R}^2$, and the relation $\ge$ applies componentwise.  Fig. \ref{fig:HV_model} shows that this region consists of an orthogonal polytope and can be regarded as the union of $n$ axis-aligned rectangles with a common vertex $r$.

\begin{defi}[Pareto Optimality \cite{DPA02}]\label{def:Pareto Optimality}
A solution ${\bf x}$ is Pareto optimal, if there does not exist any solution $\bf{y}$ dominating ${\bf x}$ in the decision space $\Omega$.
\end{defi}

Obviously, all Pareto optimal solutions obtained are non-dominated with each other.
In the current SC scenario, we will mainly consider two conflicting factors, service quality loss and expected inference error by using MOEA while in each privacy setting Pareto-optimal solutions are gradually improved by evolutions. The evolutionary process would be stopped when the HV value for Pareto-dominance set of solutions almost converges.

\section{Our Proposed Geo-MOEA Approach}\label{sec:framework}
This section describes our privacy protection approach Geo-MOEA, including basic model, domain partition, local adaptive obfuscation, partition MOEA and task matching.
\subsection{System Framework}

\begin{figure}[tb]
\centering
\includegraphics[scale=0.32]{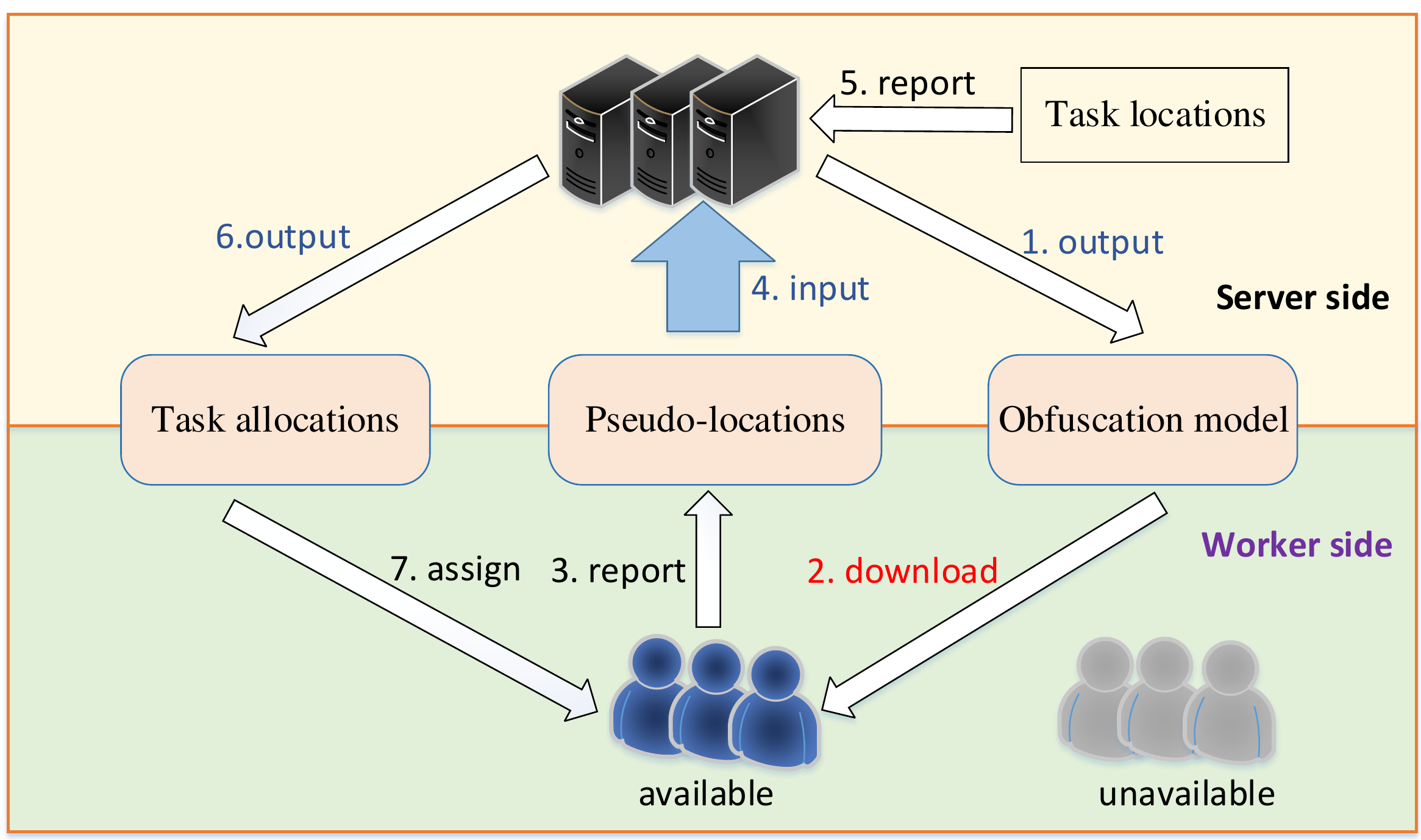}
\caption{The SC scenario of Geo-MOEA.}
\label{fig:SC_model}
\end{figure}

We consider the privacy protection problem of \emph{Spatial Crowdsourcing} (SC) worker locations in the SAT mode. Fig. \ref{fig:SC_model} describes our proposed seven-stage system model in SC scenario that consists of three parts: SC-server, requesters and workers. The SC-server is assumed to be semi-honest and intentionally deduces sensitive information of locations from the workers which has to be protected locally at worker side.

Each idle worker downloads the public obfuscation model, generates pseudo-location with \emph{Local Differential Privacy} (LDP) guarantee by inputting privacy parameters in private and sends the obfuscation matrix $f$ and the false location to SC-server. Requesters submits tasks and exposed location information to SC-Server. After receiving a task request, the SC-server determines the allocation by querying the locations reported by workers and initiates the geocast process.
It also plays the role of communications, and computations to determine task allocations are carried out at the SC-server part.  Possible disclosure of worker location and identity after his/her consent to the task is outside our scope.

As summarized by Algorithm \ref{alg:Geo-MOEA}, the main steps of Geo-MOEA include:

\begin{algorithm}[tb]
	\caption{Basic steps of Geo-MOEA}
	\label{alg:Geo-MOEA}
	\begin{algorithmic}[1]
        \STATE Dividing the domain into cells $\{X_i\}$ by binary partition on counts.
        \STATE Initializing adaptive reporting range $\mathcal{Y}_j$ composed of multiple connected PLSs.
		\STATE Generating Pareto-optimal solutions by MOEA among which the worker selects one for reporting.
        \STATE Task Matching by SC platform according to reported locations.
		\RETURN Task assignment with Geo-obfuscation strategy.
	\end{algorithmic}
\end{algorithm}

\begin{enumerate}[Step 1.]
\item[Step 1.]\textit{Domain Partition}: The SC service domain is divided into $2^s$ cells for some integer $s$ where each cell $X_i$ has almost the same count of discrete locations. The binary partition method is carried out iteratively according to the widthwise and lengthwise directions (the longer distance is preferred). This ensures that each cell obtained covers a required amount of locations and their convexity and compactness are satisfied.
\item[Step 2.]\textit{Initializing Local Adaptive Obfuscation Mechanism}: Each worker downloads in mobile networks the obfuscation model of the generation matrix from the server side, and inputs preferred privacy parameter values to initialize \emph{Protection Location Set} (PLS) partitions. 
Each PLS is restricted in one $X_i$ and meets the lower bound of inference error and geo-indistinguishability. For each PLS $\Phi_j$, the mechanism adaptively generates a reporting range $\mathcal{Y}_j$ composed of multiple connected PLSs, with which several cells $X_i$'s may be associated. Such partitions are repeated randomly at each MOEA iteration.
\item[Step 3.]\textit{Partition Optimizations by MOEA}: At each iteration the obfuscation mechanism utilizes genetic evolution to generate a new partition family and improves the Pareto-optimal solutions (graph). After achieving the convergence of graph's HV value, the MOEA outputs the set of Pareto-optimal partitions. The worker selects one solution and inputs his/her actual location, then a pseudo-location is generated by the exponential mechanism to be reported to SC-server.
\item[Step 4.]\textit{Task Matching}: According to the collected pseudo-locations, the SC platform informs the three nearest available (idle) workers around the task location. Finally, the task is assigned to the earliest responder.
\end{enumerate}

\begin{figure}[tb]
\centering
\includegraphics[scale=0.32]{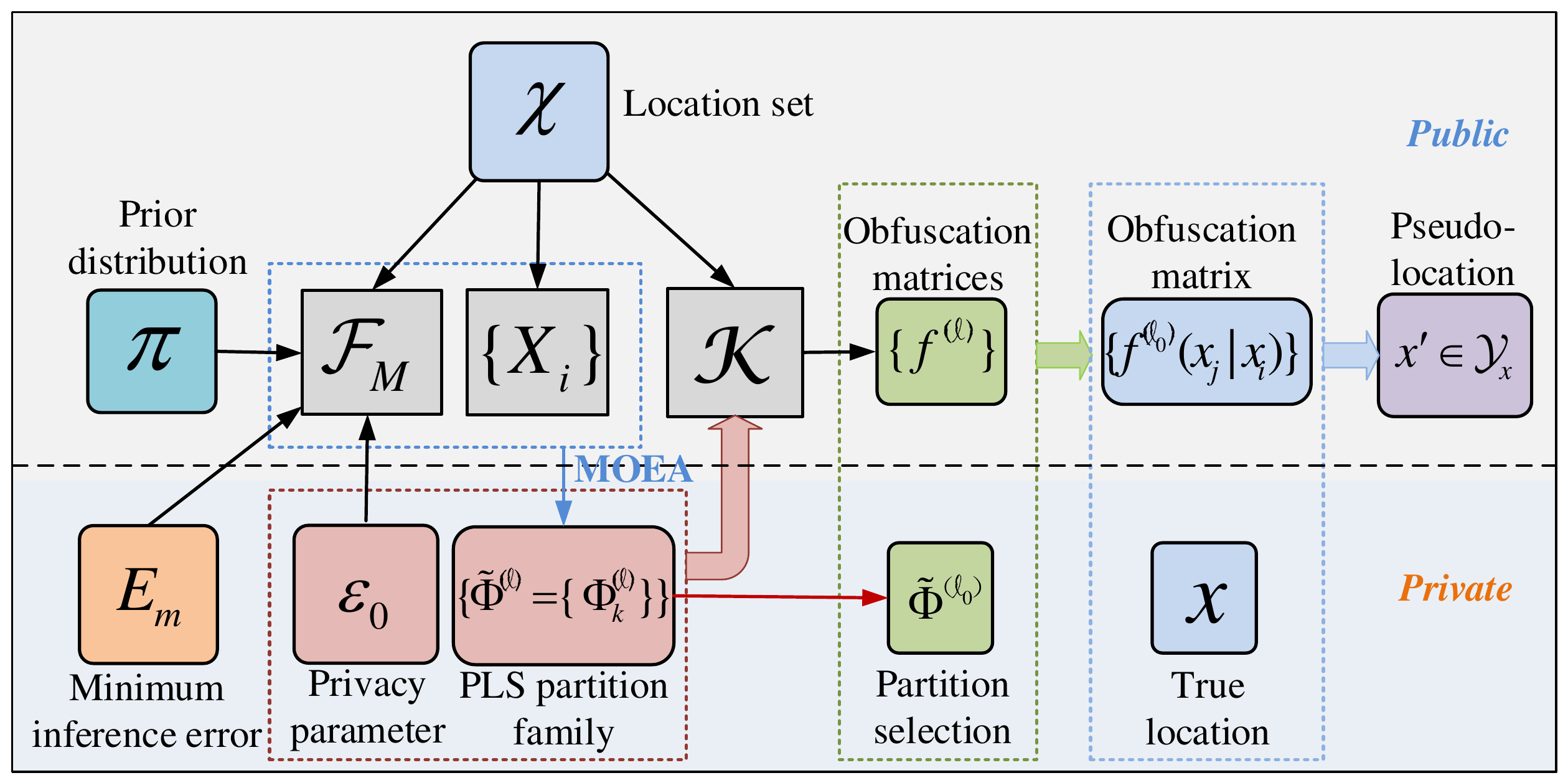}
\caption{ The privacy framework of Geo-MOEA.}
\label{fig:MF}
\end{figure}

As shown in Fig. \ref{fig:MF}, the privacy framework of Geo-MOEA 
is mainly composed of two components: the partitioning algorithm $\mathcal{F}_M$ 
(to determine a family of PLS partitions) and the differentially private mechanism $\mathcal{K}$ (to generate a pseudo-location). $\mathcal{F}_M$ has four inputs, prior distribution $\pi$, inference error threshold $E_m$, privacy parameter $\epsilon_0$ and location sets $\mathcal{X}$. The sets $X_i$'s are derived from the binary partition of $\mathcal{X}$.
For the two privacy parameters, users are allowed to control the posterior information leakage via the provisioning of privacy budget $\epsilon_0$ and specify $E_m$ to bound the expected inference error in the worst case. Each PLS contains obviously at least two locations and ensures the lower error bound.

By combining $\mathcal{F}_M$ with MOEA, we conduct multi-objective optimization in each $X_i$ to generate a PLS partition family $\{ \tilde{\Phi}^{(\ell)} = \{\Phi_k^{(\ell)} \} \}$ with Pareto optimality, where the symbol $\ell$ stands for the partition candidates for users and
the range of $k$ locally depends on $\tilde{\Phi}^{(\ell)}$.
Obviously, the output of our algorithm $\mathcal{F}_M$ is independent of the true location due to its no input. $\mathcal{F}_M$ partitions each $X_i$ into Pareto-optimal disjoint PLSs with setting \eqref{eq:loose_equality}.
 Then the mechanism $\mathcal{K}$ utilizes the exponential mechanism to calculate the distribution matrices of reporting probability $\{f^{(\ell)}(x_j| x_i)\}$ where the diameter of its PLS is assigned locally as the sensitivity in each $x_i$'s row. 
 After individual partition selection $\tilde{\Phi}^{(\ell_0)}$ that gives a public obfuscation matrix $\{f^{(\ell_0)}(x_j| x_i)\}$, a pseudo-location is produced with the input of true location.

Such a privacy framework supports Geo-MOEA to provide workers with differential and geo-distortion location privacy guarantee in all regions against Bayesian inference attacks. In particular, we develop a binary partition method adaptively producing cells, provide obfuscation mechanisms with locally adaptive reporting ranges and construct a novel multi-objective genetic algorithm with crossovers and mutations of PLS partitions.

\subsection{Domain Partition}

To ensure the local error bound $E_m$, we divide the domain into disjoint PLSs. However, the large-scale domain boosts greatly the computational complexity of searching optimal PLSs.
For this, we first partition the domain into region $X_i$'s in each of which PLSs are obtained locally later.
\begin{figure}[tb]
\centering
\includegraphics[scale=0.45]{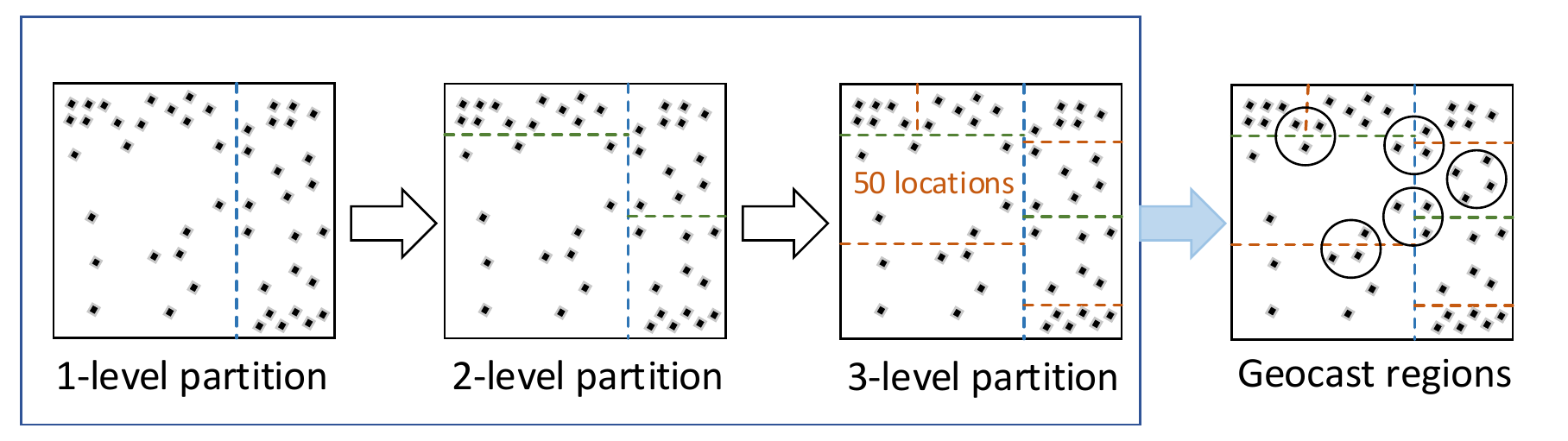}
\caption{Binary partition and geocast instances.}
\label{fig:partition}
\end{figure}


First, we circumscribe the domain with a rectangle like urban roads according to the longitude and latitude spans.
The added domain is naturally empty for locations. Next,
 the rectangle is divided into multiple rectangles $X_i$ by binary partitions. At each iteration, each rectangle is divided, along the larger edges, into two smaller rectangles including equal number of locations. The recursive partitions end until each cell achieves the expected range of location number. Each final cell is denoted by the region $X_i$. Compared to the existing spatial decompositions \cite{LWZ18}, such a partition method
 ensures the convexity and compactness in each cell.


To be specific, assume that the total number of workers is $N$ and the range of expected number of locations in each cell is $[n_0, 2n_0)$, then there are $\lfloor \log_2 \frac{N}{n_0}  \rfloor$ levels of partitions.
 Fig. \ref{fig:partition} shows the partitions of $N=400$ locations with $n_0=33$.
In the first level, since the rectangular length is larger than its width, it is divided into
two rectangles, one left and the other right, equally covering 200 locations. The deeper partitions are processed recursively, which results in $8$ cells.

\subsection{Local Adaptive Obfuscation Mechanism}\label{sec:ada-mech}

\begin{figure}[tb]
\centering
\includegraphics[scale=0.32]{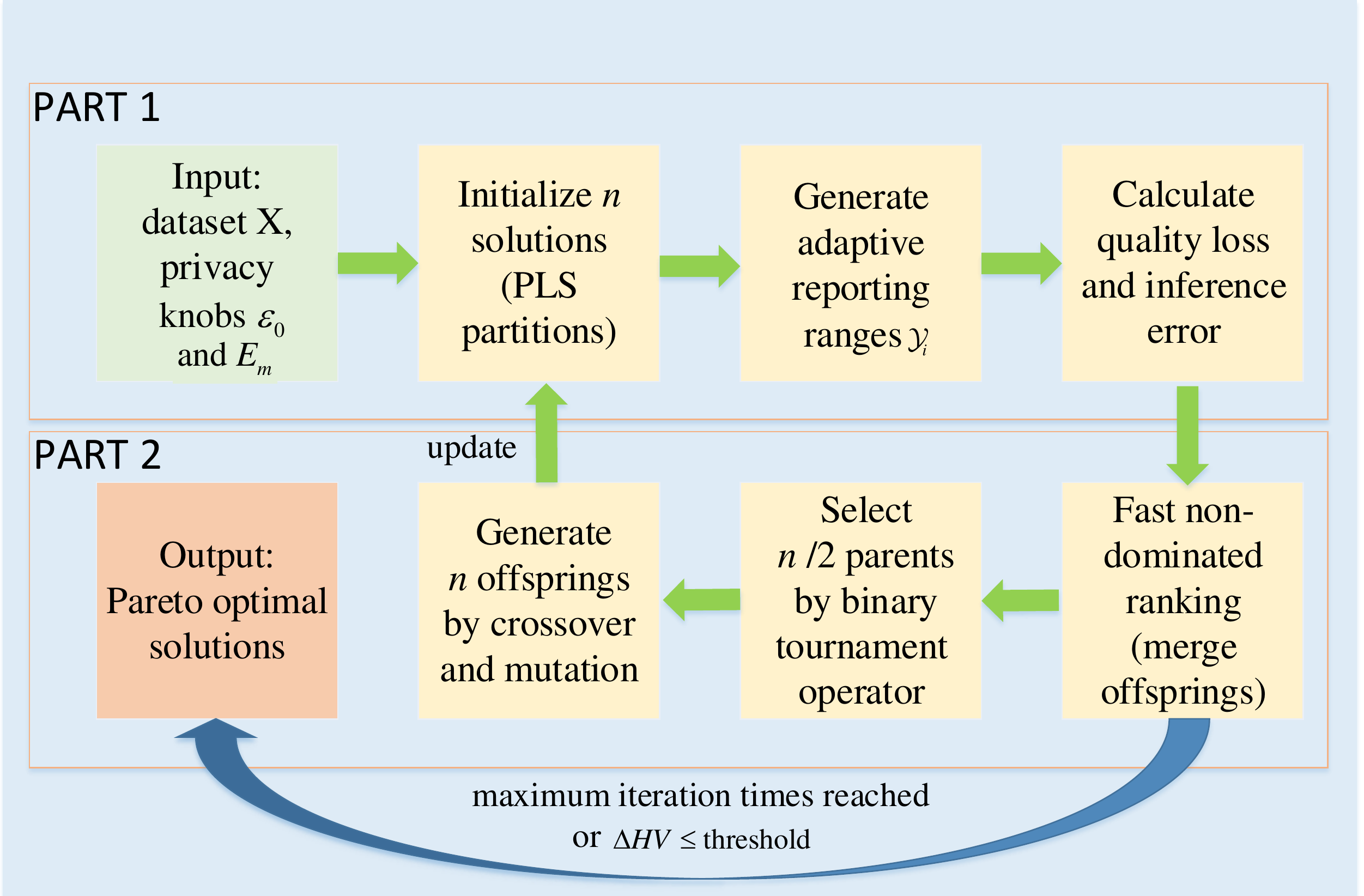}
\caption{Generating Pareto-optimality PLS partitions.}
\label{fig:MOEA_model}
\end{figure}

This section is the core part of Geo-MOEA, offering the initialization of obfuscation schemes involving adaptive reporting ranges.
The flowchart is shown in Fig. \ref{fig:MOEA_model} (PART 1).


Given the user-defined threshold, $ExpEr(x^\prime)\ge E_m$, for optimal inference attacks using any observed pseudo-location $x'$, the conditional expected
inference error is
\begin{equation}
ExpEr(x')=\mathop{\min}\limits_{\hat{x}\in \mathcal{X}}\sum_{x\in \mathcal{X}}\text{Pr}(x|x')d(\hat{x},x), \ \ \text{for}\ x'\in \mathcal{X}.
\end{equation}

Given a partition $\{\Phi_k\}$ of $\mathcal{X}$, we denote $z=\mathop{\rm argmin}\limits_{\hat{x}\in \mathcal{X}}\sum_{x\in \mathcal{X}}\
\text{Pr}(x|x')d(\hat{x},x)$,
$\text{Pr}(\Phi_k|x') = \sum_{y\in \Phi_k}\text{Pr}(y|x')$ and $\mathcal{Y}'=\{x:\text{Pr}(x|x')>0\}$.
By normalization over each $\Phi_k$ (with $\epsilon_k$-DP) in $\mathcal{Y}'$, we obtain, as in \cite{ZDC21},
\begin{equation}\label{ineq:lowerbound}
\begin{split}
&ExpEr(x')=\sum_{x\in \mathcal{Y}'}\text{Pr}(x|x')d(z,x)\\
=&\sum_{k}\sum_{x\in \Phi_k}\text{Pr}(x|x')d(z,x)\\
\ge &\sum_{k}\mathop{\min}\limits_{\widehat{x}_k\in \mathcal{X}}\sum_{x\in \Phi_k}\text{Pr}(x|x')d(\widehat{x}_k,x)\\
=& \sum_{k} \text{Pr}(\Phi_k|x') \mathop{\min}\limits_{\widehat{x}_k\in \mathcal{X}}\sum_{x\in \Phi_k}\frac{\text{Pr}(x|x')d(\widehat{x}_k,x)}{\sum_{y\in \Phi_k}\text{Pr}(y|x')}
 \\
=& 
\sum_{k} \text{Pr}(\Phi_k|x') \mathop{\min}\limits_{\widehat{x}_k\in \mathcal{X}}\sum_{x\in \Phi_k}\frac{\pi(x)f(x'|x)d(\widehat{x}_k,x)}{\sum_{y\in \Phi_k}\pi(y)f(x'|y)}
\\
\ge & 
\sum_{k} \text{Pr}(\Phi_k|x') \mathop{\min}\limits_{\widehat{x}_k\in \mathcal{X}}\sum_{x\in \Phi_k}\frac{\pi(x)f(x'|x)d(\widehat{x}_k,x)}{\sum_{y\in \Phi_k}\pi(y)e^{\epsilon_k}f(x'|x)}
\\
= & \sum_{k} \text{Pr}(\Phi_k|x') e^{-\epsilon_k} E'(\Phi_{k}), 
\end{split}
\end{equation}
where
\begin{equation}
E'(\Phi)=\mathop{\min}\limits_{\hat{x}\in \mathcal{X}}\sum_{x\in \Phi}\frac{\pi(x)}{\sum_{y\in \Phi}\pi(y)}d(\hat{x},x).
\end{equation}

Since $\sum_{k} \text{Pr}(\Phi_k|x')=1$, 
the condition
that for all $\Phi_{k}$,
\begin{equation}\label{eq:inequality}
E'(\Phi_k)\ge e^{\epsilon_k} E_m,
\end{equation}
implies the user-defined threshold, $ExpEr(x^\prime)\ge E_m$. Following this, we have the extended assertion:

\begin{thm}\label{thm:DPIVE}
Given a domain partition $\{\Phi_k\}$ and an observed pseudo-location $x^\prime$, suppose that an obfuscation mechanism satisfies $\epsilon_k$-differential privacy on each PLS $\Phi_k$. If $E'(\Phi_k)\geq e^{\epsilon_k}E_m$ for each $\Phi_k$,
 $ExpEr(x^\prime)\ge E_m$ for the optimal inference attack.
\end{thm}

While allowing adaptive privacy specification $\epsilon_k$ on each PLS, the rule of \eqref{eq:inequality} with using the identical privacy budget $\epsilon_k=\epsilon_0$ is so strict that some candidates are skipped for determining PLSs. This motivates us to search for optimized partitions $\{\Phi_k\}$ in global sense. we introduce a looser condition for ensuring the error threshold,
\begin{equation}\label{eq:loose_inequality}
\epsilon_k \leq \min\left(\ln(E'(\Phi_k)/E_m),\epsilon_0\right),
\end{equation}
which means that the case $E_m < E'(\Phi) < e^{\epsilon_0}E_m$ should be involved in search algorithm. Theoretically,

\begin{thm}\label{thm:our_inequality} Given a domain partition $\{\Phi_k\}$, if a location obfuscation mechanism achieves $\epsilon_k$-differential privacy on each PLS $\Phi_k$ with \eqref{eq:loose_inequality}, then the optimal inference using any observed pseudo-location $x'$ satisfies $ExpEr(x')\geq E_m$.

\begin{proof} Due to \eqref{eq:loose_inequality}, we have, for each $\Phi_k$,
\begin{align*}
\begin{split}
e^{\epsilon_k}\cdot E_m &\leq e^{\min(\ln(E'(\Phi_k)/E_m),\epsilon_0)}\cdot E_m\\
&\leq e^{\ln(E'(\Phi_k)/E_m)}\cdot E_m
=E'(\Phi_k).
\end{split}
\end{align*}
By \eqref{ineq:lowerbound}, $ExpEr(x')\ge\sum_{k} \text{Pr}(\Phi_k|x')e^{-\epsilon_k} E'(\Phi_{k})\ge E_m$.


\end{proof}

\end{thm}

\begin{figure*}[tb]
\centering
\includegraphics[scale=0.67]{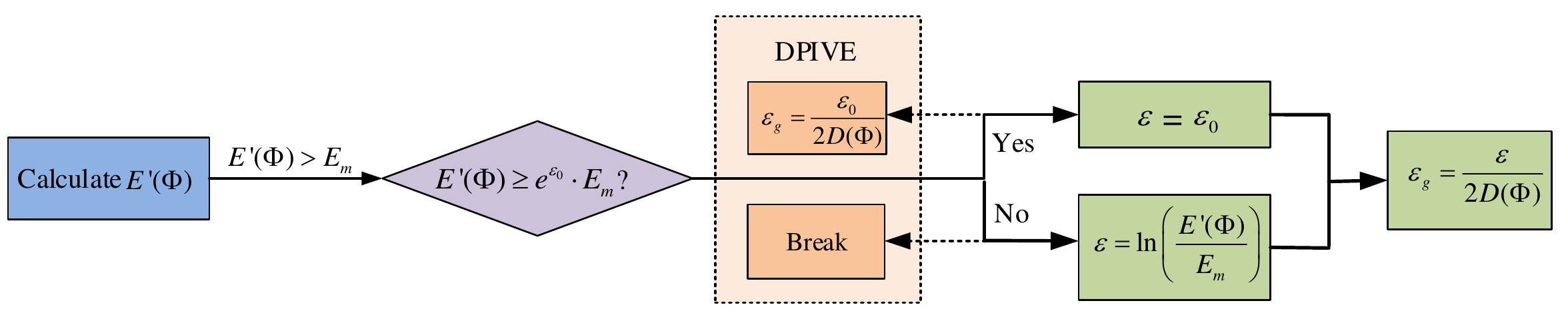}
\caption{Adaptive privacy allocation procedure for PLS candidates.}
\label{fig:eps}
\end{figure*}

Under the total privacy budget $\epsilon_0$, Theorem \ref{thm:our_inequality} supports the obfuscation mechanism to achieve $\epsilon_k$-differential privacy on each PLS  $\Phi_{k}$ using the value, as in Fig. \ref{fig:eps},
\begin{equation}\label{eq:loose_equality}
\epsilon_k = \min\left(\ln(E'(\Phi_k)/E_m),\epsilon_0\right),
\end{equation}
for minimizing quality losses with resistance to attacks.

The exponential mechanism $\mathcal{K}$ computes
the probability distribution $f(x'| x)= w_{\Phi_j}(x) \exp(\frac{-\epsilon_0 d(x,x')}{2D(\Phi_j)})$ for the true location $x\in\Phi_j$ and any possible pseudo-location $x'$ from the reporting range $\mathcal{Y}_x$, where
\begin{equation}\label{wx}
w_{\Phi_j}(x) = \left( \sum_{x^\prime \in \mathcal{Y}_x } \exp\left(\frac{-\epsilon_0 d(x,x^\prime)}{2D(\Phi_j)}\right) \right)^{-1}.
\end{equation}
Then the quality loss on $x$ is
\begin{equation}\label{QLx}
QL_x= \sum_{x^\prime \in \mathcal{Y}_x }f(x'| x)d(x',x).
\end{equation}

\begin{algorithm}[tb]
	\caption{Clustering with retreats for PLSs (Ret-C)}
	\label{alg:Ret-C}
	\begin{algorithmic}[1]
		\REQUIRE each cell $X_i=\{x_0,x_1,...,x_{n-1}\}$, prior probability $\pi$, inference error bound $E_m$, user privacy parameter $\epsilon_0$

		\STATE Get $k_i$ by QK-means algorithm for each $X_i$ separately
        \STATE Random select $k$ = $k_i$ or $k = k_i+1$
		\STATE Choose $k$ locations from $X_i$ randomly as the centers
		\STATE Init remaining locations $Q_i=X_i$, $\tilde{\Phi}_i=\{\Phi_1,\Phi_2,...,\Phi_k\}$ with $\Phi_j=\{\}(1\leq j\leq k)$\label{state:init_R}

		\STATE Remove each $x \in Q_i$ in ascending order of $\min_jd(x, \Phi_j)$ to its closest $\Phi_j$

		\STATE Once \eqref{eq:inequality-eps0} is satisfied for some $\Phi_j$ in $\tilde{\Phi}_i$ \textbf{then}
        calculate each of the $\Phi_j$'s historical values $\epsilon_g = \epsilon_j/(2D)$ with \eqref{eq:loose_equality}, and retain only the set having the largest $\epsilon_g$ while the other points return to join new clusters until $Q_i=null$
        \STATE \textbf{if} $\epsilon_g>0$ holds for all $\Phi_j$ and $Q_i=null$, \textbf{break} for $\tilde{\Phi}_i$
		\RETURN A partition for disjoint PLSs $\tilde{\Phi}=\bigcup_i \tilde{\Phi}_i$
	\end{algorithmic}
\end{algorithm}

\begin{thm}\label{thm:QL-epsg}
Given the user's location $x$ and its reporting range $\mathcal{Y}_x$, the quality loss of the exponential mechanism on $x$ decreases monotonously with increasing $\epsilon_g=\epsilon/(2D)$, where $D$ denotes the diameter of the PLS containing $x$.
\begin{proof}
For the mechanism $\mathcal{K}$ carried out on the true location $x$, the quality loss $QL_x$ can be rewritten shortly as
\begin{equation}\label{QL-epsg}
L(\epsilon_g)= \frac{\sum_{j} d_j \exp(-\epsilon_g d_j)}{\sum_{j} \exp(-\epsilon_g d_j)}.
\end{equation}
Then the first derivative is
\begin{equation}\label{QL-epsg-prime}
\begin{split}
L'(\epsilon_g)&= \frac{ \left(\sum_{j}d_j e^{-\epsilon_g d_j}\right)^2-\left(\sum_{j}d_j^2 e^{-\epsilon_g d_j}\right)   \left( \sum_{j} e^{-\epsilon_g d_j} \right) }{\left(\sum_{j} \exp(-\epsilon_g d_j)\right)^2}\\
&=\frac{ \left(\sum_{j}a_j b_j\right)^2-\left(\sum_{j}a_j^2 \right)   \left( \sum_{j} b_j^2 \right) }{\left(\sum_{j} \exp(-\epsilon_g d_j)\right)^2}\le 0,
\end{split}
\end{equation}
where the last relation is due to Cauchy-Schwarz inequality with $a_j=d_j e^{-\epsilon_g d_j/2}\ge 0$ and $b_j=e^{-\epsilon_g d_j/2}\ge 0$. Hence, $L(\epsilon_g)$ is monotonically decreasing with $\epsilon_g$.
\end{proof}
\end{thm}

Theorem \ref{thm:QL-epsg} yields that for each PLS the larger $\epsilon_g$ produces the smaller quality losses.
In the PART 1, based on the most general condition, $E'(\Phi)> E_m$, as shown in our designed Algorithm \ref{alg:Ret-C}.
We first execute QK-means algorithm from \cite{ZDC22} on each cell $X_i$ to find its optimal number $k_i$ of disjointed PLSs with satisfying
  \begin{equation}\label{eq:inequality-eps0}
E'(\Phi_k)\ge e^{\epsilon_0} E_m.
\end{equation}
 The modifications include that Ret-C has no recursions on clustering centers and sets $\epsilon_k$ by Eq. \eqref{eq:loose_equality} in each PLS $\Phi_k$ as shown in Fig. \ref{fig:eps}.

Any two PLSs do not intersect with each other. Each PLS is involved in only one cell $X_i$ and meets the threshold $E_m$.  The mechanism repeats such PLS partitions to initialize a  population of $n$ solutions over $\mathcal{X}$ ($n$ families of partitions).

For each PLS $\Phi_j$ in a partition, the mechanism construct a reporting range $\mathcal{Y}_j$ by adding the nearest PLS to $\Phi_j$ recursively according to their centers' distance, until that at least two PLSs are included and the number of locations is not less than $50$. In this way, each reporting range $\mathcal{Y}_j$ is not simply the involved cell $X_i$, but sometimes intersects with several cells.
However, such initialized solutions are not optimal in general for the tradeoff of privacy level and service quality. Step 3 is required for further optimization.

\subsection{Partition MOEA and Task Matching}

This section introduces the multi-objective genetic algorithm into optimizing the PLS partition population and generates a pseudo-location for task matching. Specifically, our innovated random combinations and variations of cluster centers in each cell lead to
the crossovers and mutations of PLS partitions, respectively, which expands largely the solution space of the privacy problem.

The mechanism aims to generate higher unconditional expected inference error (ExpErr) and smaller service quality loss (QLoss) that conflict with each other. Following the line of MOEA, we describe the Pareto sets with regarding QLoss and minus ExpErr as two objective functions and use the unary Hypervolume (HV) indicator to measure the convergence and diversity of solutions.
HV \cite{TSZ21} is a usual performance metric for Multi-Objective Optimization (MOO) problems that is the only unary indicator strictly monotonic with Pareto dominance. In our setting, we first find the reference point $Q$ whose coordinates are defined as the maximum values of two objectives for each Pareto-dominated solution set, respectively. Then the HV value is defined as the area of the rectangular region surrounded by point $Q$ and each point in Pareto-dominated solution set with subtracting the overlapped area.
We adopt the inverse number of ExpErr since two objectives in MOO prefer larger values (or both smaller).

The PART 2 in Fig. \ref{fig:MOEA_model} shows the genetic evolution of PLS partition populations.
After calculating the two metrics for each population, the mechanism makes fast non-dominated ranking for each population and determines the crowded degree in each level. This generates $n$ solutions sorted by levels. Then, based on $n/2$ parents (in the PLS partition family) obtained by a binary tournament selection operator \cite{DPA02}, the mechanism adopts crossover and mutation operators to generate $n/2$ offsprings, respectively. For crossover we randomly select $k_i$ or $k_i+1$ cluster centers from $5$ parents in each cell $X_i$ to generate a new partition, while randomly replacing half cluster centers in each cell for mutation.
Afterwards,
all $2n$ solutions join the new fast non-dominated ranking and half are filtered out. Such a MOEA using genetic algorithm proceeds recursively until the maximum iteration times is attained or the increase value
 $\Delta$HV vanishes.

Given the set of Pareto optimal recommendations, the worker selects a preferred solution with personalized threshold and
carry out the exponential mechanism to produce a pseudo-location that is reported to SC-server. For each true location $x\in\Phi_j$, the utility (scoring) of output  $x'\in\mathcal{Y}_j$ can be defined as $u(x,x')=-d(x,x')$. The sensitivity of $u$ for each PLS $\Phi_j$ is its diameter \cite{ZDC22}.


After a worker uploads the pseudo-location, the server improves the worker's information.
Once a task request is received, the server geocasts the task to the three idle workers closest to the task according to the collected pseudo-locations, see Fig. \ref{fig:partition}. The worker who approves first is qualified to execute the task. The task and the worker may be really located in different cells $X_i$ and even are far away from each other with a small probability.
The metrics of worker travel distance (WTD) denoting their actual distance measures the service availability to a certain extent.

According to Theorem \ref{thm:Geo-MOEA}, the mechanism satisfies the $\epsilon_0$-differential privacy on each PLS $\Phi$. Moreover, for any two locations from different PLSs whose reporting ranges $\mathcal{Y}_j$'s intersect with each other,
Geo-MOEA preserves weak differential privacy with a deviation factor on the coefficient as stated in Theorem \ref{thm:region-symm}.

\begin{thm}\label{thm:Geo-MOEA}
Assume disjoint PLSs $\{\Phi_j\}$, then the exponential mechanism $\mathcal{K}$ in Geo-MOEA satisfies $\epsilon_0$-differential privacy on each $\Phi_j$.

\begin{proof}
For any $x,y \in \Phi_j$ and $x'\in\mathcal{Y}_j$, we have by  \eqref{eq:loose_equality},

\begin{equation}
\begin{split}
\frac{f(x'|x)}{f(x'|y)}&=
\frac{w_{\Phi_i}(x) \exp\left(\frac{-\epsilon_i d(x,x')}{2D(\Phi_i)}\right)}{w_{\Phi_i}(y) \exp\left(\frac{-\epsilon_i d(y,x')}{2D(\Phi_i)}\right)}\\
&\le\frac{w_{\Phi_i}(x)}{w_{\Phi_i}(y)}
e^{\frac{\epsilon_i}{2D(\Phi_i)}|d(x,x')- d(y,x')|}
\\
&\le \frac{\sum\limits_{t\in\mathcal{Y}_i}\exp\left(\frac{-\epsilon_i (d(x,t)-D(\Phi_i))}{2D(\Phi_i)}\right)}{\sum\limits_{t\in\mathcal{Y}_i}\exp\left(\frac{-\epsilon_i d(x,t)}{2D(\Phi_i)}\right)}
e^{\frac{\epsilon_i d(x,y)}{2D(\Phi_i)}}
\\
&\le \exp\left(\frac{\epsilon_i}{2}\right)\exp\left(\frac{\epsilon_i}{2}\right)\le \exp(\epsilon_0).
\end{split}
\end{equation}
where $\epsilon_g=\epsilon_j/(2D(\Phi_j))$.
\end{proof}
\end{thm}

This shows that the mechanism $\mathcal{K}$ achieves $(\epsilon_g,D(\Phi_j))$-geo-indistinguishability
within each PLS $\Phi_j$.

\begin{thm}\label{thm:region-symm}
Assume that $\mathcal{Y}_i$ and $\mathcal{Y}_j$ intersecting with each other are the reporting ranges for disjoint PLSs, $\Phi_i$ and $\Phi_j (i\neq j)$, respectively. Then the exponential mechanism $\mathcal{K}$  satisfies $\left(\frac{D(\mathcal{Y}_i)}{D(\Phi_ i)}+\frac{D(\mathcal{Y}_j)}{D(\Phi_j)}\right)\frac{\epsilon_0}{2}$-differential privacy on $\Phi_i\cup\Phi_j$ with a deviation: For any $x\in\Phi_i$, $y \in \Phi_j$ and $x'\in\mathcal{Y}_i\cap\mathcal{Y}_j$,

\begin{equation}
\frac{f(x'|x)}{f(x'|y)}
\leq \frac{|\mathcal{Y}_j|}{|\mathcal{Y}_i|} \exp\left(\frac{\epsilon_0}{2}\left(\frac{D(\mathcal{Y}_j)}{D(\Phi_j)}+\frac{D(\mathcal{Y}_i)}{D(\Phi_i)}\right)\right).
\end{equation}

\begin{proof}
\begin{equation}
\begin{split}
\frac{f(x'|x)}{f(x'|y)}
&=\frac{\exp\left(\frac{-\epsilon_i d(x,x')}{2D(\Phi_i)}\right)}{\sum\limits_{s\in\mathcal{Y}_i}\exp\left(\frac{-\epsilon_i d(x,s)}{2D(\Phi_i)}\right)}\cdot
\frac{\sum\limits_{t\in\mathcal{Y}_j}\exp\left(\frac{-\epsilon_j d(y,t)}{2D(\Phi_j)}\right)}{\exp\left(\frac{-\epsilon_j d(y,x')}{2D(\Phi_j)}\right)}\\
&=\frac{\sum\limits_{t\in\mathcal{Y}_j}\exp\left(\frac{\epsilon_j}{2D(\Phi_j)}\left(d(y,x')- d(y,t)\right)\right)}
{\sum\limits_{s\in\mathcal{Y}_i}\exp\left(\frac{\epsilon_i}{2D(\Phi_i)}\left(d(x,x')- d(x,s)\right)\right)}\\
&\leq\frac{|\mathcal{Y}_j|}{|\mathcal{Y}_i|} \exp\left(\frac{\epsilon_j}{2}\frac{D(\mathcal{Y}_j)}{D(\Phi_j)}+\frac{\epsilon_i}{2}\frac{D(\mathcal{Y}_i)}{D(\Phi_i)}\right)\\
&\leq \frac{|\mathcal{Y}_j|}{|\mathcal{Y}_i|} \exp\left(\frac{\epsilon_0}{2}\left(\frac{D(\mathcal{Y}_j)}{D(\Phi_j)}+\frac{D(\mathcal{Y}_i)}{D(\Phi_i)}\right)\right).
\end{split}
\end{equation}
\end{proof}
\end{thm}

The assertions obtained above for variable reporting ranges provide strictly theoretical support on the differential privacy protection guarantee in the large-scale SC scenario.

\begin{figure}[tb]
\centering
\subfigure[3000 locations (NYTaxi)]{
		\includegraphics[scale=0.3]{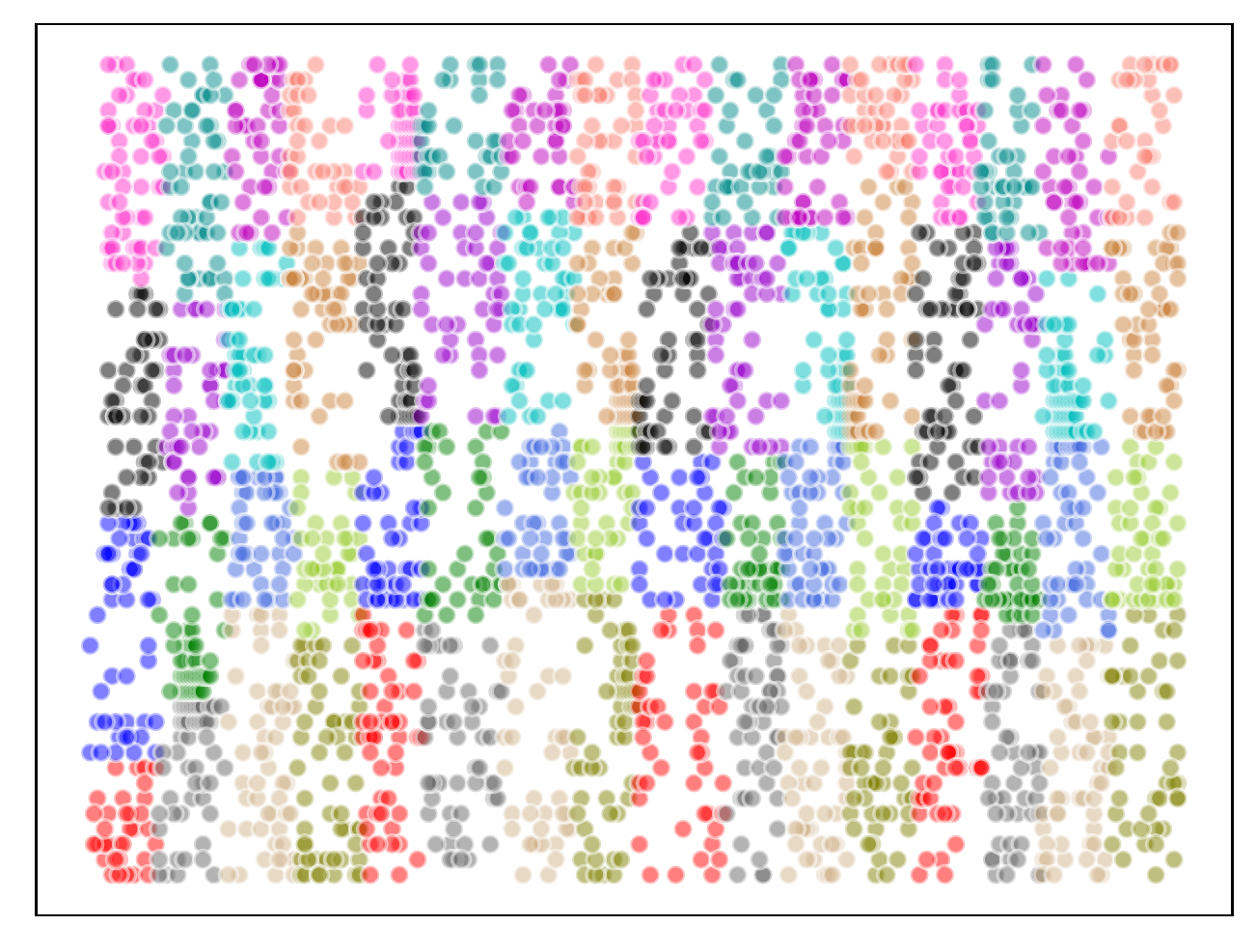}
	}
\subfigure[1000 locations (Gowalla)]{
		\includegraphics[scale=0.3]{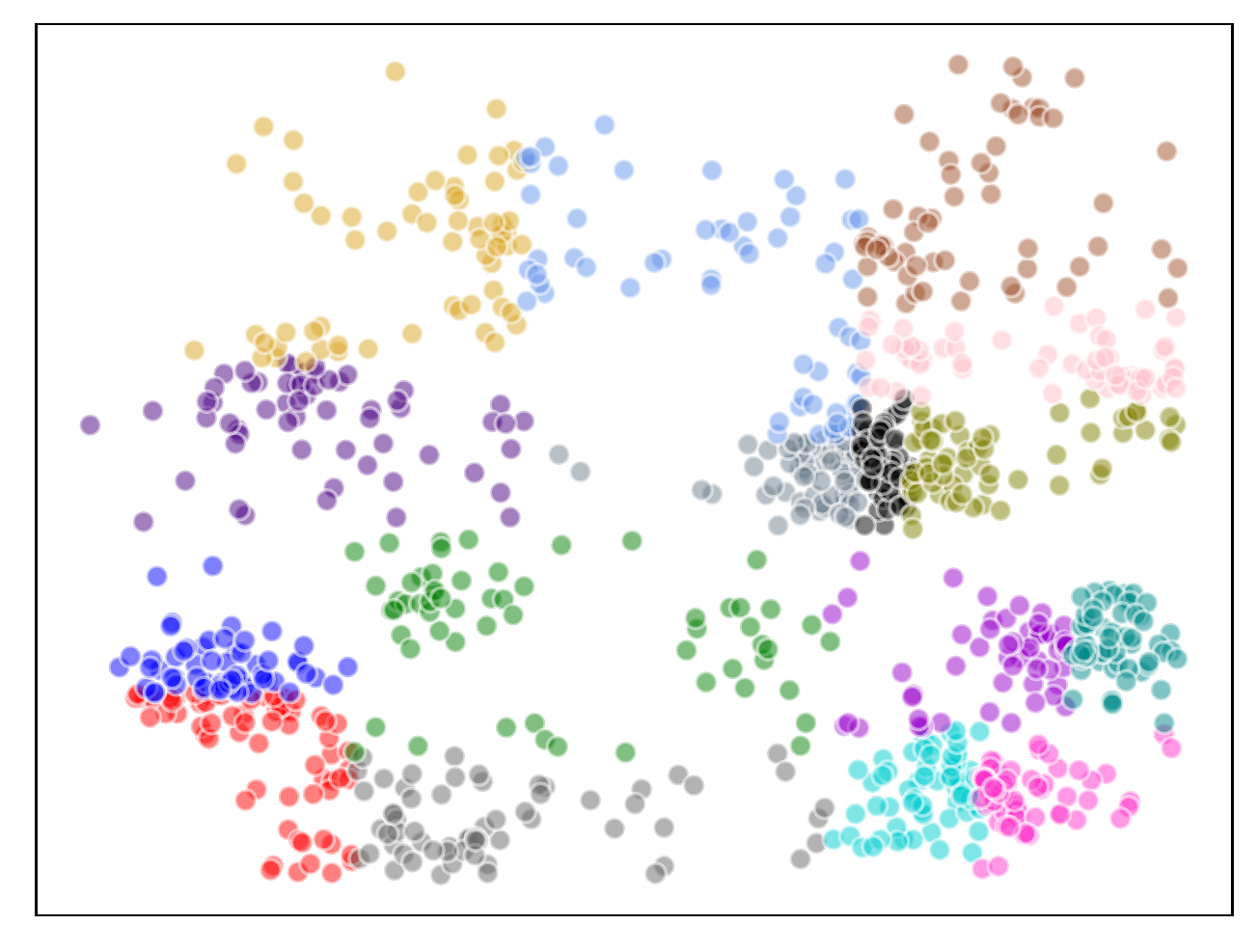}
	}
\caption{Distribution charts with binary partitions.}
\label{fig:datasets}
\end{figure}

\section{Performance Analysis}\label{sec:evaluation}
This section compares our Geo-MOEA approach with some previous mechanisms on the metrics of location privacy and service quality, then presents Pareto analysis, application analysis and visualization analysis. The results show that our mechanisms greatly improves service availability on the basis of ensuring individual location privacy.

\subsection{Experimental Methodology}
\textbf{Datasets.} Since SC services are mainly located in urban domains, we investigate its distribution characteristics in cities and sample two real datasets,
NYTaxi is a dataset of taxi locations in New York. We extract $3000$ positions picked up (longitude from $-73.982738$ to $-73.982742$, latitude from $40.750562$ to $40.760480$) on May 21, 2013. 
Since most positions are distributed on the roads, we randomly and uniformly blur each position into a circle centered at the pickup position and with radius (\emph{Blur Radius, BR}) of $80$ meters.
Gowalla is a dataset of social network check-in locations. We extract
 $1000$ locations (longitude from $-179.9910$ to $-179.9641$, latitude from $89.9232$ to $89.9330$). Gowalla involves more relatively sparse regions.
 Fig. \ref{fig:datasets} presents their distribution charts with binary partitions. In each chart, all cells $X_i$ are marked in different colors.
We simulate a prior distribution uniformly in which each value is sampled randomly in [0.00028, 0.00038] and [0.0005, 0.0015] on two datasets, respectively.

\begin{figure*}[tb]
\centering
\subfigure[$E_m=0.10,\epsilon_0=0.50$]{
		\includegraphics[scale=0.28]{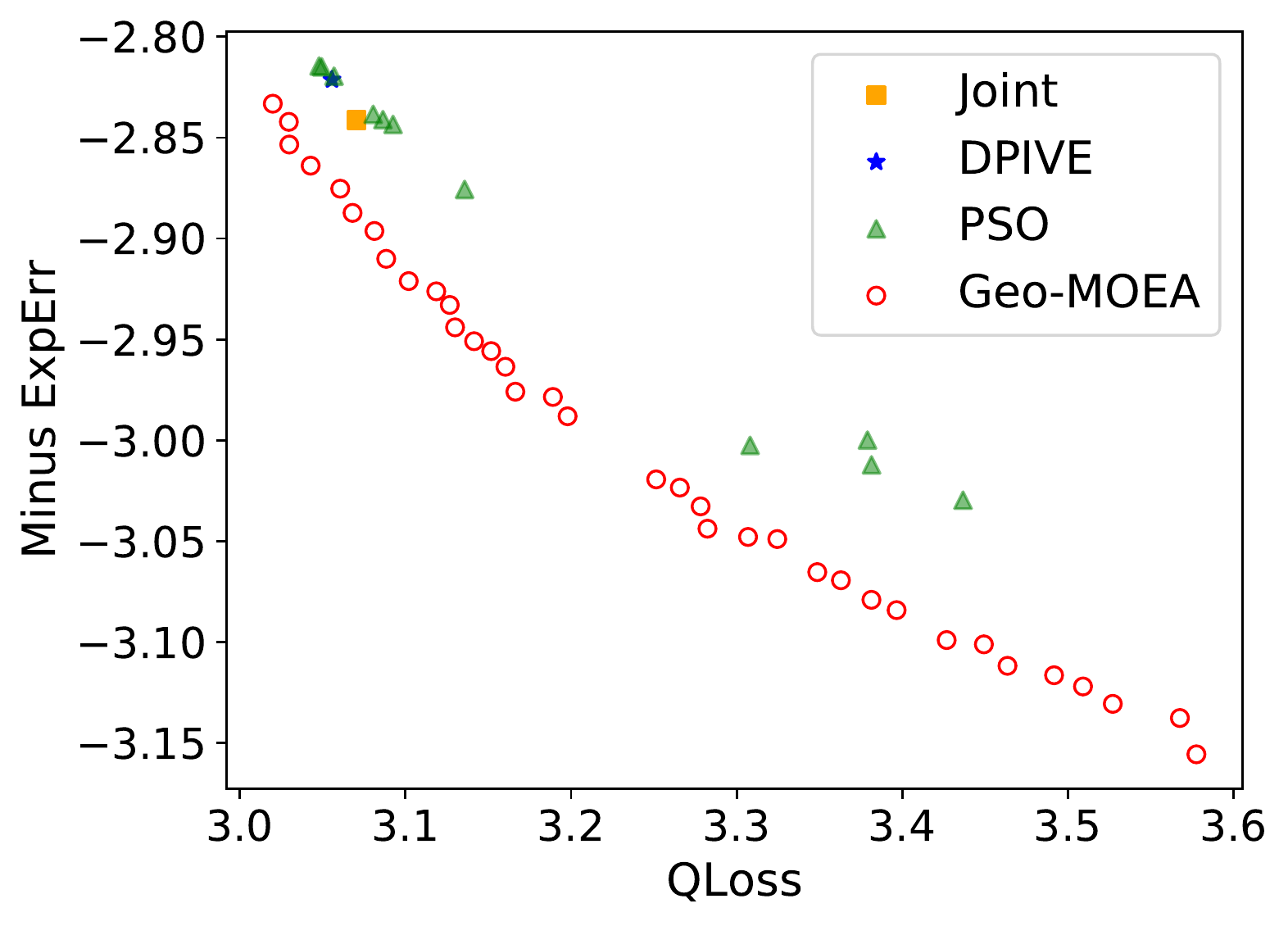}
	}
\subfigure[$E_m=0.10,\epsilon_0=1.00$]{
		\includegraphics[scale=0.28]{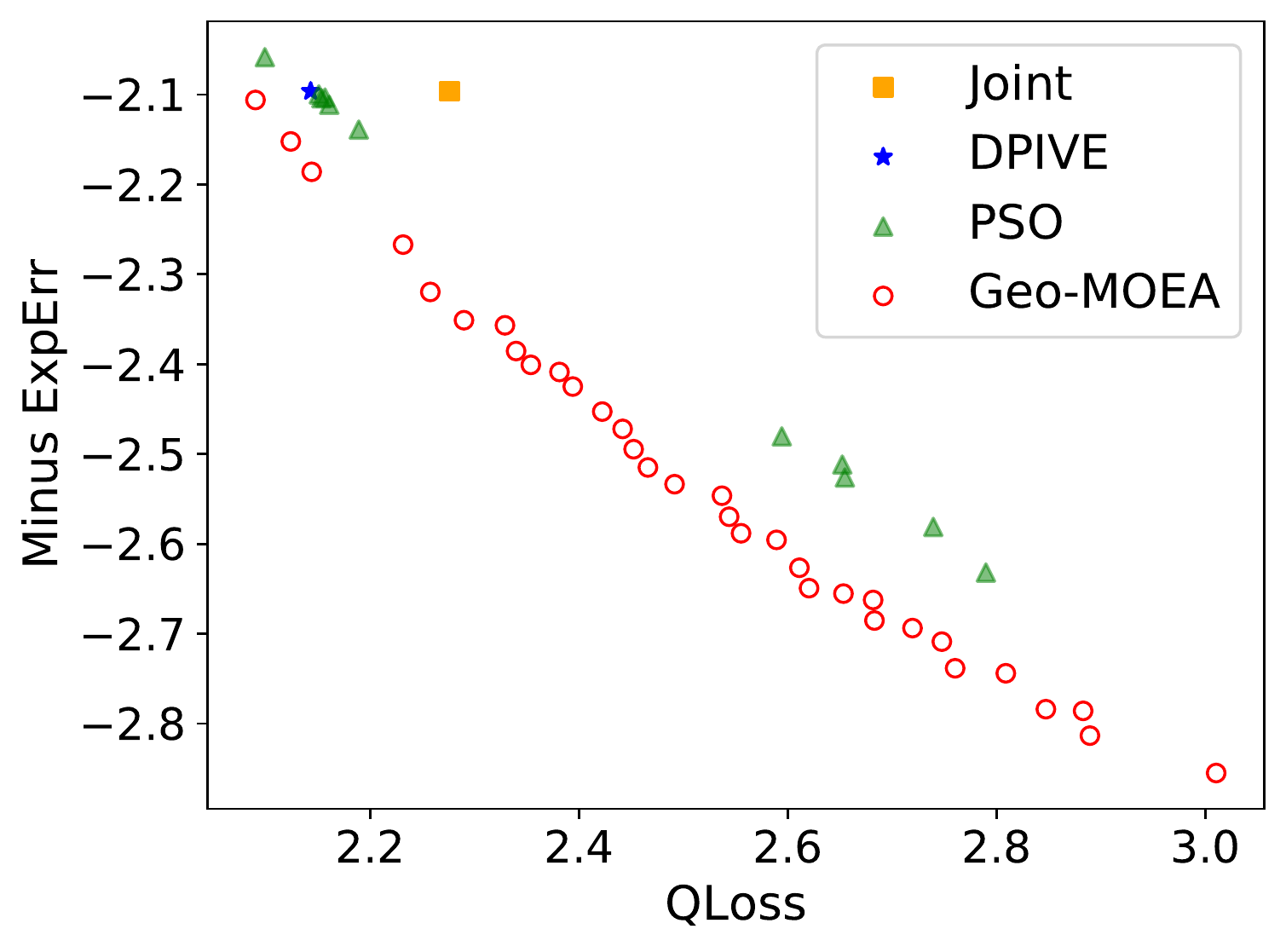}
	}
\subfigure[$E_m=0.10,\epsilon_0=1.50$]{
		\includegraphics[scale=0.28]{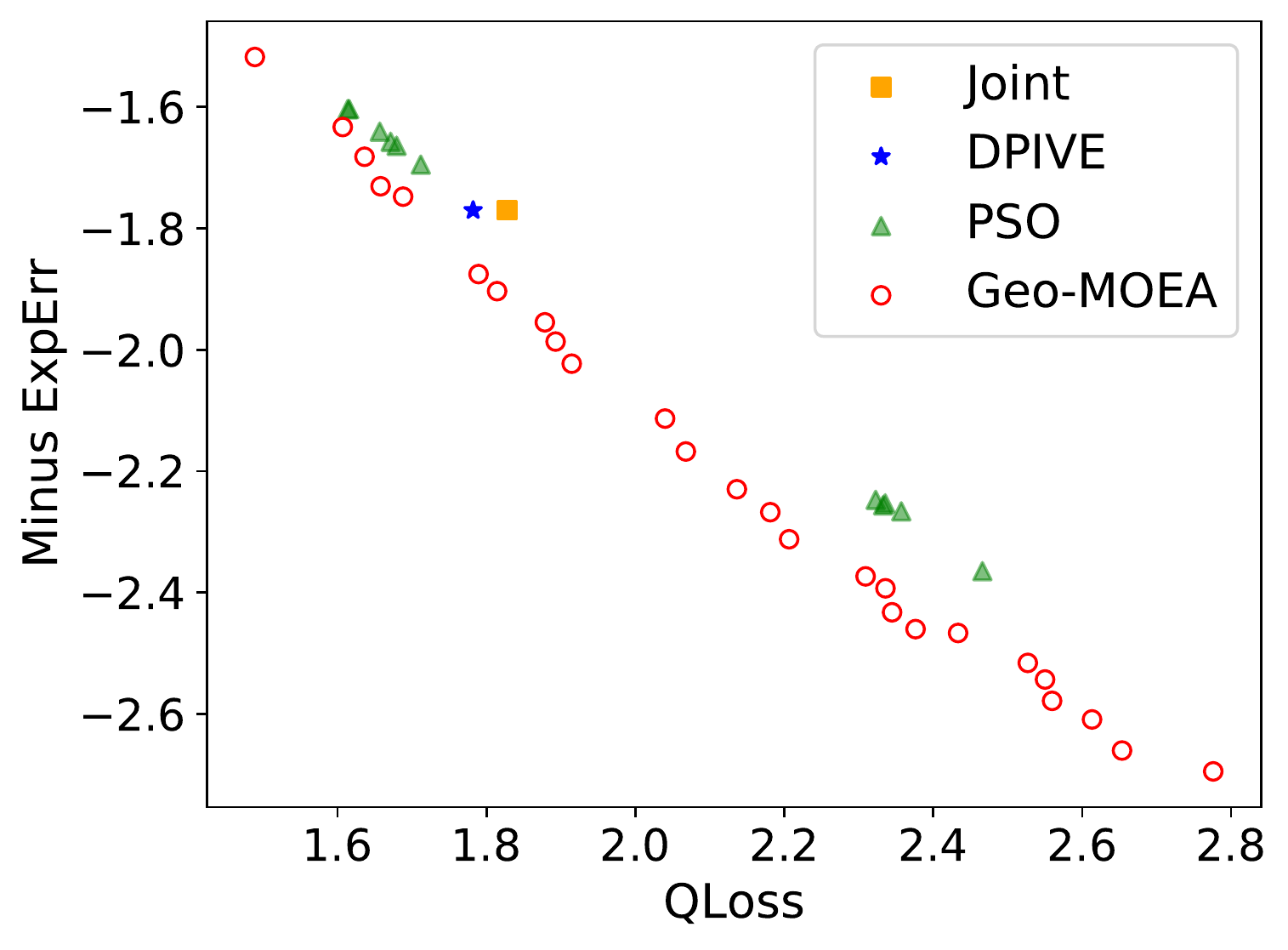}
\label{fig:fig3}
	}

\subfigure[$E_m=0.15,\epsilon_0=0.50$]{
		\includegraphics[scale=0.28]{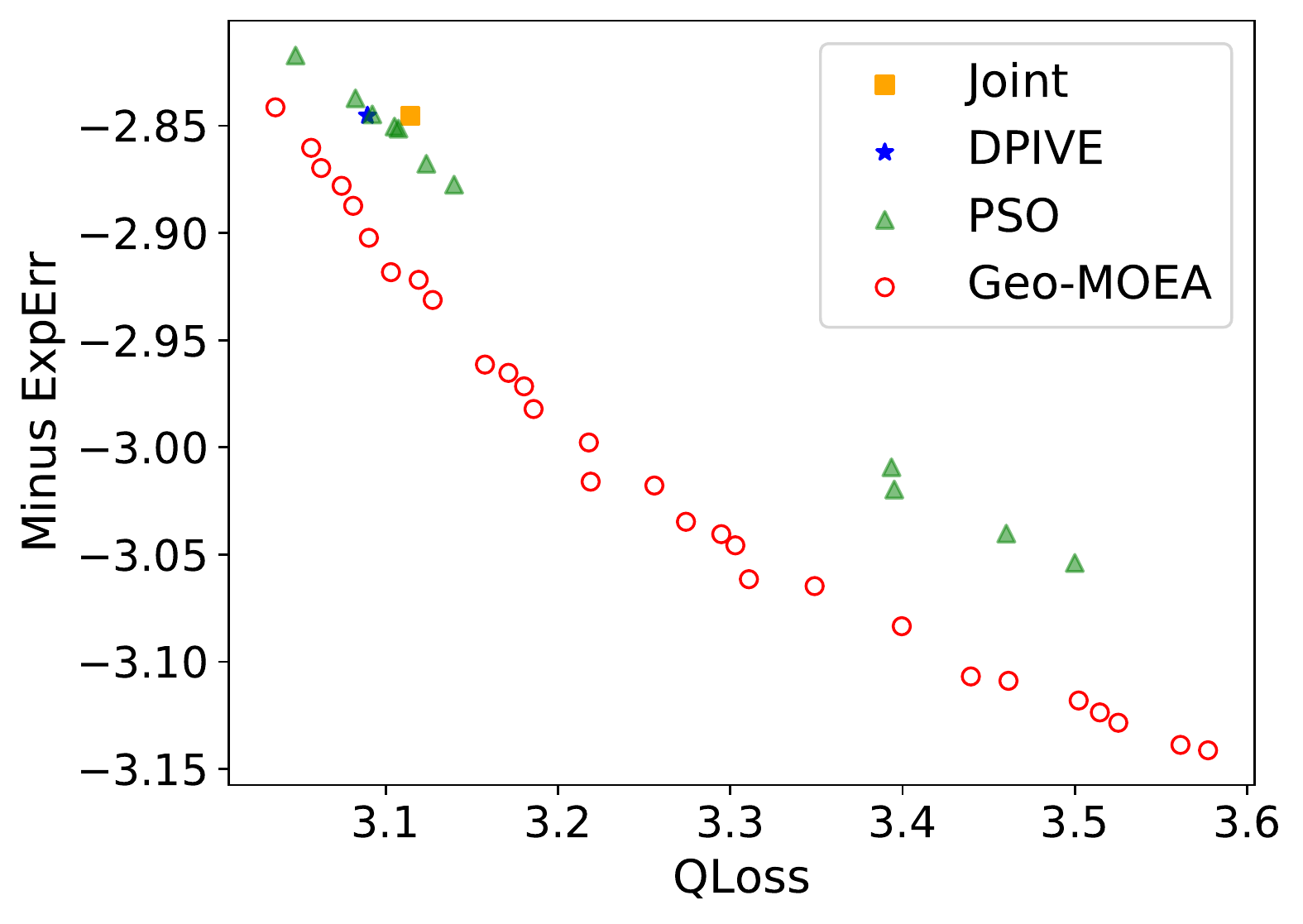}
	}
\subfigure[$E_m=0.15,\epsilon_0=1.00$]{
		\includegraphics[scale=0.28]{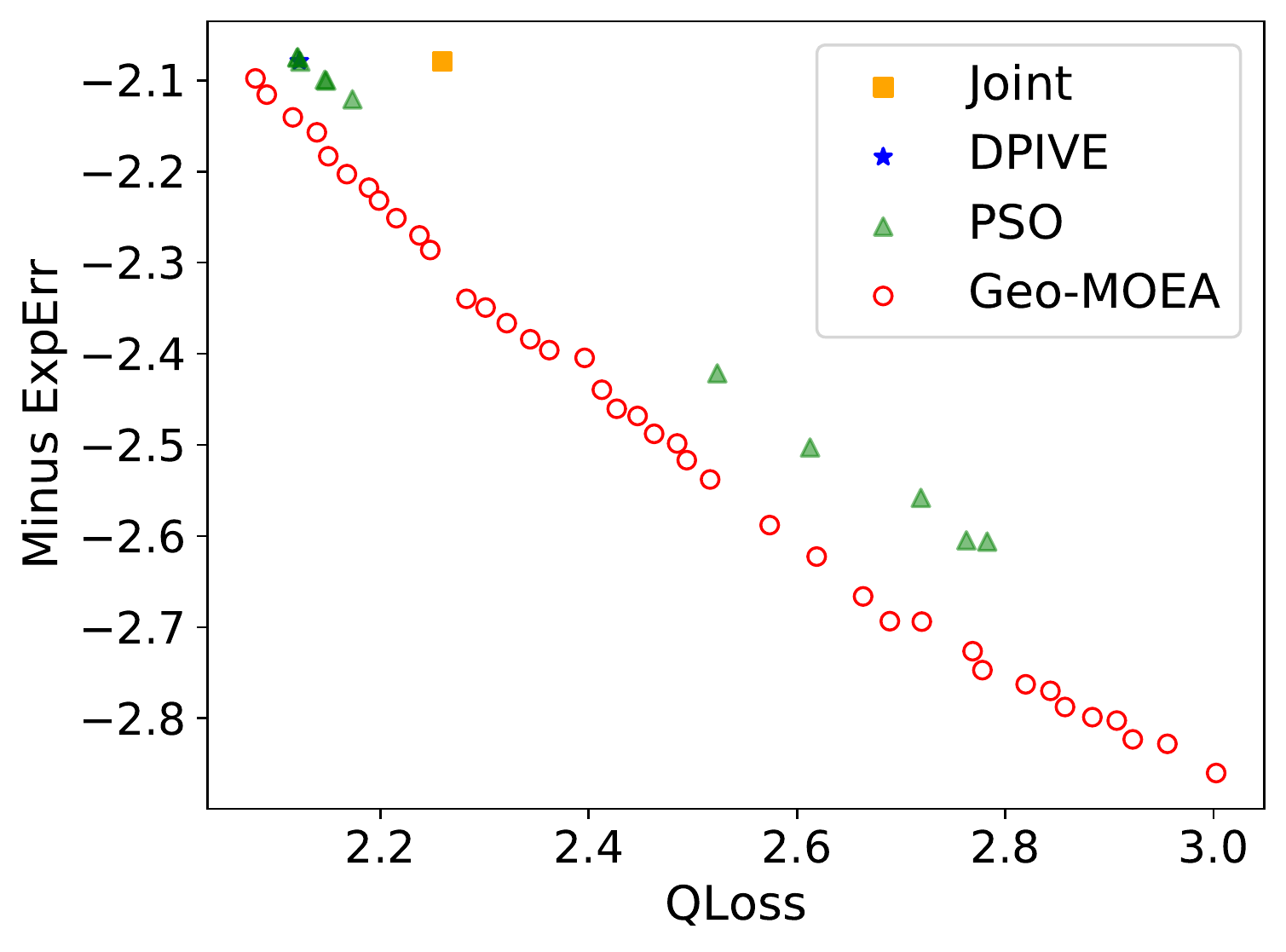}
	}
\subfigure[$E_m=0.15,\epsilon_0=1.50$]{
		\includegraphics[scale=0.28]{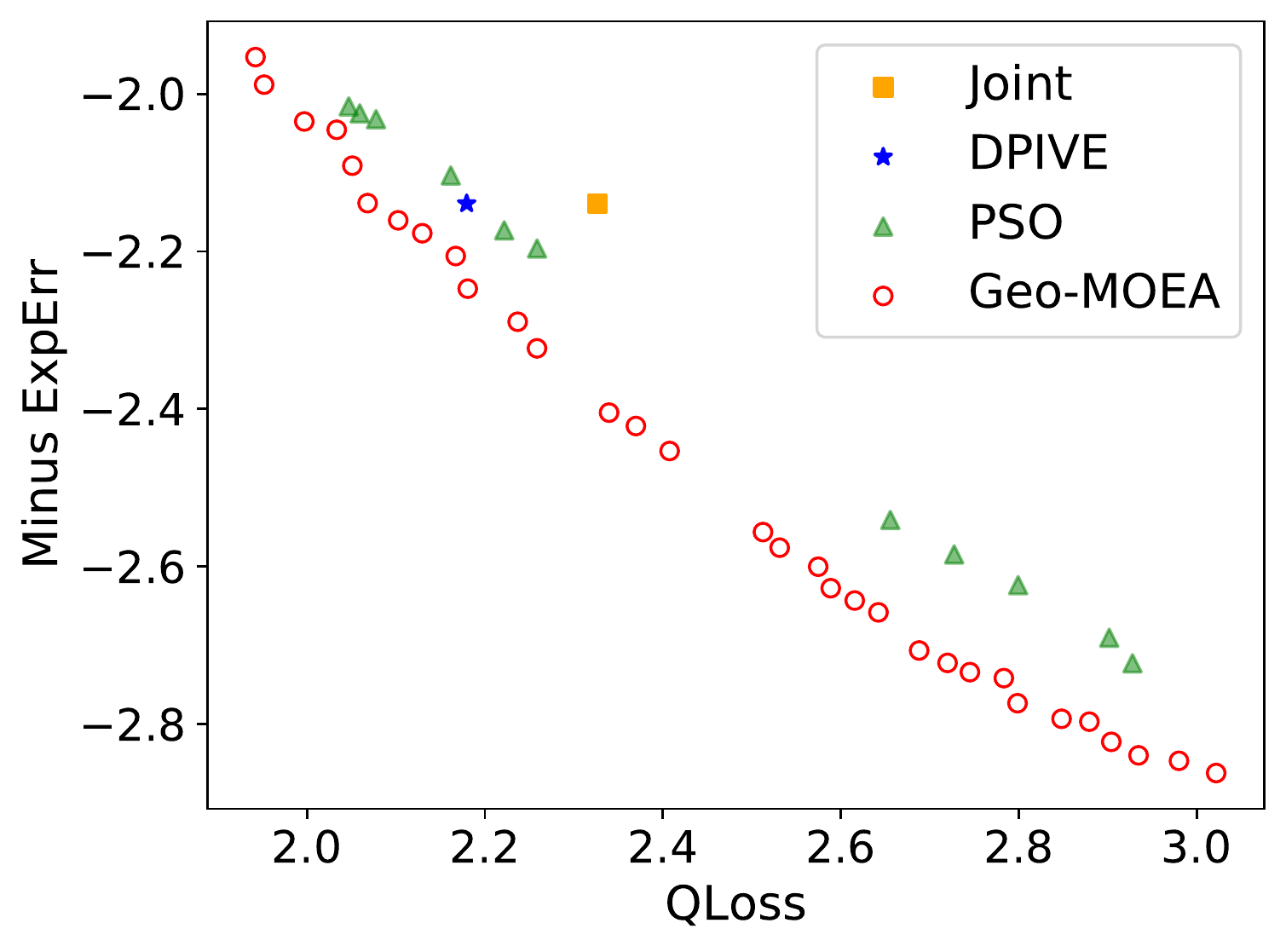}
	}

\subfigure[$E_m=0.20,\epsilon_0=0.50$]{
		\includegraphics[scale=0.28]{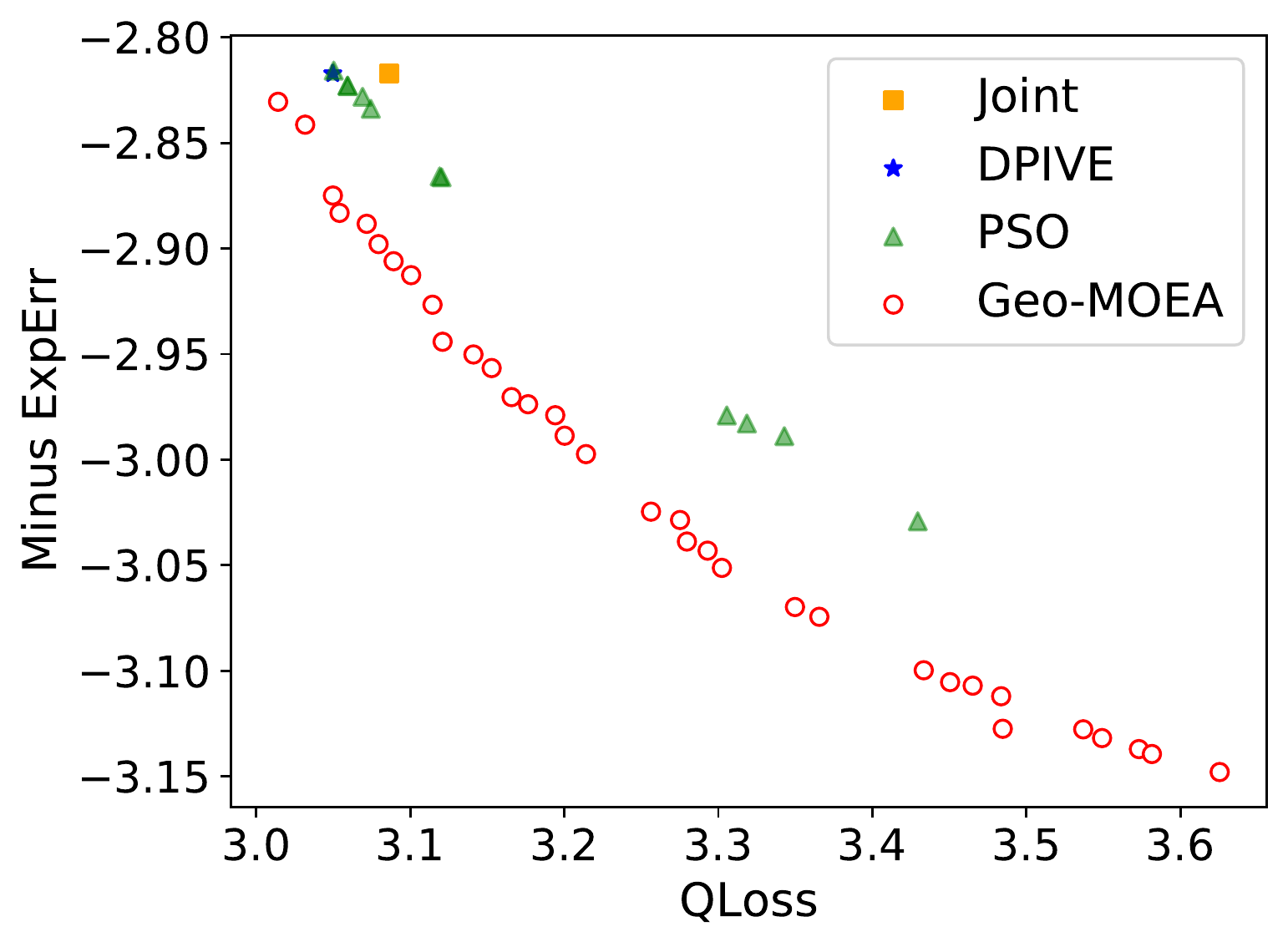}
	}
\subfigure[$E_m=0.20,\epsilon_0=1.00$]{
		\includegraphics[scale=0.28]{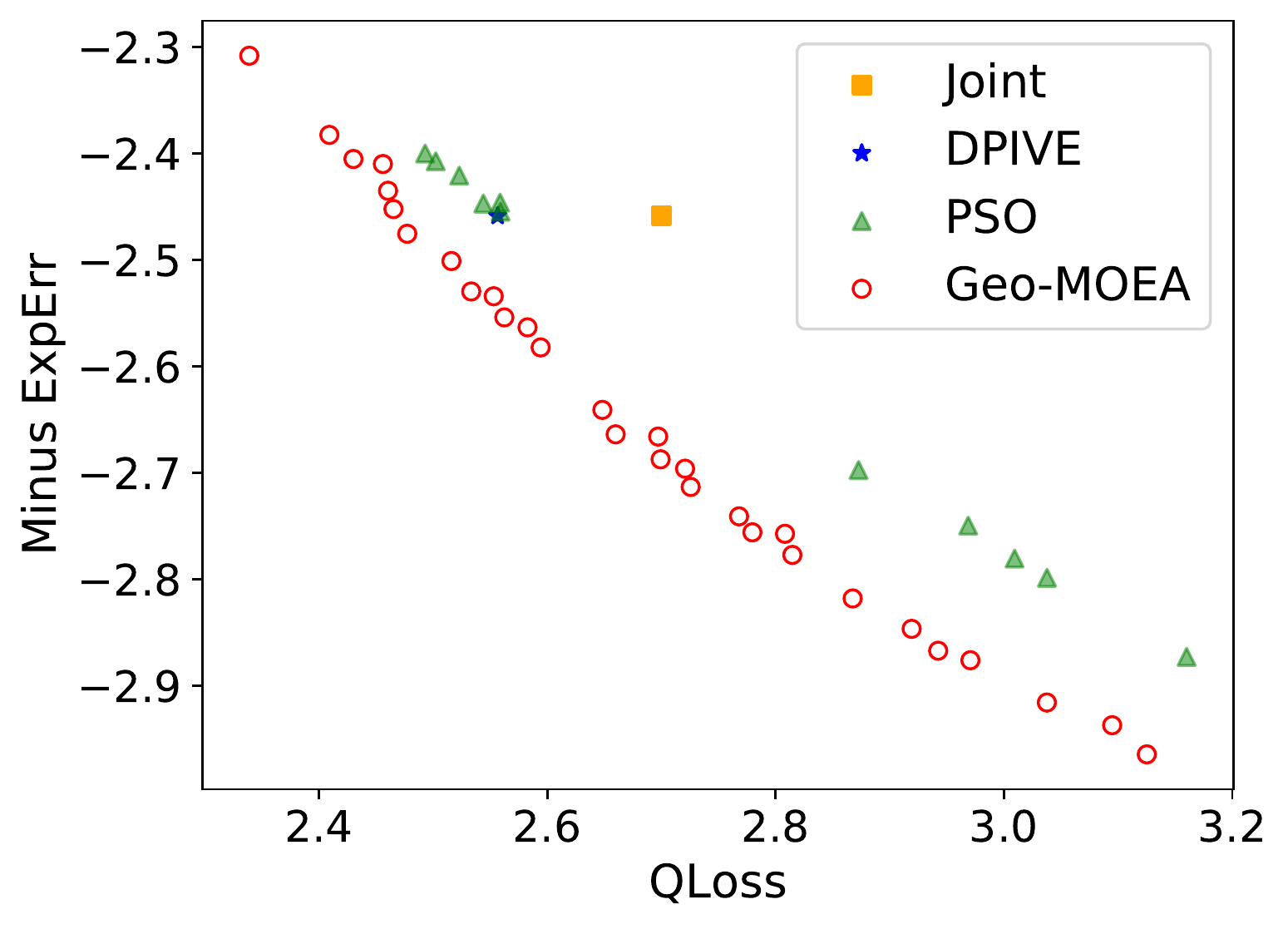}
	}
\subfigure[$E_m=0.20,\epsilon_0=1.50$]{
		\includegraphics[scale=0.28]{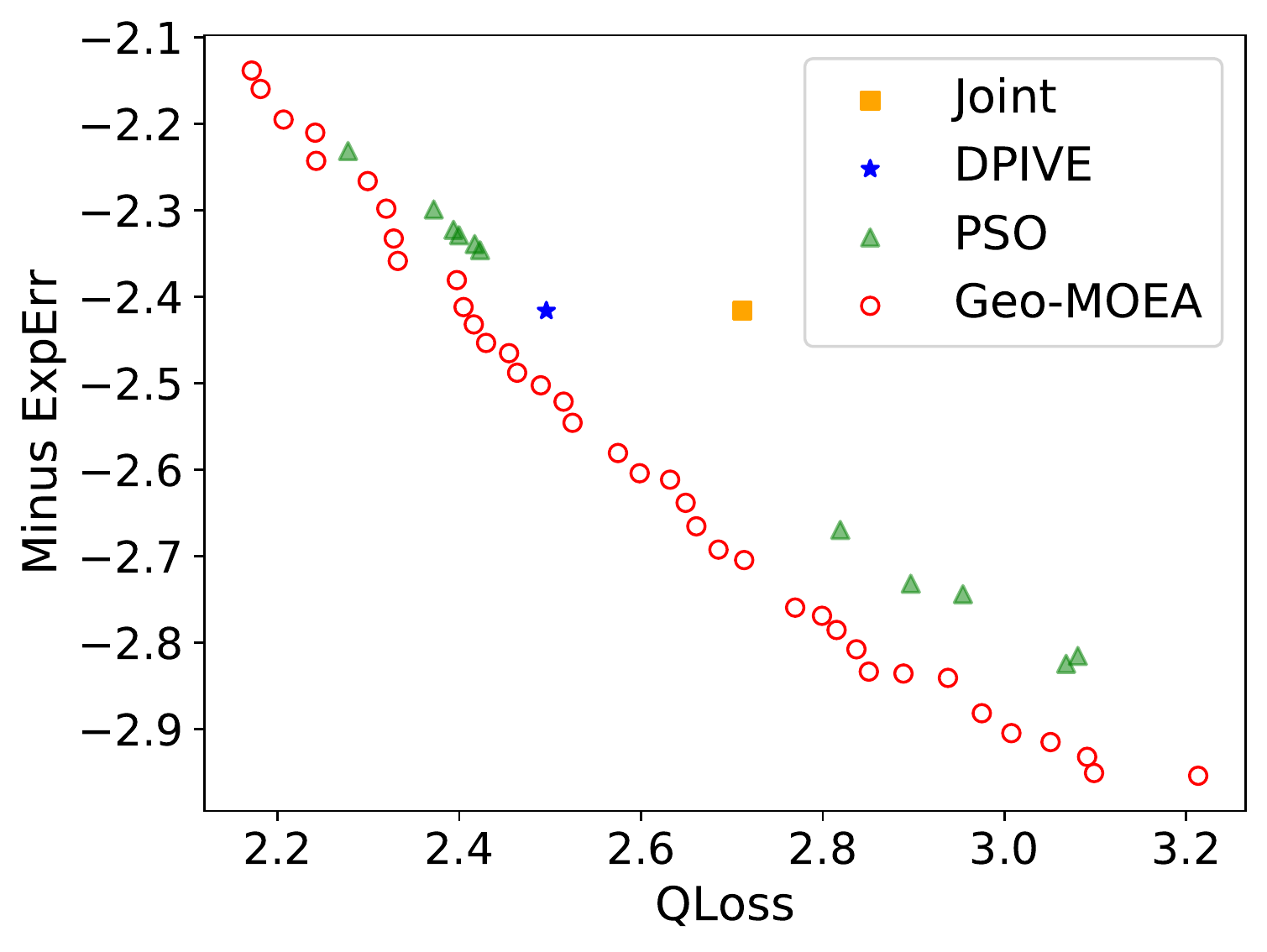}
	}
\caption{Comparison of Geo-MOEA with other mechanisms on NYTaxi dataset.}
\label{fig:Comparison1}
\end{figure*}

\begin{figure*}[tb]
\centering
\subfigure[$E_m=0.10,\epsilon_0=0.50$]{
		\includegraphics[scale=0.28]{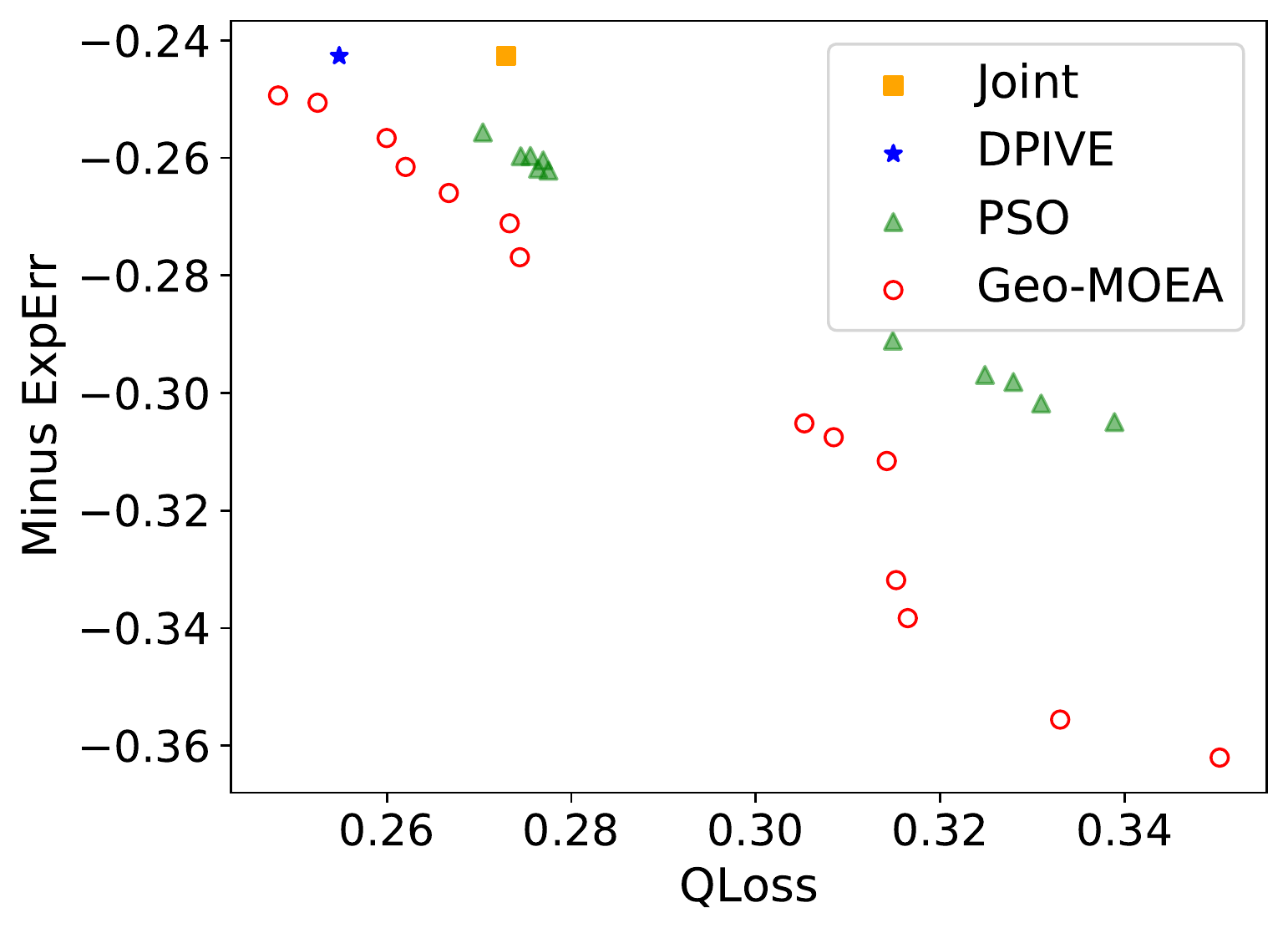}
	}
\subfigure[$E_m=0.10,\epsilon_0=1.00$]{
		\includegraphics[scale=0.28]{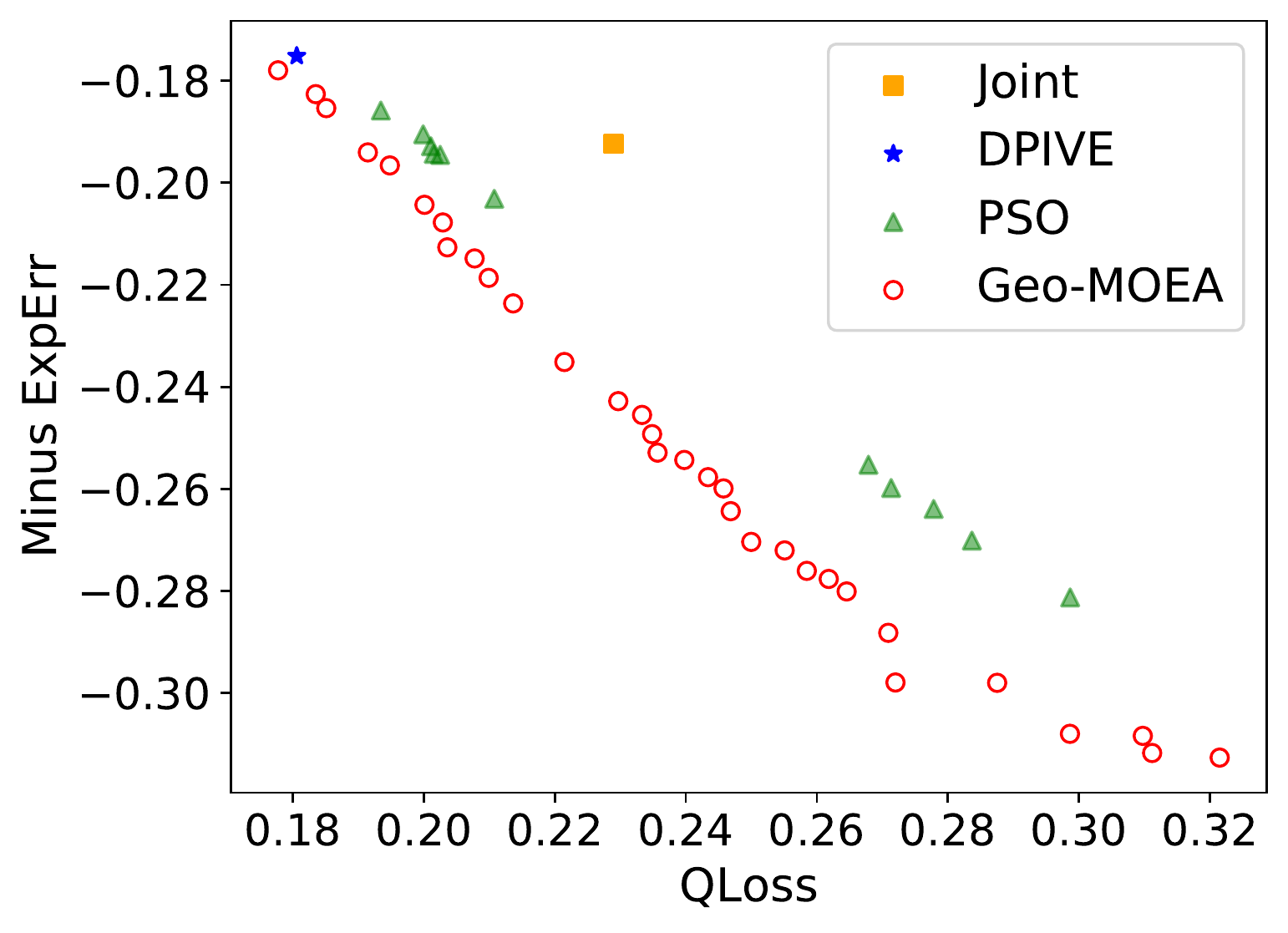}
	}
\subfigure[$E_m=0.10,\epsilon_0=1.50$]{
		\includegraphics[scale=0.28]{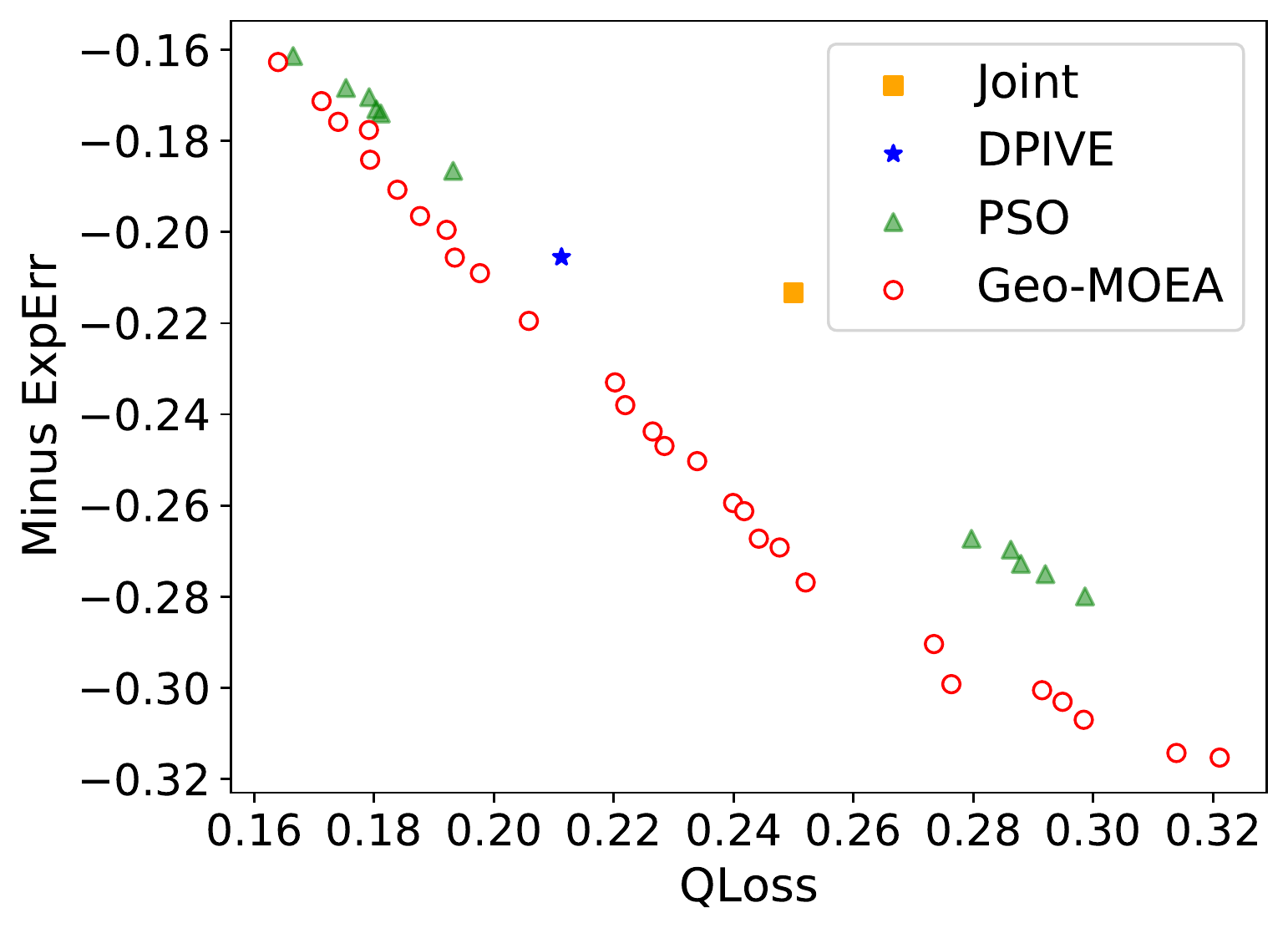}
	}

\subfigure[$E_m=0.15,\epsilon_0=0.50$]{
		\includegraphics[scale=0.28]{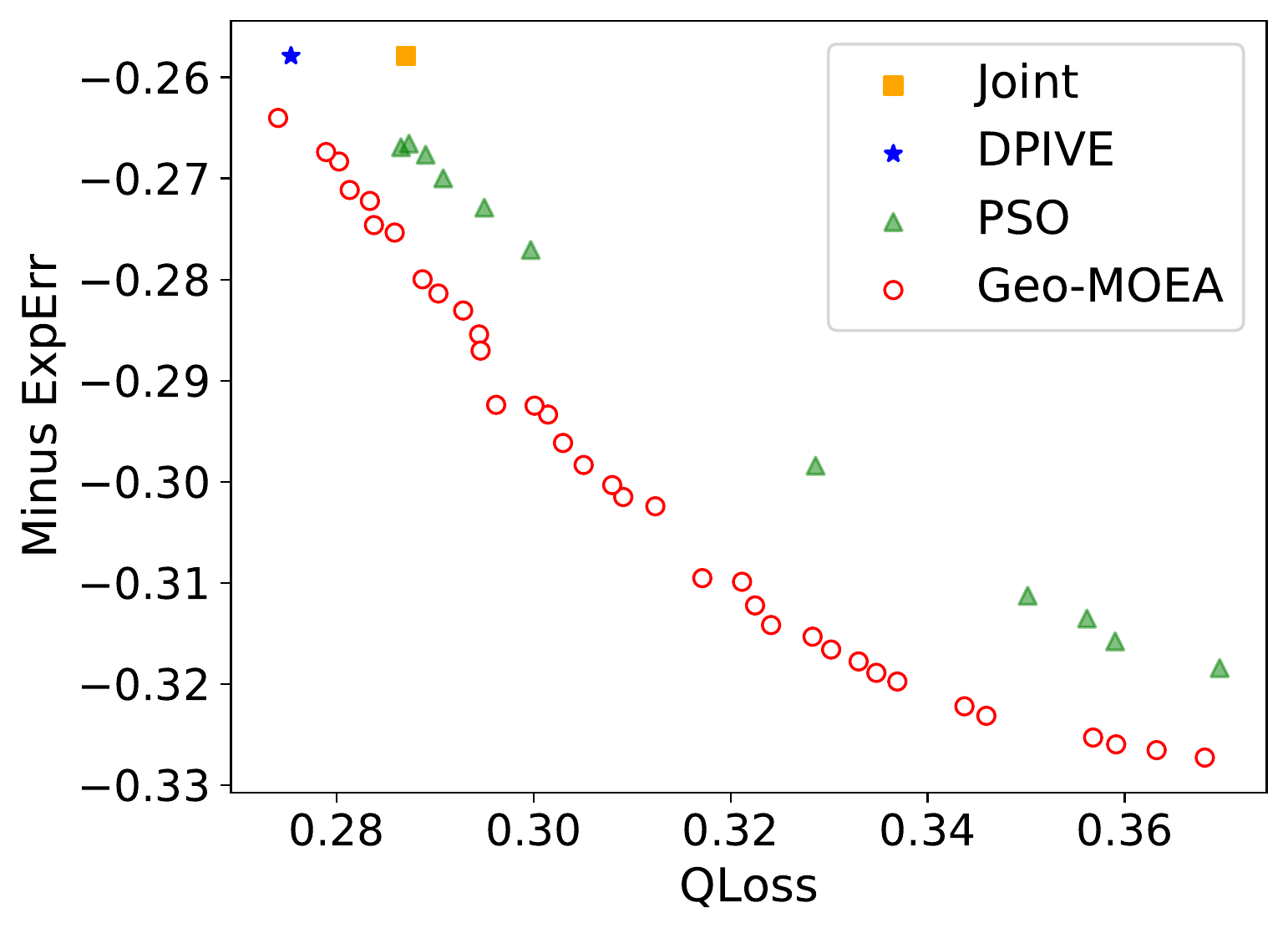}
	}
\subfigure[$E_m=0.15,\epsilon_0=1.00$]{
		\includegraphics[scale=0.28]{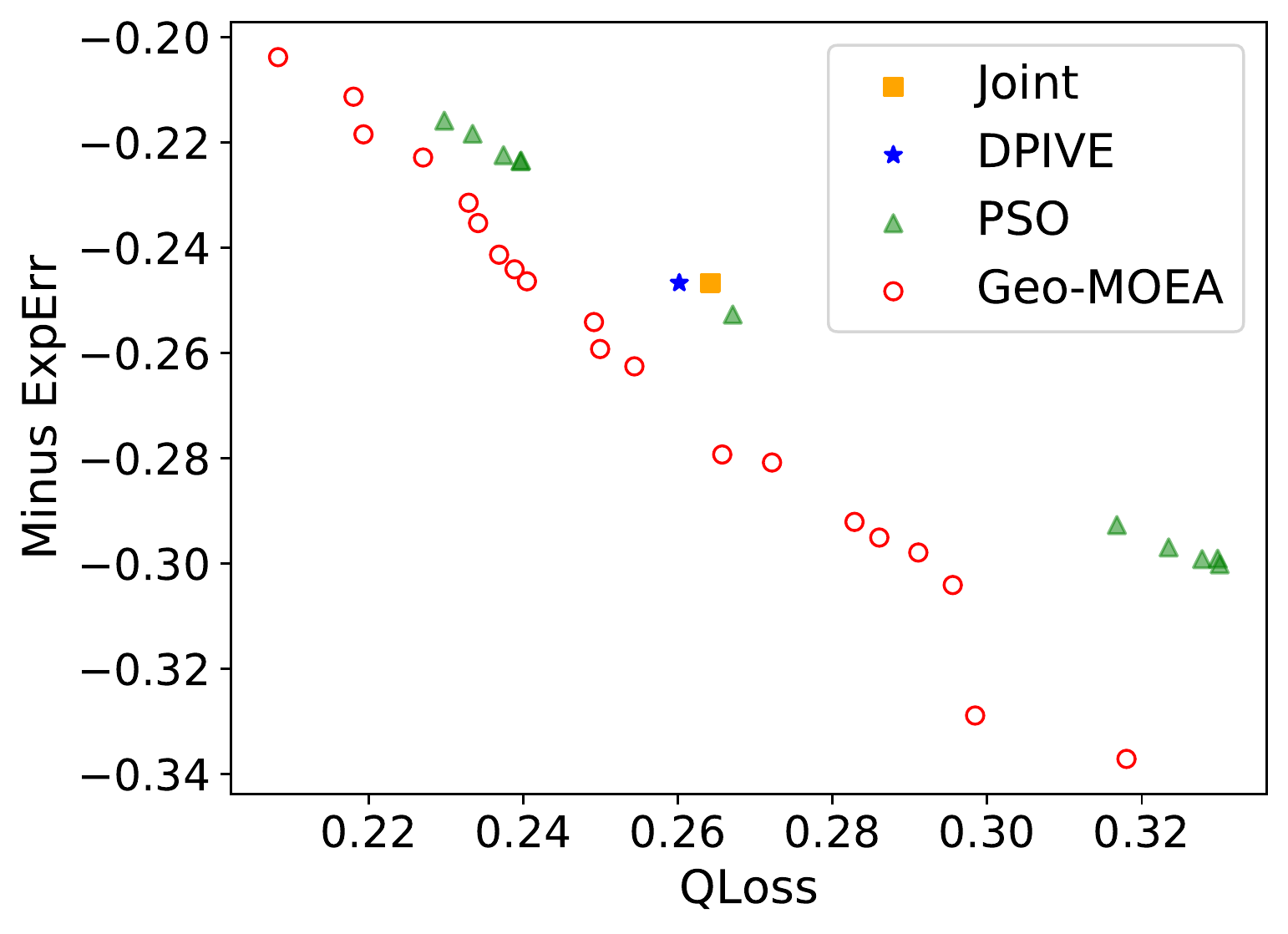}
	}
\subfigure[$E_m=0.15,\epsilon_0=1.50$]{
		\includegraphics[scale=0.28]{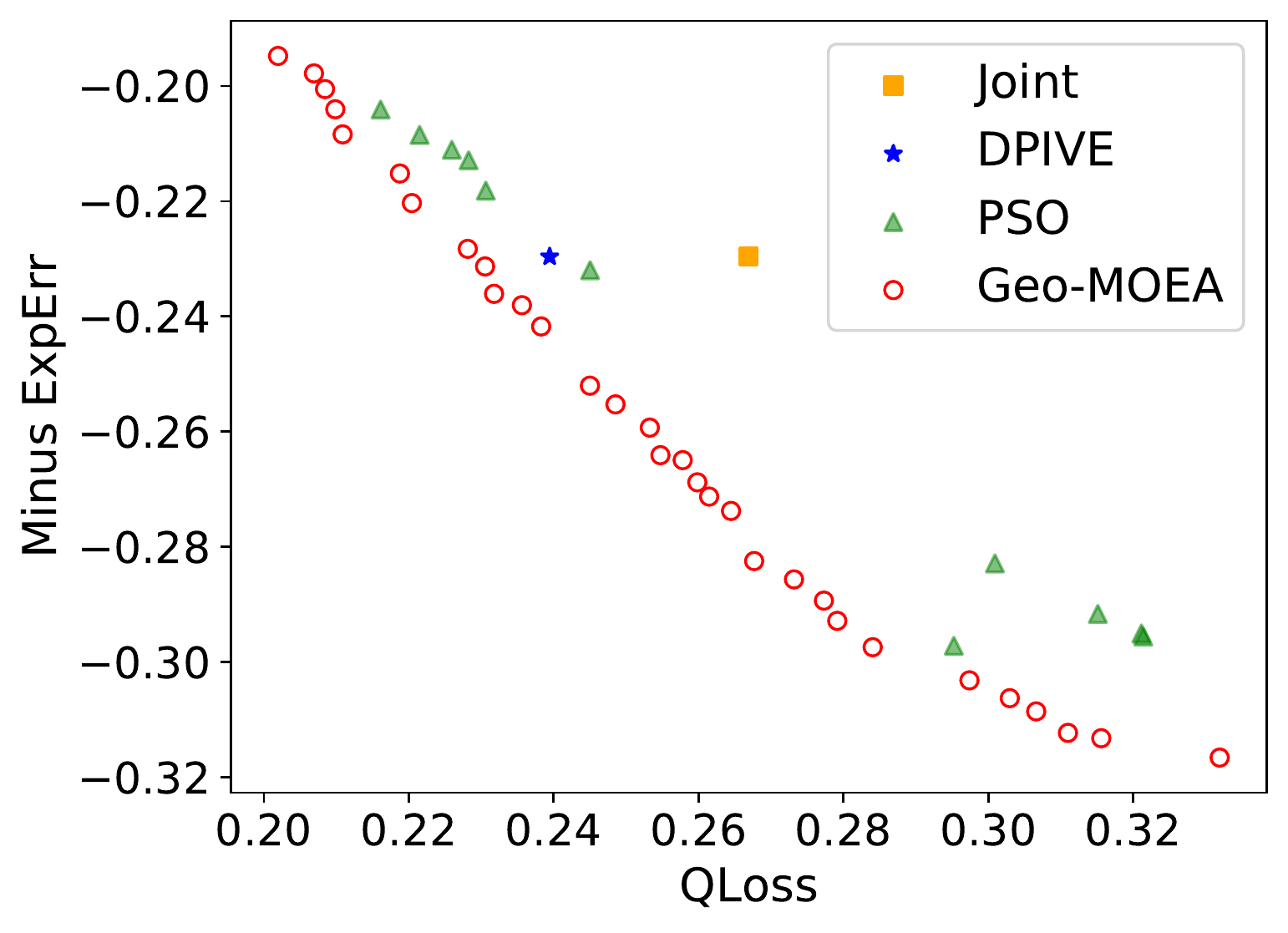}
	}

\subfigure[$E_m=0.20,\epsilon_0=0.50$]{
		\includegraphics[scale=0.28]{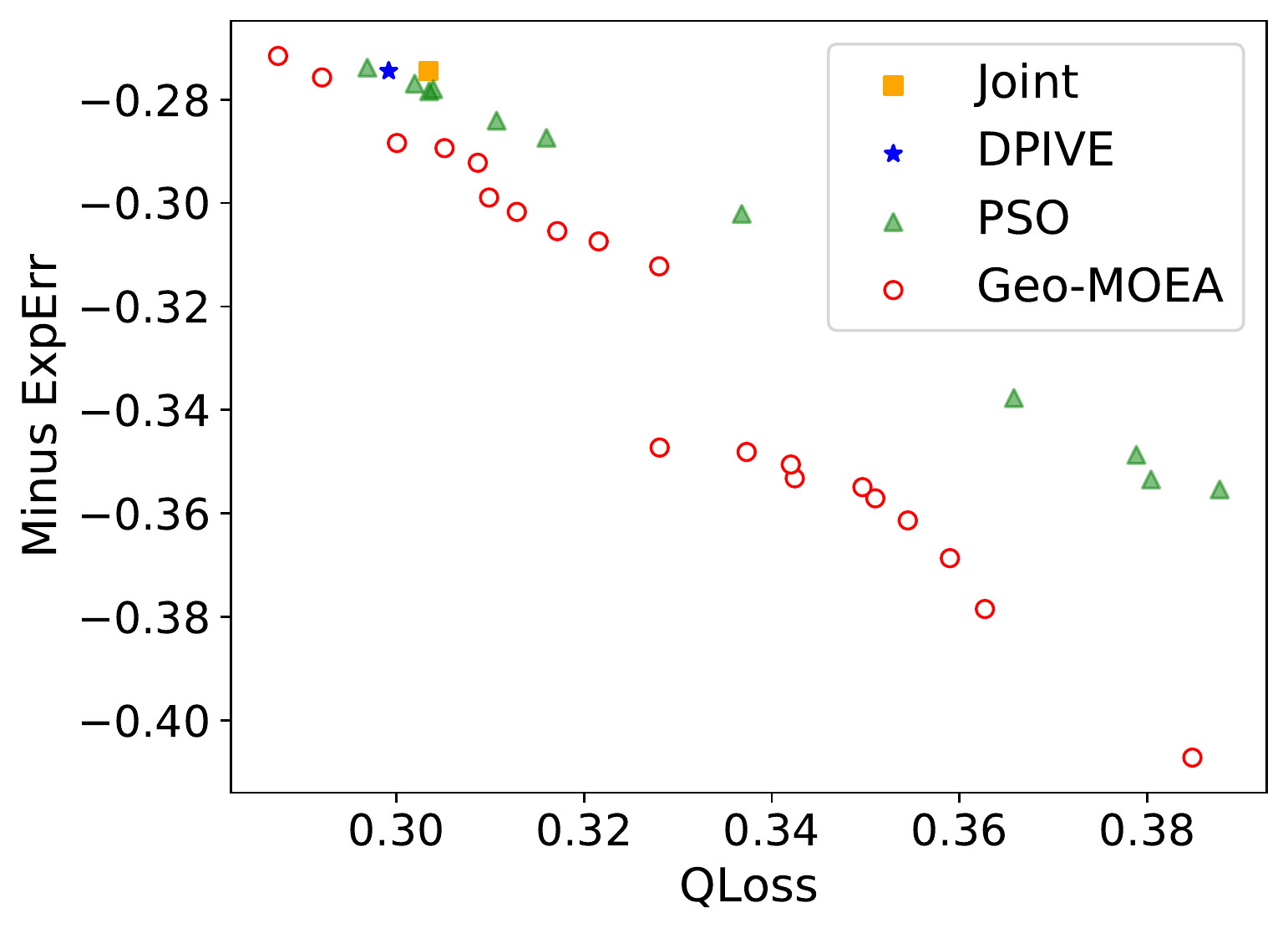}
	}
\subfigure[$E_m=0.20,\epsilon_0=1.00$]{
		\includegraphics[scale=0.28]{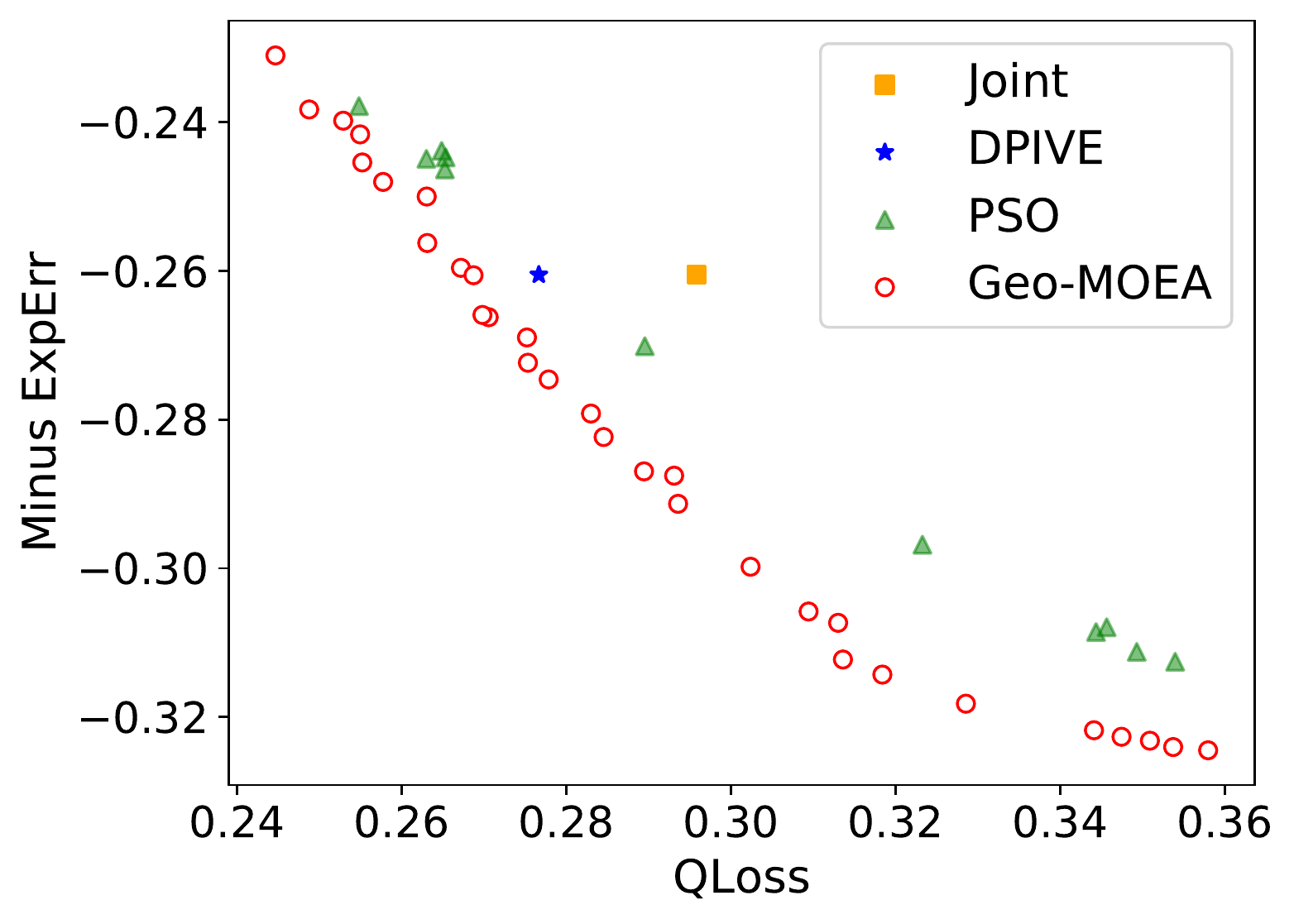}
\label{fig:fig8}
	}
\subfigure[$E_m=0.20,\epsilon_0=1.50$]{
		\includegraphics[scale=0.28]{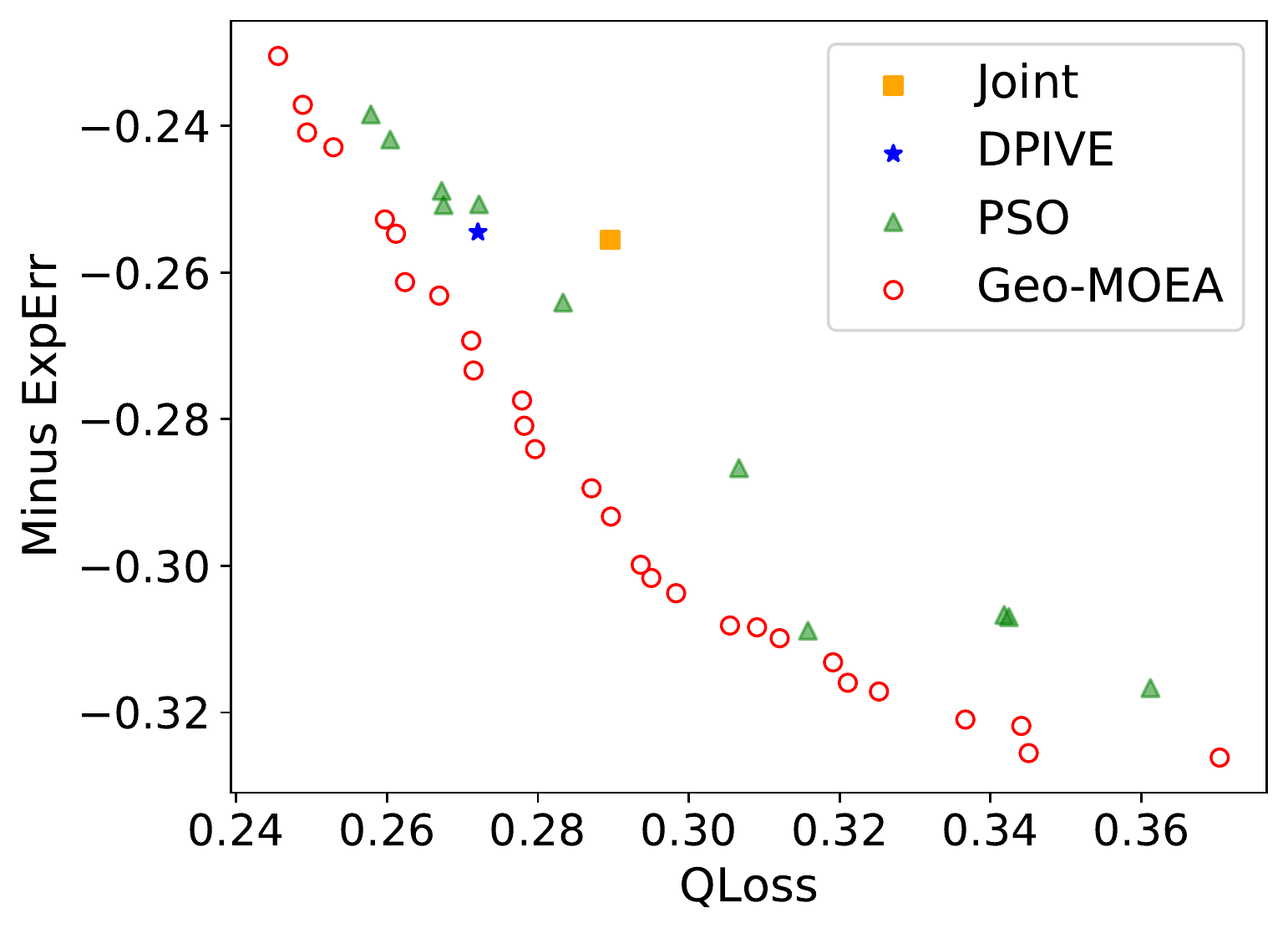}
	}
\caption{Comparison of Geo-MOEA with other mechanisms on Gowalla dataset.}
\label{fig:Comparison2}
\end{figure*}

\textbf{Parameter settings.} The minimum expected number of locations in each cell is $n_0=33$. Geo-MOEA gives two privacy control knobs: 1) differential privacy parameter $\epsilon_0\in \{0.10,0.20,\ldots,1.50\}$ and 2) threshold of expected inference error $E_m\in \{0.050,0.075,\ldots,0.300\}$. The default setting is given by $\epsilon_0=1.0$ and  $E_m=0.1$. Its maximum iteration times is $500$. We sample randomly $200$ tasks on the distributed locations with $2000$ and $900$ idle workers in each dataset, respectively. We consider two types of idle workers' distributions: (1) $80\%$ of idle workers are densely located in 20\% space while the remaining $20\%$ workers are distributed in the rest of $80\%$ space (denoted by ``$1:4$ mode"). (2) the idle workers are uniformly distributed in each dataset (denoted by ``uniform mode"). We assume that different tasks have no spatio-temporal confliction
to each other. This means that they can share a single worker. Their WTDs are computed on average. On comparisons of the metrics WTD, for the Pareto solutions of Geo-MOEA, only the extreme solution with the smallest QLoss is taken into account.

\begin{table} \small
\caption{Comparison of Geo-MOEA with other mechanisms on various requirements}
\label{tlb:goal}
        \centering
		\begin{tabular}{rcccc}
			\toprule
\quad & Local DP & 
$E_m$ &  Scalability  &  Genetic evolution \\
			\midrule
                        Joint \cite{Sho15}      & \Checkmark & \XSolid & \XSolid   & \XSolid
\\
						DPIVE \cite{ZDC22}       & \Checkmark & \Checkmark & \XSolid   & \XSolid
\\
                        PSO \cite{EK95}       & \Checkmark & \Checkmark & \Checkmark  & \XSolid
                        \\
                        Geo-MOEA    & \Checkmark & \Checkmark & \Checkmark  & \Checkmark
                        \\
			\bottomrule
		\end{tabular}
	\end{table}

\subsection{Privacy Protection Goals}
The privacy protection goals of the experiments for applications are as follows.
(1) preserving $\epsilon_0$-DP inside each PLS based on geo-indistinguishability and weak DP in the large-scale data scenario.
(2) achieving distortion privacy with lower bound $E_m$ against Bayesian attack even on isolated locations. and
(3) improving the data utility for applications.

We mention that our proposed Geo-MOEA is able to defeat Bayesian adversary attacks locally and globally. Locally, the lower error threshold $E_m$ controls the (conditional) expected inference error ExpErr for each location involving the isolated regions. Globally, the unconditional expected inference error ExpErr quantifies the degree of resistance to Bayesian inference attack, which provides a method to pick a solution from the Pareto optimal recommendations.

Let us make comparisons on various requirements with DPIVE \cite{ZDC22}, Joint \cite{Sho15} and  PSO \cite{EK95} (the original Particle Swarm Optimization), see Table \ref{tlb:goal}. Only Geo-MOEA meets all requirements. These mechanisms can satisfy the (local) lower inference error bound $E_m$ except Joint.
DPIVE and Joint are not suitable for large-scale location domain scenarios while the PSO algorithm produces bad solutions due to it linear-weighted combination approach for multi-objective optimization.
Geo-MOEA involving expected inference errors can effectively resist Bayesian attacks via prior knowledge, and meets the requirements of large-scale location domain. In essence, the three existing mechanisms adopt single-objective optimization while our proposed Geo-MOEA takes into account the trade-off between two conflicting aspects: distortion privacy level and service quality.


\subsection{Pareto Analysis}
In this section, we compare Geo-MOEA with DPIVE \cite{ZDC22}, Joint  \cite{Sho15} and  PSO \cite{EK95}, especially to verify the efficiency in balancing service quality loss and average expected inference error as well as some extreme solutions.
Since the lower bound $E_m$ ensures the conditional expected inference error locally,
we pay more attention to the global performance on the metrics, QLoss, Minus ExpErr, and HV.

The DPIVE carried out in each cell $X_i$ separately is a single-objective optimization algorithm that aims to minimize the quality loss while ensuring the lower bound $E_m$ of inference error. 
Joint is the first mechanism that uses linear programming to combine two privacy notions of expected inference error and geo-indistinguishability. We also carry out Joint in each cell $X_i$ separately and use the global expected inference error of DPIVE as
the minimum desired distortion privacy level $d_m$, via adjusting $\epsilon_0$ to obtain the same expected inference error.

PSO converts our problem
to single-objective optimization and proceeds without mutations and crossovers \cite{EK95}.
 The fitness function defined by $F(\alpha)=\alpha \cdot QLoss - (1-\alpha)\cdot ExpErr$ with $\alpha \in \{0.0,0.1,\ldots,1.0\}$ as in \cite{WYH19}, which combines both objectives to optimize the obfuscation schemes.


Figs. \ref{fig:Comparison1} and  \ref{fig:Comparison2} present a series of comparisons on the conflicting metrics of QLoss and ExpErr with the convergent Pareto dominance solutions on the two datasets.
Each DPIVE solution is obviously dominated by some solutions of Geo-MOEA in the (a)-(i) settings. On the aspect of QLoss, DPIVE increases by $7.2\%$ and $8.8\%$ on average,  and particularly $19.9\%$ in case Fig. \ref{fig:fig3} and  $24.9\%$ (reversely, a 20.0\% reduction) in case Fig. \ref{fig:fig8}, respectively on the two datasets, compared to the extreme (left) mechanism with the smallest quality loss in the Pareto-recommendations generated by Geo-MOEA. The reasons include that, the reporting range of DPIVE is restricted in local small-scale cell while more PLSs are located in the corners, and DPIVE minimizes the average diameter of PLSs for optimizing QLoss while ignoring the fitness of ExpErr.
Joint increases by $11.5\%$ and $22.3\%$ on average and and a maximum increase of $24.9\%$ and $52.4\%$, respectively on the two datasets, compared to Geo-MOEA.
Moreover, the Pareto solutions of Geo-MOEA are basically located at the lower left side of PSO solutions, which means that Geo-MOEA solutions can well Pareto dominate those of PSO globally. Indeed, the PSO algorithm also initializes its population, but lacks the crossover and mutation processes that help Geo-MOEA expand the solution space.

\begin{figure}[tb]
\centering
\subfigure[HV, NYTaxi]{
		\includegraphics[scale=0.21]{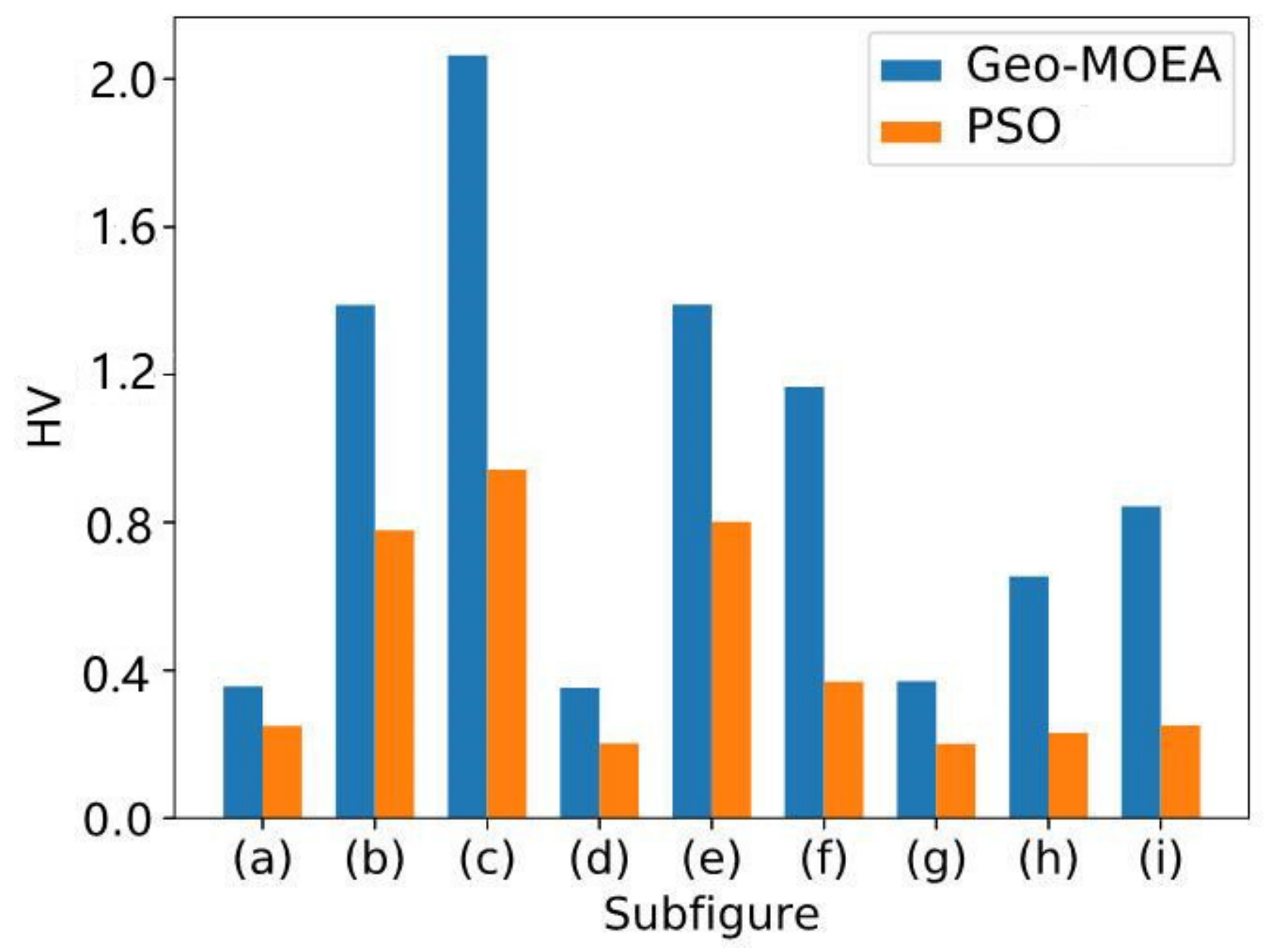}
	}
\subfigure[HV, Gowalla]{
		\includegraphics[scale=0.21]{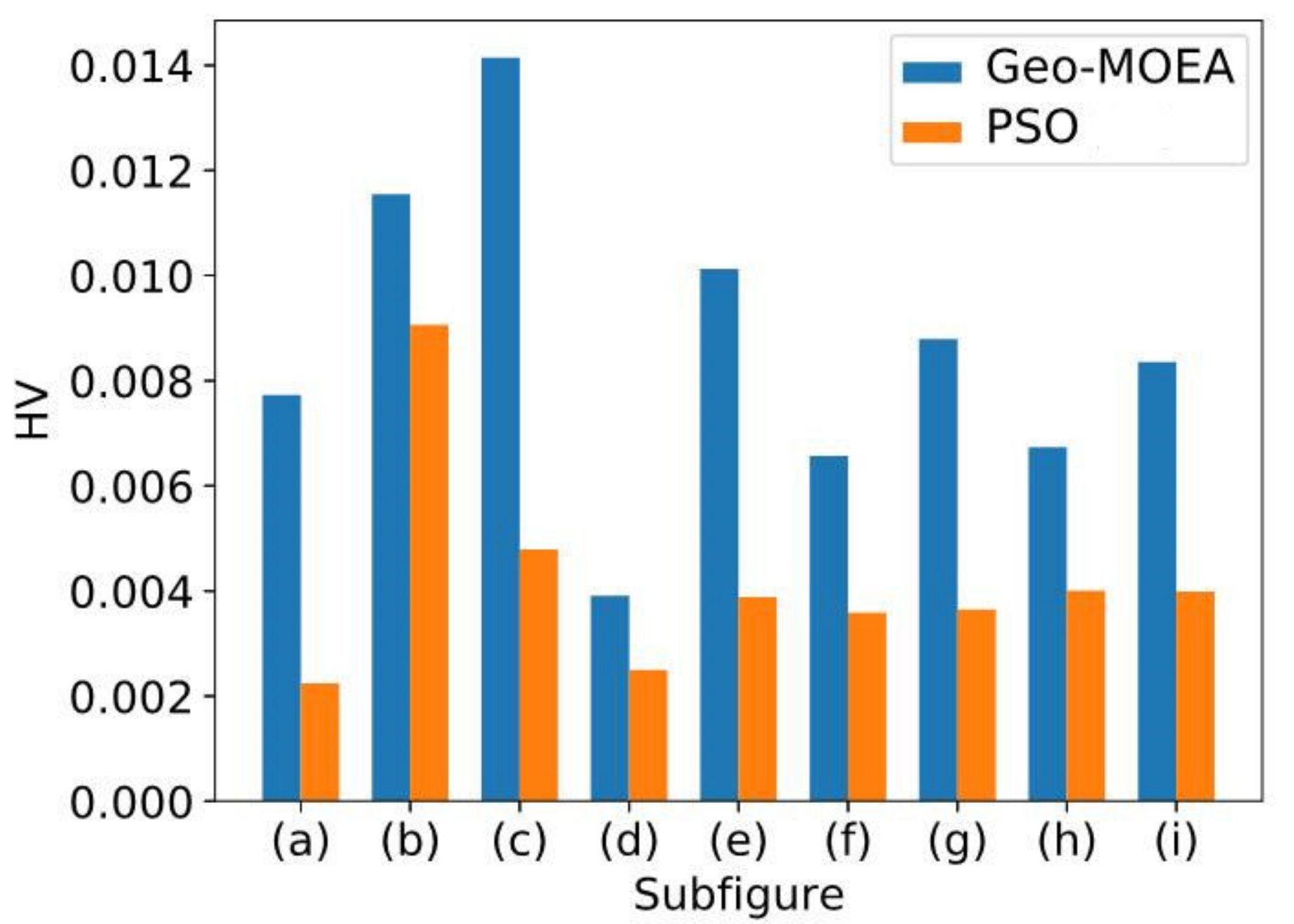}
	}
\caption{Comparisons of Geo-MOEA with PSO on HV.}
\label{fig:HV}
\end{figure}

As a result, the HV of Geo-MOEA solutions is twice as large as that of PSO solutions on average and even $3.3$ and $2.4$ times, respectively on the two datasets as shown in Fig. \ref{fig:HV}. Hence, Geo-MOEA brings more optimized solutions, faster convergence and greater diversity while achieving privacy protection goals.

\begin{figure}[tb]
\centering
\includegraphics[scale=0.5]{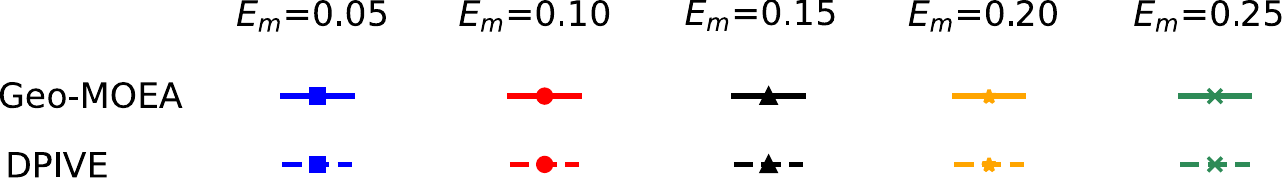}

\subfigure[QLoss, NYTaxi]{
		\includegraphics[scale=0.25]{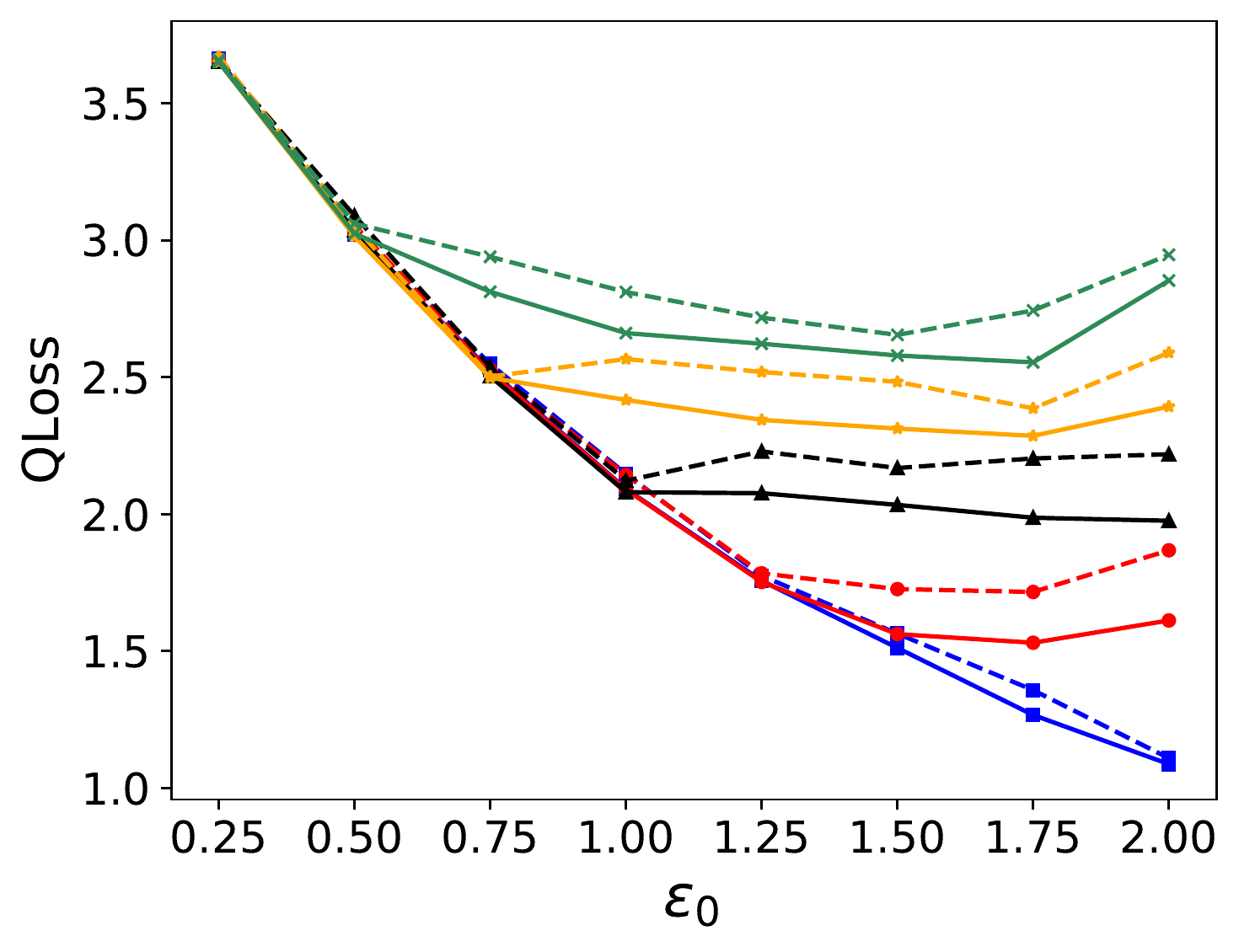}
	}
\subfigure[QLoss, Gowalla]{
		\includegraphics[scale=0.25]{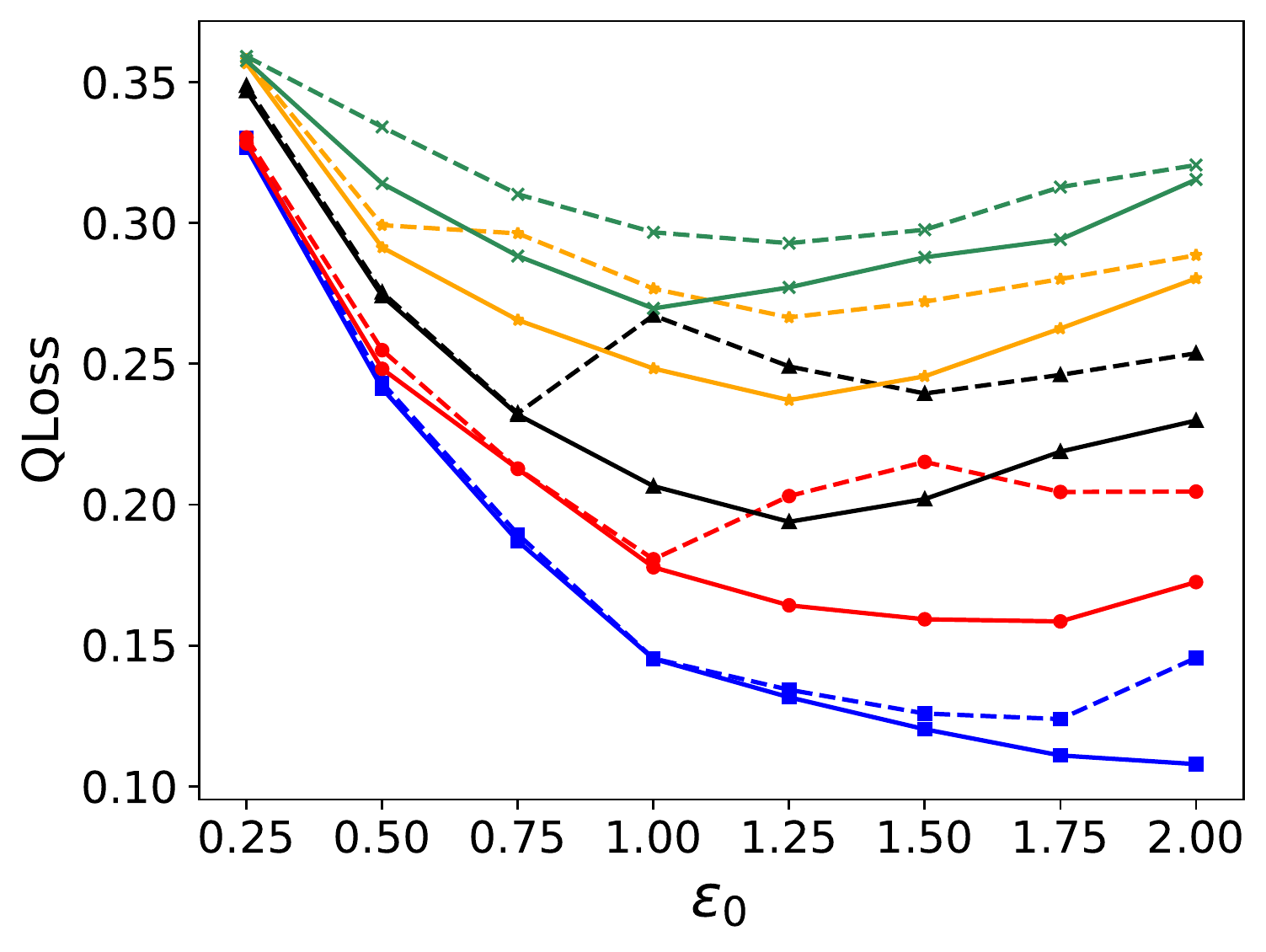}
	}
\caption{Comparisons on QLoss with varying $\epsilon_0$.}
\label{fig:QLoss1}
\end{figure}

Fig. \ref{fig:QLoss1} shows that, compared with the extreme Geo-MOEA, DPIVE has an average increase of $3.9\%$ and $7.7\%$, and a maximum increase of $13.7\%$ and $25.9\%$, respectively on the two datasets. The larger the $E_m$ is, the earlier the branch point of curve appears. Once \eqref{eq:inequality-eps0} is satisfied, the gap between the curves expands soon due to the possible retreat for larger $\epsilon_g$. When $E_m$ is very small, there are no possibilities to retreat so that both curves are close to each other. However, as $\epsilon_0$ becomes large, it is not so easy to satisfy \eqref{eq:inequality-eps0} and there is larger search space for Geo-MOEA with $\epsilon<\epsilon_0$.


\subsection{Engineering Applications}

The distance WTD that the worker travels from the actual location to the allocated task determines the efficiency of mechanism application. For convenience, we compute WTD by the Euclidean distance simply while the actual travel distance depends on the road networks.
 We conduct comparative experiments on the two datasets with varying privacy parameter $\epsilon_0$. Under each parameter setting, we average the WTDs on random tasks for comparisons of four schemes.
The notation Non-privacy means Geo-MOEA without privacy protection, that is, the SC-server geocasts the three idle workers closest to the task directly based on the real locations and their average WTD is referred to.

\begin{figure}[tb]
\centering
\includegraphics[scale=0.4]{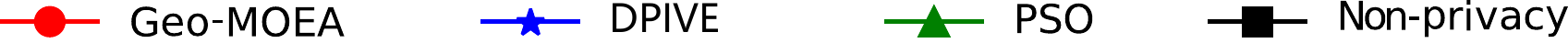}

\subfigure[WTD, NYTaxi (1:4)]{
		\includegraphics[scale=0.25]{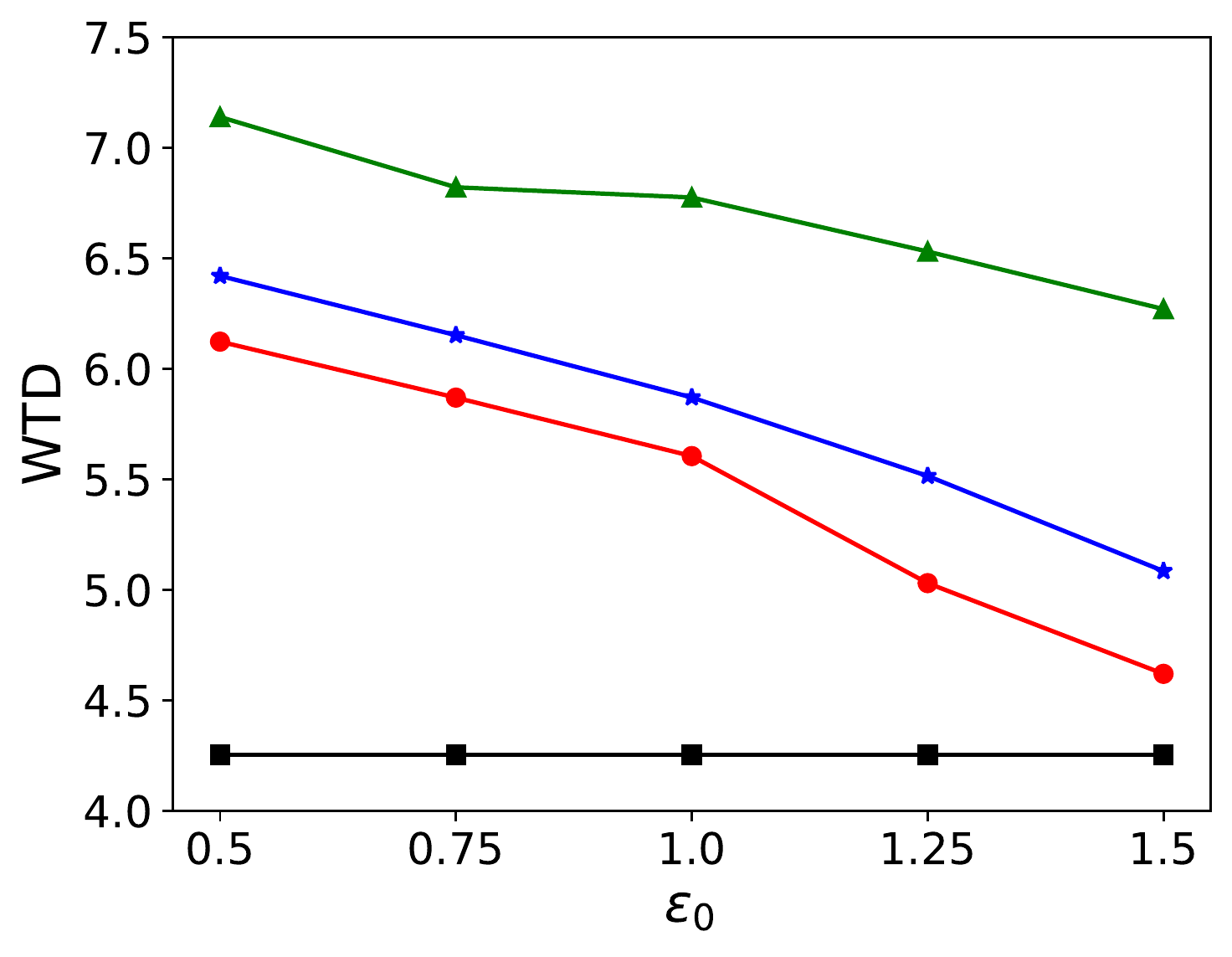}
\label{fig:WTD_3000_1}
	}
\subfigure[WTD, Gowalla (1:4)]{
		\includegraphics[scale=0.25]{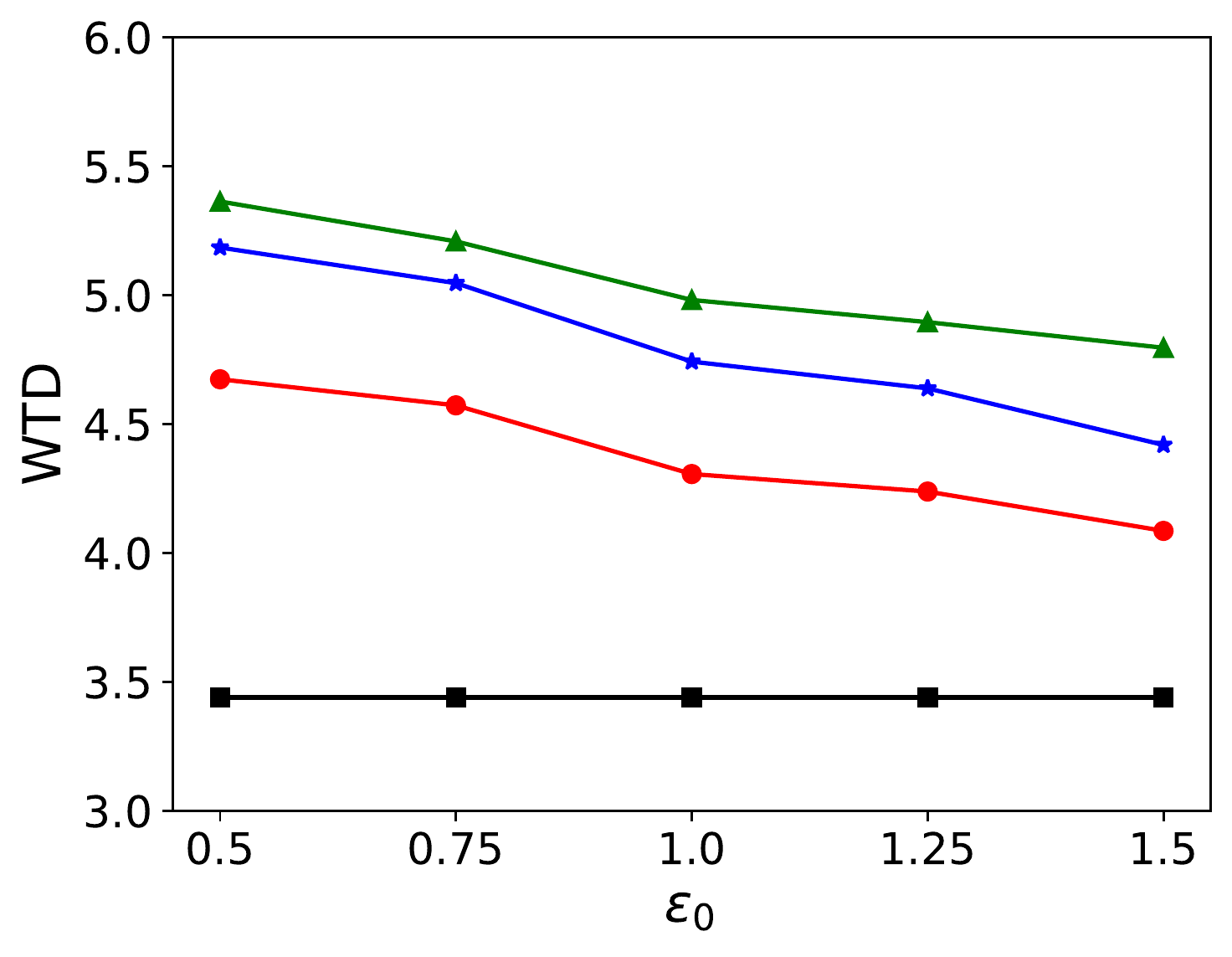}
\label{fig:WTD_1000_1}
	}

\subfigure[WTD, NYTaxi (uniform)]{
		\includegraphics[scale=0.25]{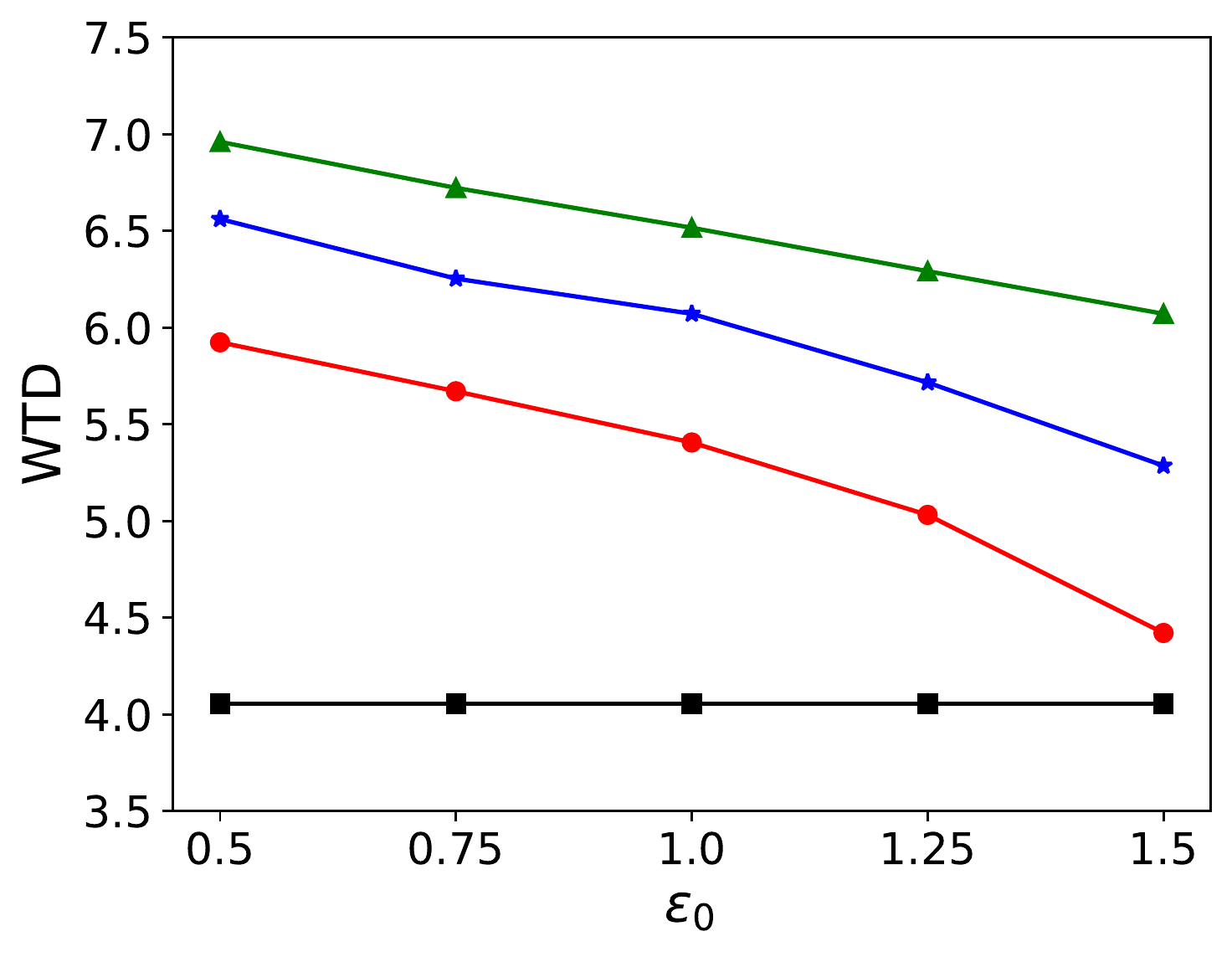}
\label{fig:WTD_3000_2}
	}
\subfigure[WTD, Gowalla (uniform)]{
		\includegraphics[scale=0.25]{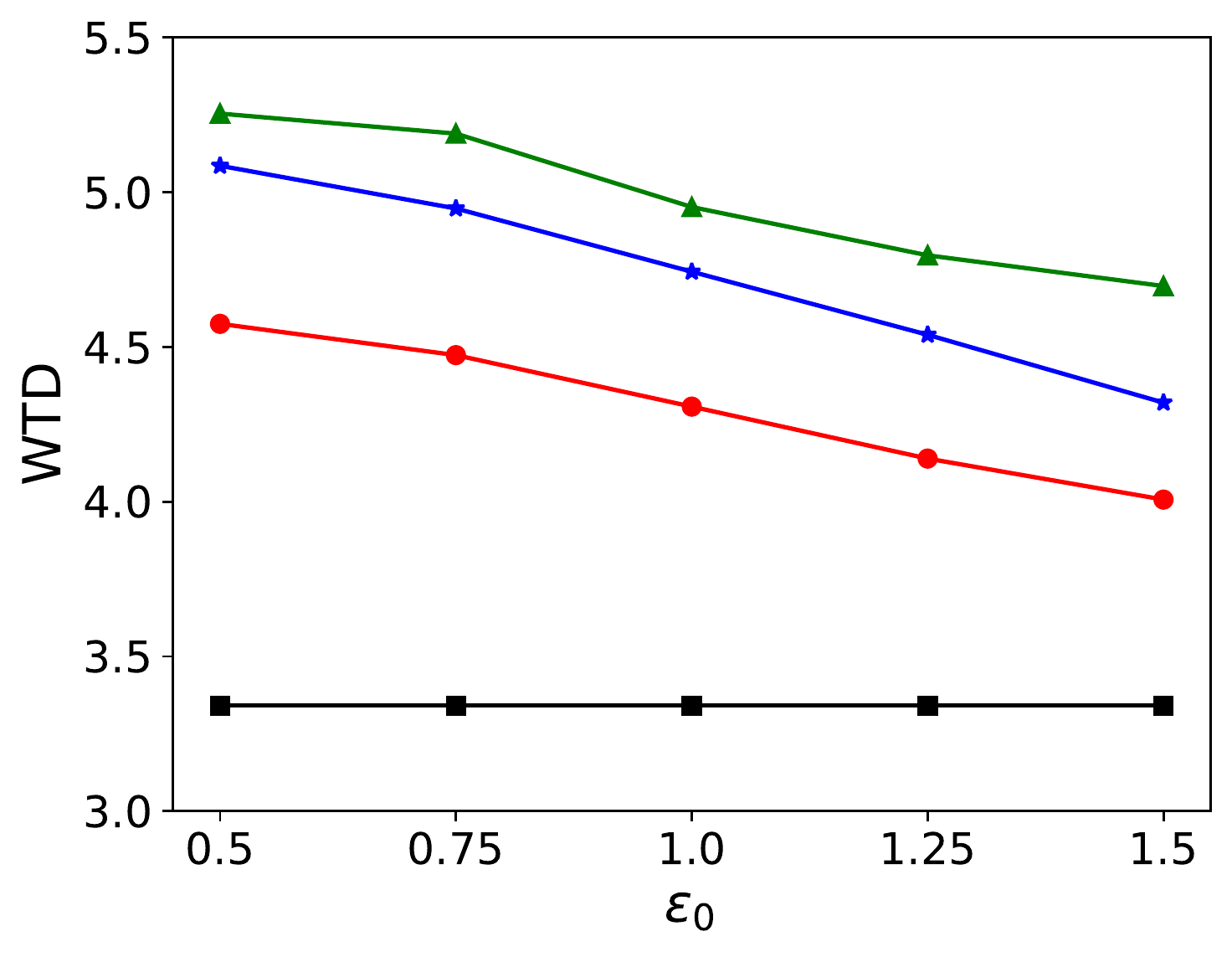}
\label{fig:WTD_1000_2}
	}
\caption{Comparisons on WTD with varying $\epsilon_0$.}
\label{fig:WTD}
\end{figure}

As shown by Fig. \ref{fig:WTD_3000_1} and Fig. \ref{fig:WTD_1000_1} for the $1:4$ mode, on comparisons to the extreme mechanism with the smallest quality loss among the Parero-optimality recommendations generated by Geo-MOEA,
DPIVE has an average increase of $6.8\%$ and $9.8\%$, and a maximum increase of $10.1\%$ and $10.9\%$, respectively, on the two datasets, while PSO has an average increase of $23.8\%$ and $15.4\%$ and a maximum increase of $35.7\%$ and $17.4\%$, respectively. In Fig. \ref{fig:WTD_3000_2} and Fig. \ref{fig:WTD_1000_2} for the uniform mode, DPIVE has an average increase of $11.6\%$ and $9.8\%$, and a maximum increase of $16.3\%$ and $11.2\%$, respectively, on the two datasets, while PSO has an average increase of $18.9\%$ and $15.7\%$ and a maximum increase of $27.2\%$ and $17.2\%$, respectively. Geo-MOEA performs multi-objective genetic algorithms, which helps to generate extreme mechanisms reducing largely the quality losses. On the other hand, the privacy protection for Geo-MOEA leads to higher WTD ($30.5\%$, $26.4\%$, $22.5\%$ and $22.1\%$ on average in the four sub-figures, respectively) than the Non-privacy case due to the privacy level and the spare distribution of idle workers particularly for the $1:4$ mode.
This shows that our mechanism can protect well worker's location privacy while improving the availability of existing SC mechanisms in  large-scale  scenario.

\begin{figure}[tb]
\centering
\subfigure[HV-WTD, NYTaxi]{
		\includegraphics[scale=0.175]{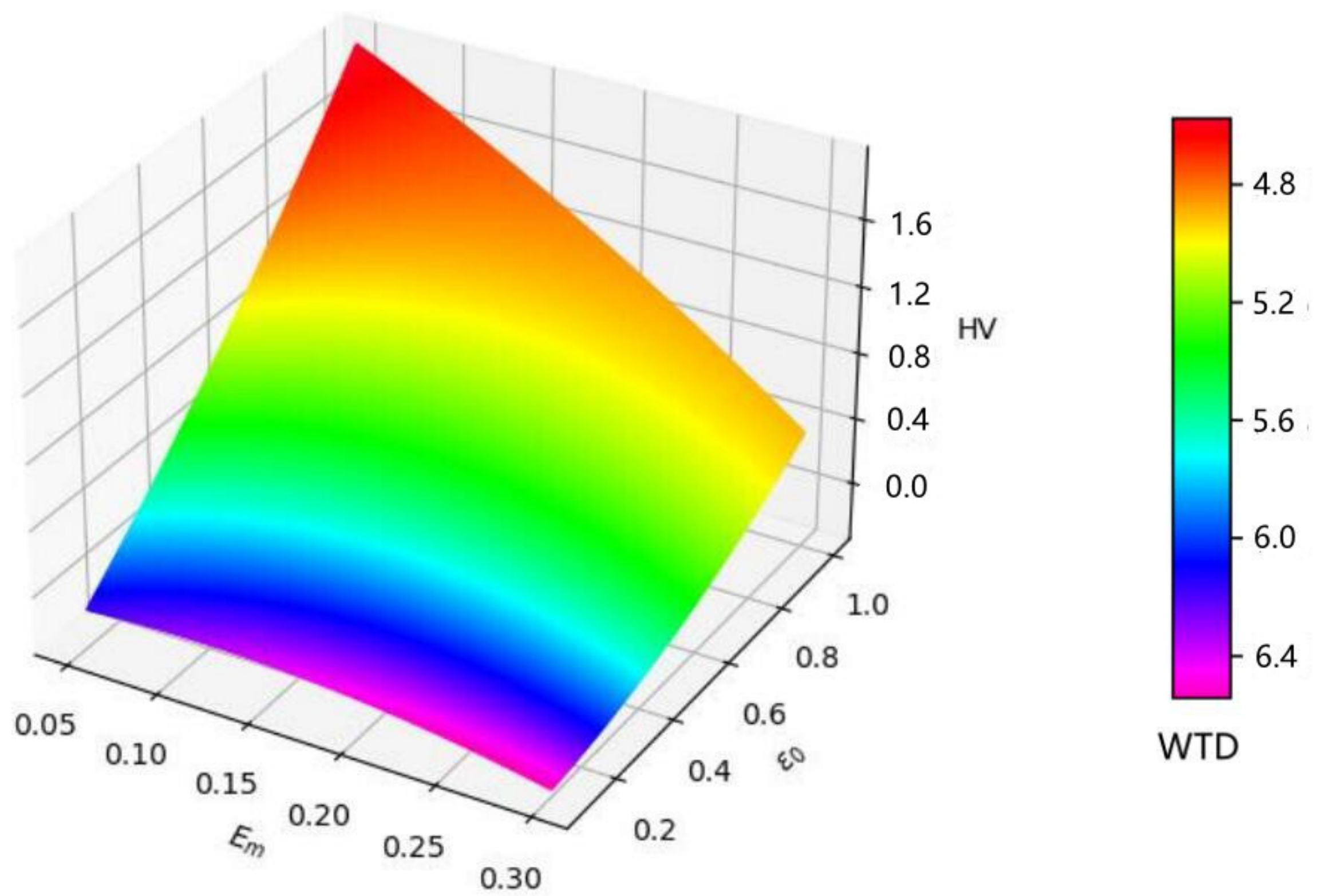}
	}
\subfigure[HV-WTD, Gowalla]{
		\includegraphics[scale=0.175]{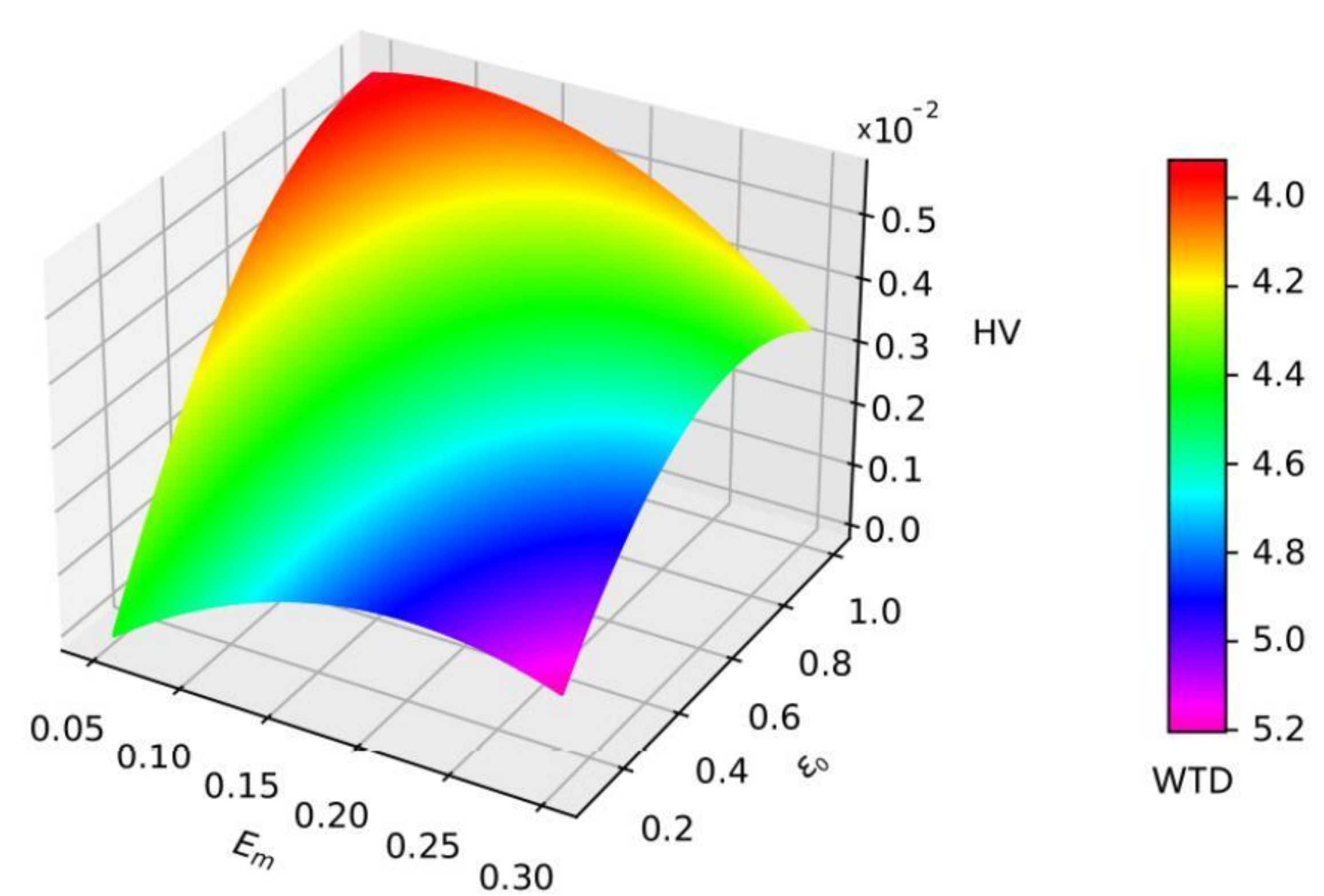}
	}
\caption{Visualized metrics via surface fitting.}
\label{fig:vis}
\end{figure}

Further, we make analysis in terms of visualization.
For the extreme Geo-MOEA mechanism with the smallest quality loss, we make visualization analysis on the experiments under $110$ settings of privacy knobs, $E_m =0.050,0.075, \ldots, 0.300$ and $\epsilon_0 =0.1,0.2,\ldots,1.0$ for each dataset. We adopt surface fitting by the metric HV and mark the surface in variable colors by fitting on the metric WTD. The idle workers are located randomly in the uniform mode as mentioned above.
As shown in Fig. \ref{fig:vis}, for fixed $E_m$, the larger $\epsilon_0$ leads to higher HV and the more expanded solution space while generating smaller WTD naturally due to the lower level of privacy requirement.
When $\epsilon_0$ remains unchanged, WTD presents an upward trend for increasing $E_m$.
For both datasets, the metrics of HV and WTD reach respective extreme values at the case of $E_m = 0.05, \epsilon_0=1.0$. The surfaces are marked in green in the middle part around $\epsilon_0 =0.6$  and more green part are expressed for Gowalla dataset due to its local denser property.
Each curved surface provides a state of high HV and low WTD, which indicates that the Geo-MOEA algorithm reaches an extreme state. In this setting, the PLSs are small as well as injected noises and the multi-objective solution space is large, which makes the Pareto-optimal solutions more diverse.


Moreover, in most crowdsourcing applications, workers are not located in fixed locations. That is, their locations for most change dynamically.
For this, we can set up some discrete locations $\{x_i\}$ such as check-ins and bus stations. The closest $x_i$ to the worker is regarded as her true location for adopting Geo-MOEA.
In addition, the workers can join the platform and leave the platform dynamically by reporting.

\subsection{Efficiency Evaluation}
In real life, besides the quality of task assignment, the time cost of generating mechanisms should also be taken into account. The execution of Geo-MOEA mainly includes two stages. The first stage (stage A) is the utilization of DPIVE to divide the domain into regions $X_i$ with generating its corresponding $k_i$. In the second stage (stage B), the Pareto-optimal solutions are produced by using MOEA.

The testing environment consists of Intel(R) Xeon(R) Silver 4210R CPU 2.40GHz, 132GB SSD, Red Hat 4.8.5-44, and python version 3.8.
Based on this, the stage A takes about $160$ seconds and $100$ seconds, respectively on the two datasets, while the stage B takes about $200$ seconds and $140$ seconds in various settings of privacy parameters, respectively. Since the configuration of our running environment is relatively low, the whole running time for each privacy setting can be reduced to $20$ seconds for some advanced platforms.
In the mobile device applications, the Pareto-optimal solutions for various privacy settings can be renewed on the service platform every half or an hour for effective SC.



\section{Conclusion and Future work}\label{sec:conclusion}
This paper implements the geo-indistinguishable location privacy protection of \emph{Spatial Crowdsourcing} (SC) workers in mobile networks.
Our proposed Geo-MOEA achieves to generate a pseudo-location among an adaptive reporting range with distortion privacy guarantee in the large-scale data domain.
Together with a binary partition method,
a multi-objective genetic algorithm for \emph{Protection Location Set} (PLS) partitions is introduced into the \emph{Local Differential Privacy} (LDP) protection, which makes the first attempt to realize the optimized tradeoff between quality loss and inference error. Moreover, the most general condition for qualified PLSs is presented, which helps theoretically to design a clustering algorithm with retreats for more search space, particularly reducing quality loss. Experimental results confirm strongly that Geo-MOEA outperforms the existing 
schemes in terms of service quality. In the future work, we plan to explore multi-task assignment problems with LDP that involves higher computational complexity and various requirements on communication environments. The specific environments include complex road networks, workers' reputation, task location privacy, and the wide applications in cyber-physical systems.

\bibliographystyle{IEEEtran}
\bibliography{Geo_MOEA}

\vskip -2\baselineskip plus -1fil
\begin{IEEEbiography}[{\includegraphics[width=1in,height=1.25in,clip,keepaspectratio]{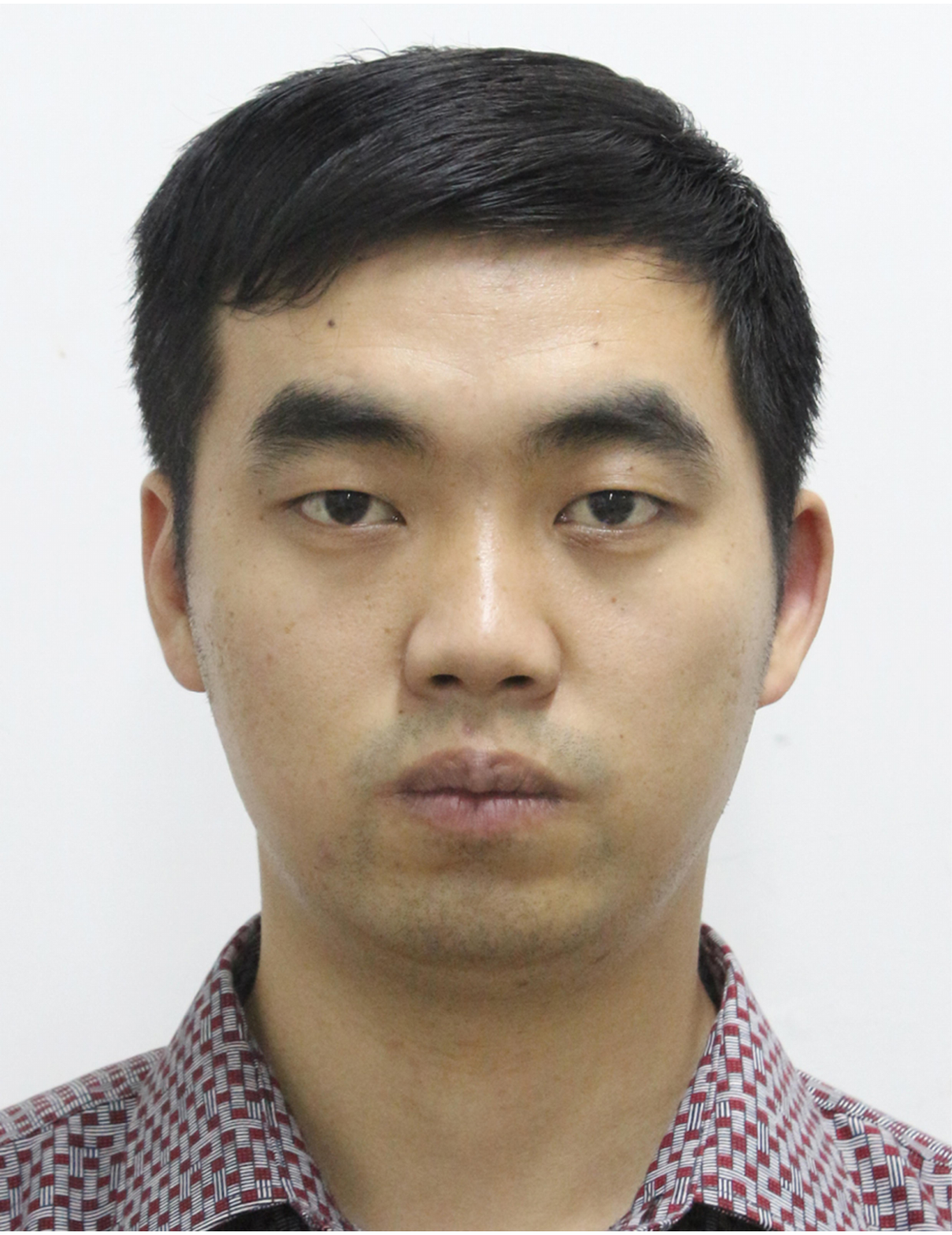}}]{Shun Zhang}
was born in Anhui Province, China, in  1982. He received his PhD degree
in applied mathematics from Beijing Normal University in 2012. He was a visiting scholar at Friedrich-Schiller-Universitat Jena, Germany, from 2014 to 2015. He is currently an associate professor at Anhui University. He has published more than 30 scientific papers. His research interests include privacy preservation, computational complexity, and quantum secure multi-party computation.
\end{IEEEbiography}

\vskip -2\baselineskip plus -1fil
\begin{IEEEbiography}[{\includegraphics[width=1in,height=1.25in,clip,keepaspectratio]{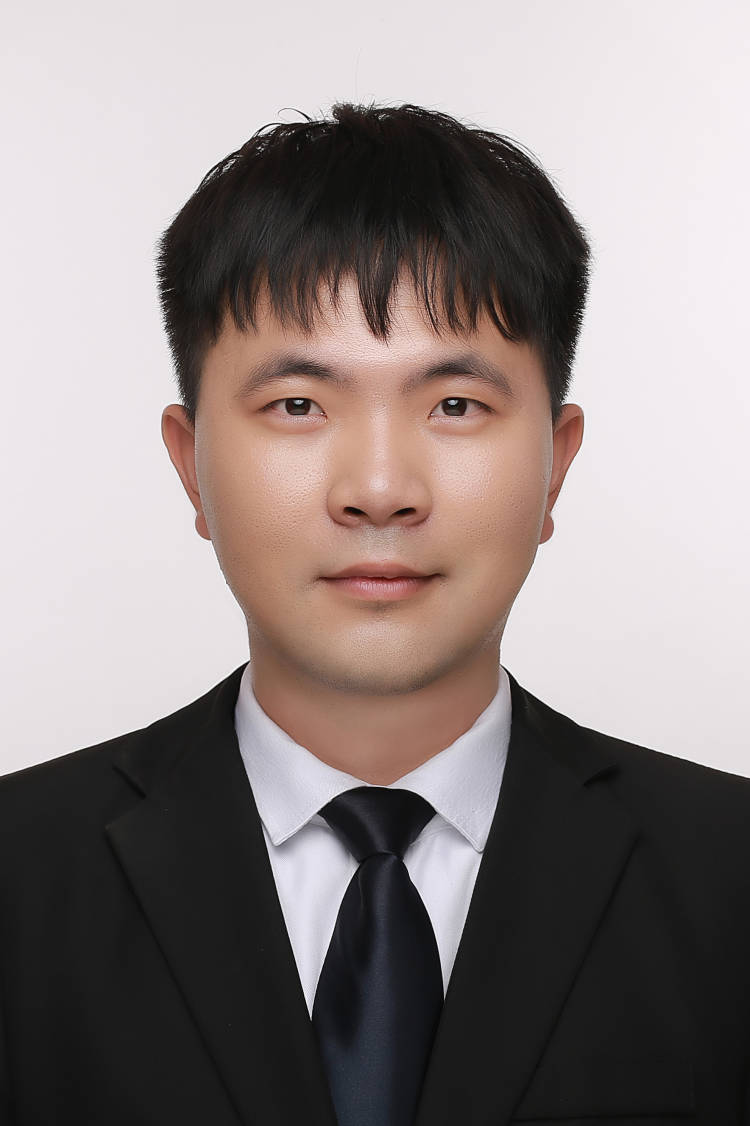}}]{Tao Zhang}
was born in Anhui Province, China, in 1998. He is currently a master student in Anhui University. His main research interests include differential privacy, location privacy.
\end{IEEEbiography}

\vskip -2\baselineskip plus -1fil
\begin{IEEEbiography}[{\includegraphics[width=1in,height=1.25in,clip,keepaspectratio]{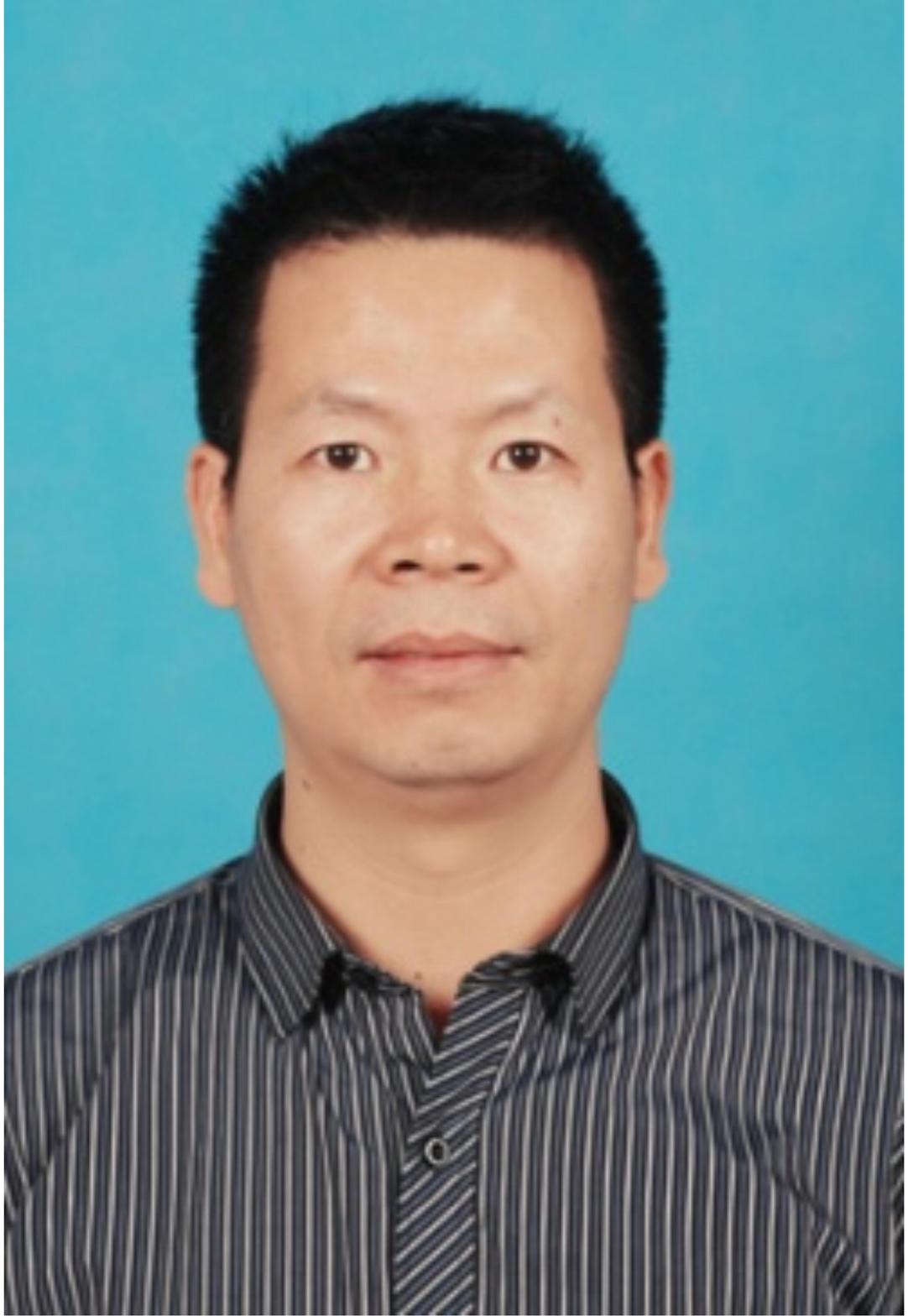}}]{Zhili Chen}
was born in Fujian Province, China, in 1980. He received his PhD degree in computer science from University of Science and Technology of China in 2009. He is currently a professor and Ph.D. supervisor at East China Normal University. He has published more than 40 papers. His main research interests include privacy preservation, secure multiparty computation, information hiding, spectrum auction and game theory in wireless communications.
\end{IEEEbiography}

\vskip -2\baselineskip plus -1fil
\begin{IEEEbiography}[{\includegraphics[width=1in,height=1.25in,clip,keepaspectratio]{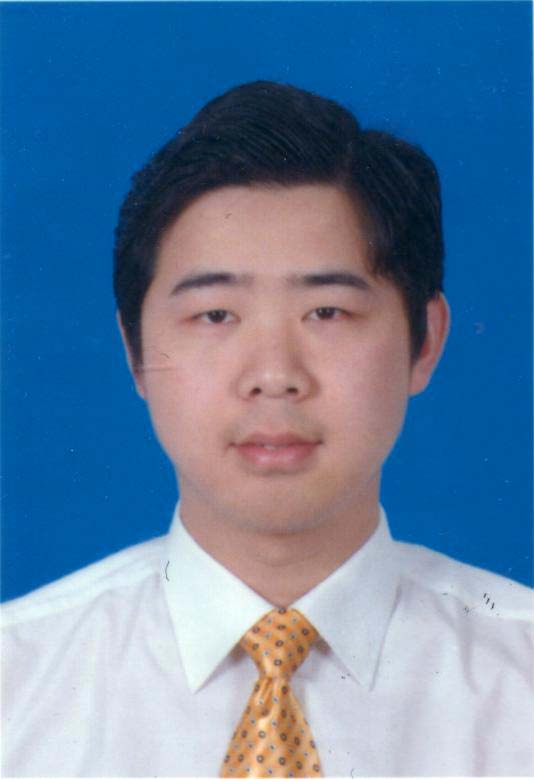}}]{N. Xiong}
is currently a Distinguished Professor at National Engineering Research Center for E-Learning, Central China Normal University (CCNU), Wuhan, Hu Bei Province, 430079, China. He is also with the Department of Computer Science, Georgia State University, Atlanta, GA 30302, USA. He received his PhD degree in School of Information Science, Japan Advanced Institute of Science and Technology (JAIST) on March 1, 2008. His research interests include Deep Learning, Reliable Networks, Software Engineering, and Big Data Analytics.\\
    Dr. Xiong works in CCNU for many years, and obtained many research funding and many industrial projects. He published over 600 journal paper with 200+ IEEE journal papers. He also creates a company about design and analysis for complex reliable software systems, and obtains over 10 patents.
\end{IEEEbiography}

\end{document}